\documentclass[pre,aps,superscriptaddress,twocolumn]{revtex4}

\usepackage[dvips]{graphics}
\usepackage{graphicx}
\usepackage{amsfonts}
\usepackage{amssymb}
\usepackage{amsmath}
\usepackage{subfigure}
\usepackage{color}

\begin{document}

\title{
Dynamics of Vortex Formation in Merging Bose-Einstein Condensate Fragments}

\author{R. Carretero-Gonz{\'a}lez}
\affiliation{
Nonlinear Dynamical Systems Group\footnote{URL: {\tt http://nlds.sdsu.edu/}},
Department of Mathematics and Statistics,
and Computational Science Research Center,
San Diego State University, San Diego CA, 92182-7720, USA}

\author{B.P. Anderson}
\affiliation{
College of Optical Sciences, University of Arizona,
Tucson, Arizona 85721, USA}

\author{P. G.\ Kevrekidis}
\affiliation{Department of Mathematics and Statistics, University of
Massachusetts, Amherst MA 01003-4515, USA}

\author{D.J. Frantzeskakis}
\affiliation{Department of Physics, University of Athens, Panepistimiopolis, Zografos,
Athens 15784, Greece}

\author{
C.N. Weiler$^2$
}

\begin{abstract}
We 
study the formation of vortices in a Bose-Einstein condensate (BEC) 
that has been prepared by allowing isolated and independent condensed fragments
to merge together. We focus on the experimental 
setup of Scherer {\it et al.} [Phys. Rev. Lett. {\bf 98}, 110402 (2007)], 
where three 
BECs are created in a magnetic trap that is segmented into three regions by a repulsive optical potential; 
the 
BECs merge together as the optical potential is removed. First, we study the two-dimensional case, 
in particular we examine
the effects of the relative phases of the different fragments 
and the removal rate of the optical potential on the vortex formation. 
We find that many vortices are created by instant removal of the optical potential 
regardless of relative phases, and that fewer vortices are created if the intensity 
of the optical potential is gradually ramped down and the condensed fragments gradually merge. 
In all cases, 
self-annihilation of vortices of opposite charge is observed. 
We also 
find that for sufficiently long barrier ramp times, the initial relative phases between 
the fragments leave a clear imprint on the resulting topological configuration. 
Finally, we study 
the 
three-dimensional system and 
the formation of vortex lines and vortex rings due to
the merger of the 
BEC fragments; our results illustrate
how the relevant vorticity is manifested for appropriate phase
differences, as well as how it may be masked by the planar projections
observed experimentally.

\end{abstract}

\date{Submitted to {\em Phys.~Rev.~A}, October 2007}

\maketitle

%%%%%%%%%%%%%%%%%%%%%%%%%%%%%%%%%%%%%%%%%%%%%%%%%%%%%%%%%%%%%%%%%%%%%%%%%%%%%%
\section{Introduction.}
%%%%%%%%%%%%%%%%%%%%%%%%%%%%%%%%%%%%%%%%%%%%%%%%%%%%%%%%%%%%%%%%%%%%%%%%%%%%%%

The formation, stability and dynamics of vortex-like structures has
been a long-standing theme of interest in many areas of physics, including classical fluid mechanics
\cite{batchelor}, superfluidity and superconductivity \cite{Donnelly,Tilley,zurek}, and cosmology \cite{kibble}.
Moreover, in the past decade, there has been a tremendous growth of 
excitement in this topic in the 
branches of atomic and optical physics. 
This has been propelled by considerable experimental and theoretical advances in the 
fields 
of nonlinear optics \cite{desyatnikov}
and Bose-Einstein condensates (BECs) in dilute alkali
vapors \cite{fetter,us} (see also Ref.~\cite{pismen}).

Focusing more specifically on the rapidly growing area
of BECs \cite{review}, one can recognize that the study of vortices
has been central to the relevant literature. In particular, as concerns the experimental efforts, 
the original 
observations of single \cite{cornell1,Holland} and multiple vortices \cite{dalib1} 
was soon followed by the 
realization of 
robust lattices of large numbers of vortices \cite{kett1}. Subsequent 
studies turned to higher-charged structures such as vortices of
topological charge $S=2$ and even $S=4$ \cite{kett2}
and illustrating dynamical instability of these
topological objects \cite{kett3}.
On the other hand, the abundance of experimental results
has stirred an intense theoretical interest
in the conditions under which such vortices and vortex lattices
would be robust and experimentally observable. Most often, 
vortex existence and stability issues were examined
in the framework of the standard parabolic confining potential
(typically produced by magnetic traps). In that framework, 
and in the two-dimensional (2D) case, 
vortices of charge $S=1$ were found to be stable, while vortices of higher charge 
($S=2,3$) were shown to be potentially 
unstable \cite{pu} 
(depending on the atom numbers). 
Later, similar results were found for vortices of $S=4$ 
\cite{kawa}, while the availability of more 
substantial computational resources has more recently led 
to similar conclusions in the fully three-dimensional (3D) case 
\cite{jukka1,jukka2}. The 
studies of Refs.~\cite{motto} and \cite{ueda} examined the various
scenarios of break-up of higher-charge vortices during dynamical 
evolution simulations for repulsive and attractive interactions 
respectively. Furthermore, Ref.~\cite{carr} considered such vortices riding
on the background of not just the ground state, but also of higher,
ring-like, excited states of the system.
It should also be mentioned that these 
advancements have motivated the development of mathematically rigorous tools in order 
to study the spectrum of such vortex modes. Such methods involve
the use of the Evans function 
\cite{kollar}, or the use of the index
theorem evaluating the number of potentially unstable eigendirections
\cite{todd}.

Although the existence and stability of fundamental and higher charge 
vortices has been examined extensively as indicated 
above, the {\it formation} of such vortex structures is far 
less studied. In particular, while seminal interference experiments 
(demonstrating that BECs are coherent matter waves) were reported as early as a decade ago \cite{science}, 
the role of interference between BECs in vortex generation was experimentally studied
only recently 
\cite{bpa_prl}. This work proposed and examined the formation of vortices resulting 
from the interference and controlled merging of three condensed fragments, where
the fragments were essentially independent BECs separated by
an optical potential barrier. Such a process has close ties to elements 
of topological defect formation in phase transitions, 
as proposed by Kibble \cite{kibble} and Zurek \cite{zurek}.
Interestingly, the experimental work was almost concurrent with a theoretical study 
investigating a simpler elongated barrier separating two independent BECs \cite{ric}; 
in the latter 
setting, the interference forms a dark soliton whose bending and 
subsequent breakup due to the manifestation of the transverse modulational instability also
result in vortices.

In the present work, 
we expand on these considerations and 
study in detail, by means of systematic numerical simulations, 
the formation of vortices 
in a setting closely matching the one of the experiment in Ref.~\cite{bpa_prl}.
Our purpose is to get a deeper insight into this interference-induced vortex formation mechanism, 
investigating fundamental features, such as the number and lifetime of ensuing vortices. 
This is done upon studying in detail the parametric dependences influencing the relevant experimental 
observations. 
Specifically, we quantify the above mentioned features as functions of the elimination/ramping-down 
time of the optical barrier between the fragments, or the initial relative phases between the original 
independent fragments. 
Our investigation chiefly refers to a 2D setting, but we
also illustrate how the 
results are generalized in the pertinent 3D case. 
Notice that our considerations are motivated  
not only by their direct bearing on the experiments of Ref.~\cite{bpa_prl}, but also
by their relevance 
to studies of spontaneous symmetry breaking during phase transitions \cite{zurek,kibble,sanders}.

Our presentation is structured as follows. In Sec.~\ref{SEC:setup}, 
we briefly summarize the setup of our computational experiments.
In Sec.~\ref{SEC:num}, we study the interference 
of three BEC fragments in the 2D and 3D setup; 
special attention is payed at the role of the relative phases between 
the different condensates, and the ramp-down time of the
laser sheet barrier responsible for separating the three
fragments. In Sec.~\ref{SEC:expt}, 
we briefly 
comment on the 
relation between numerical and experimental results.
Finally, in Sec.~\ref{SEC:conclusions}, we
summarize our findings, 
present 
our conclusions, 
and discuss some possible extensions of this work.

\section{Setup}
\label{SEC:setup}

\begin{figure}[t]
\begin{center}
\includegraphics[width=8.5cm]{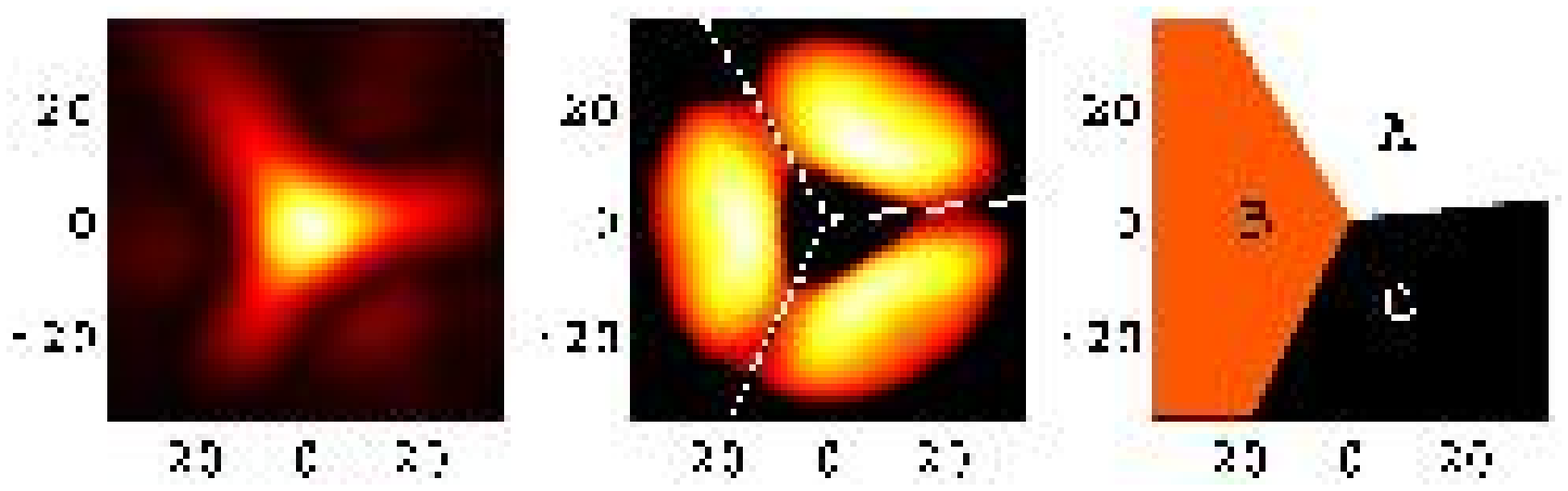}\\[2.0ex]
\includegraphics[width=8.5cm]{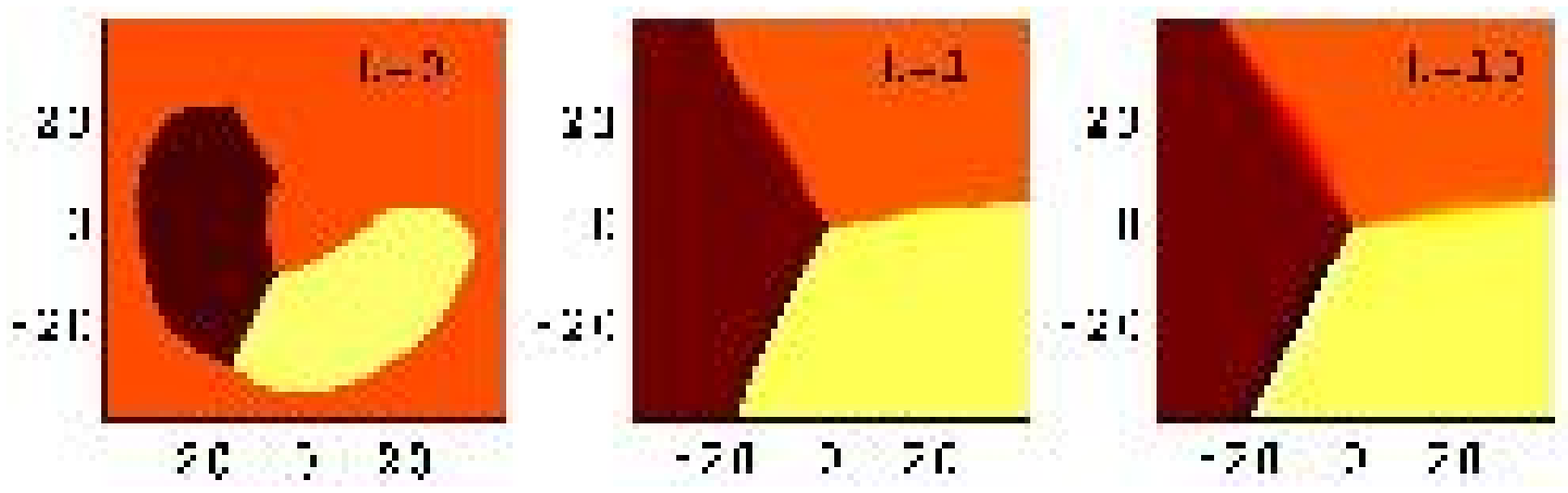}
\end{center}
\vskip-0.4cm
\caption{(Color online)
Top-left: intensity profile of the optical potential responsible for segmenting the
potential well into three local minima.
Top-center: Thomas-Fermi approximation used as an initial condition
for our relaxation method to obtain the ground state of the system.
Top-right: regions A, B, and C with respective phases
$\phi_1$, $\phi_2$, and $\phi_3$.
The bottom row of panels depicts the evolution of the phase
during our relaxation (imaginary time relaxation) towards
the initial steady state with different phases for the
fragments. This example shows the phase for the case
$\phi_k=2\pi k/3$ at the times indicated. In all panels, the field of view is approximately 70 $\mu$m per side.
The axis numbers indicate $x$ and $y$ coordinates relative to the center of the unsegmented harmonic trap.}
\label{pot}
\end{figure}

We consider a BEC at a temperature close to zero, where quantum or thermal fluctuations are negligible 
(note that 
finite temperature effects
are briefly 
discussed at the end of Sec.~\ref{SEC:num}).
This system can accurately be described by a mean-field theoretical model, namely 
the Gross-Pitaevskii equation (GPE) \cite{review}:
\begin{eqnarray}
i \hbar\frac{\partial \psi}{\partial t} =
\left[ -\frac{\hbar^2}{2m} \nabla^2 + V({\bf r};t) + g |\psi|^2 \right] \psi,
\label{gpe}
\end{eqnarray}
where $\psi=\psi({\bf r},t)$ is the condensate wavefunction
(with $n({\bf r},t)\equiv|\psi({\bf r},t)|^2$ 
being the atomic density of the condensate),
$m$ is the atomic mass, the coupling constant $g=4\pi\hbar^2 a_s/m$ 
measures the strength of inter-atomic interactions and $a_s$ is the $s$-wave
scattering length.
The potential $V({\bf r};t)$ in the GPE is taken to be of the form, 
\begin{equation}
V({\bf r};t)=V_{\rm MT}({\bf r})+\alpha(t)\,V_{\rm L}({\bf r}), 
\label{poteq}
\end{equation}
where the two components in the right-hand side of Eq.~(\ref{poteq}) are 
a harmonic magnetic trap, 
$V_{\rm MT}({\bf r})=\frac{1}{2}m(\omega_x^2 x^2 +\omega_y^2 y^2 +\omega_z^2 z^2)$, 
with trapping frequencies $\omega_x = \omega_y = 2\pi
\times 7.4$ Hz and $\omega_z = 2\pi
\times 14.1$ Hz, and
the three-armed time-dependent optical barrier, $\alpha(t) V_{\rm L}({\bf r})$,  
used in the experiments of Ref.~\cite{bpa_prl}. 
This three-armed potential induces a separation of the
ground state of the
condensate into three different fragments 
(see top-left and top-center panels in Fig.~\ref{pot} and Fig.~\ref{fig_expt}(a)). 
%
%laser intensity profile measured in the experiments of Ref.~\cite{bpa_prl}
%
Note that the function $\alpha(t)$ in 
Eq.~(\ref{poteq})
describes the ramping down of the optical barrier.
The maximum initial barrier energy for the potential
is taken to be $\alpha_0\equiv\alpha(0)=26\,k_B$ nK \cite{bpa_prl}, where $k_B$ is Boltzmann's constant.

In the numerical simulations, 
the ground state of the system is obtained by relaxation
(imaginary time integration) initiated with the Thomas-Fermi (TF)
approximation (see top-center panel in Fig.~\ref{pot})
\begin{eqnarray}
\psi({\bf r},0) =\sqrt{\max\{0,\mu-V({\bf r})\}} \times \phi({\bf r}),
\label{TF}
\end{eqnarray}
where $\mu=8\,k_B$ nK \cite{bpa_prl} 
is the chemical potential, and $\phi({\bf r})$ contains the
chosen phase of the different wells separated into the three regions A, B and C
depicted in the top-right panel of Fig.~\ref{pot}.
Let us denote by $\phi_1$, $\phi_2$, and $\phi_3$ the initial phases in
regions A, B and C, respectively. 
It is important to note that while in the experiments of Ref.~\cite{bpa_prl}
the different fragments have uncorrelated (random) phases, in our
numerical simulations we are able to control their initial phases
and, more importantly, their relative phases.
Our numerical experiments,
emulating the experimental sequence of Ref.~\cite{bpa_prl}, are
performed with the output of imaginary time relaxation
used as initial condition for the full dynamics of Eq.~(\ref{gpe}).
The bottom row of Fig.~\ref{pot} depicts the evolution of the
phase for the case $\phi_k=2\pi k/3$ during 
relaxation.
As can be observed from the figure, 
the initial ``seed'' ($t=0$) starts with
sharp phase boundaries around the localized fragments. As the
relaxation procedure evolves, the phase boundaries become smoother
and adjust to the boundaries for the top-right panel of the figure.
The phase profile seems to settle after 10~ms of imaginary time
relaxation. In our simulations, 50~ms of imaginary time 
relaxation was used to ensure proper convergence to the steady-state solution.

\begin{figure}[ht]
\begin{center}
\includegraphics[width=8.5cm]{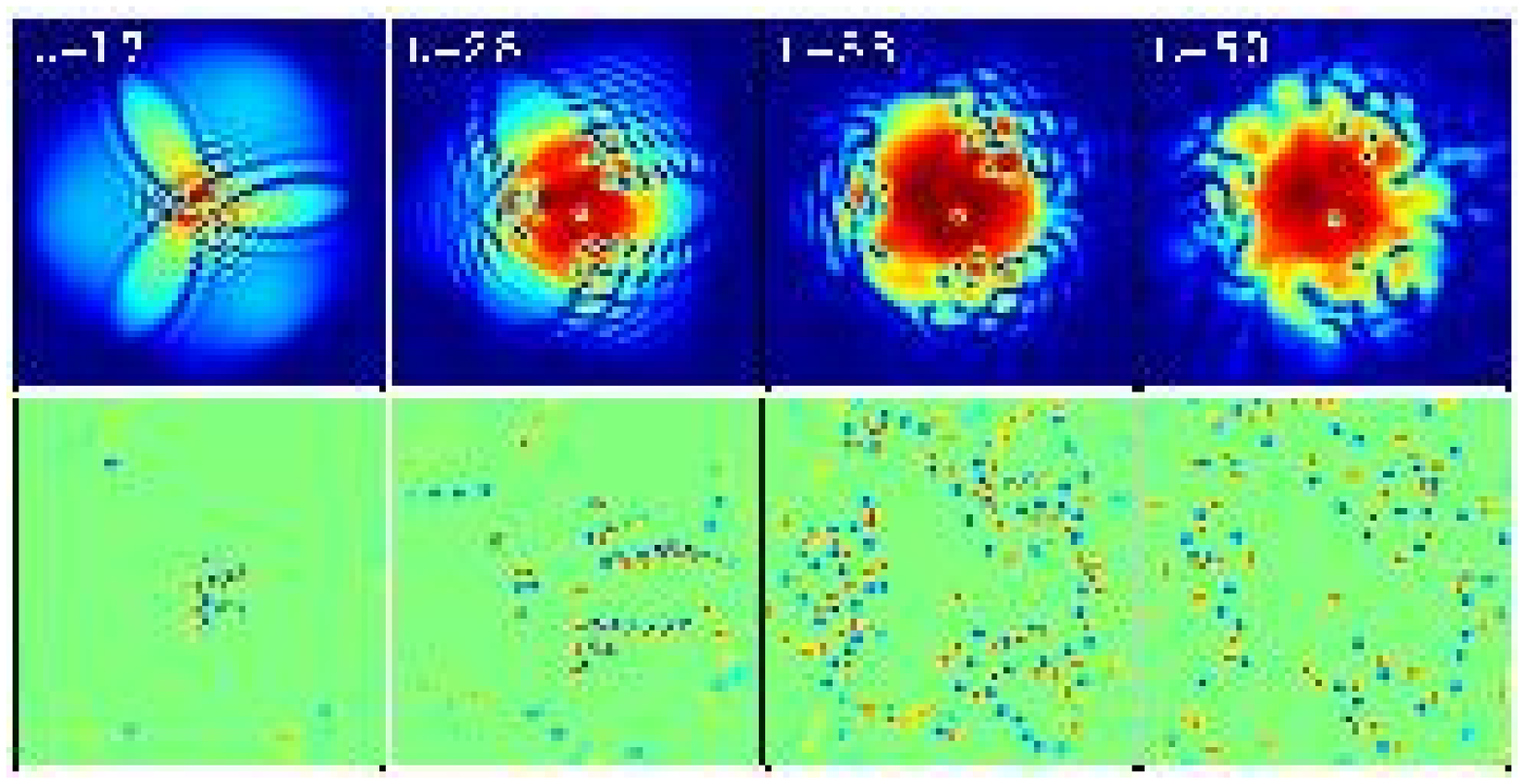}
\includegraphics[width=8.5cm]{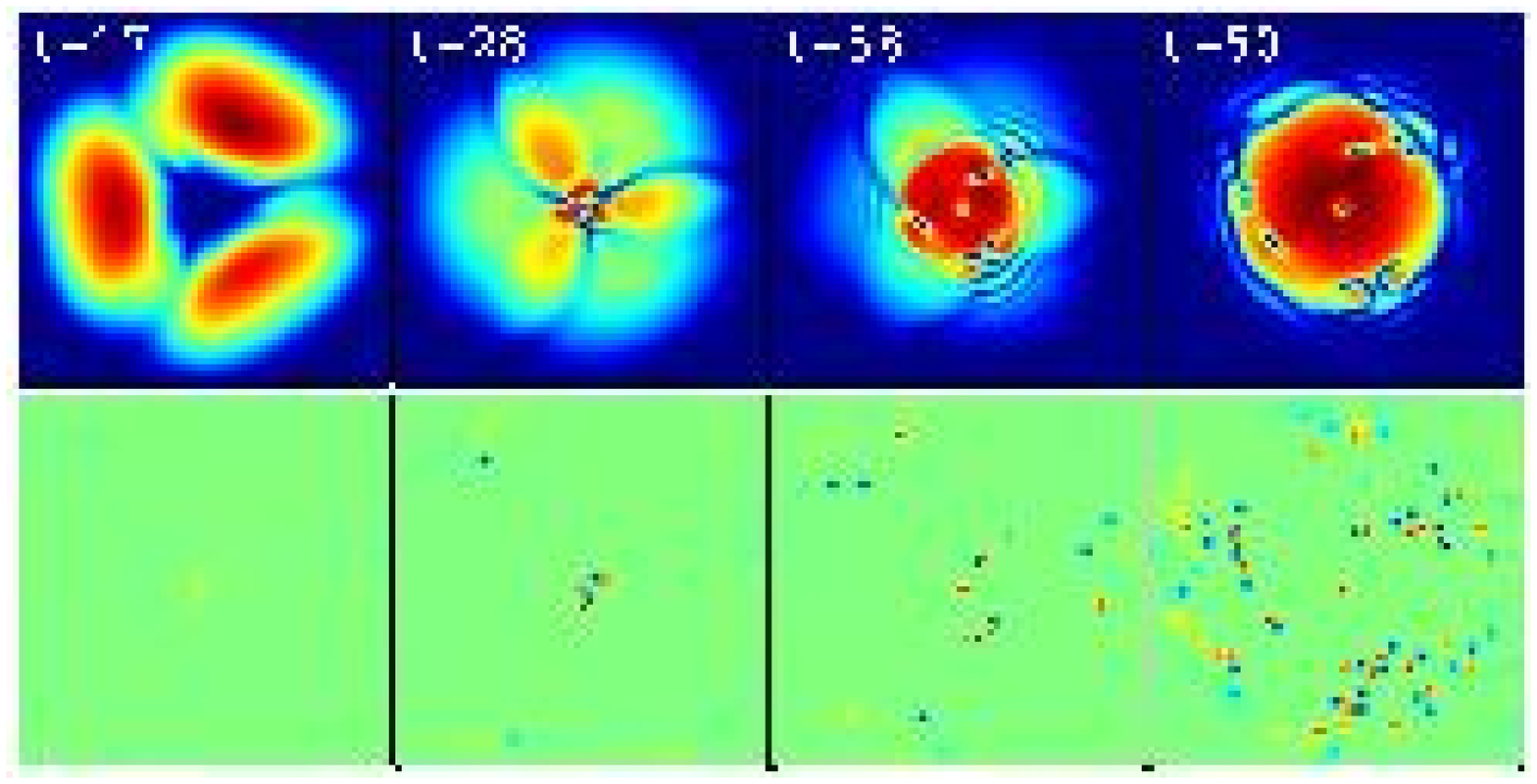}
\includegraphics[width=8.5cm]{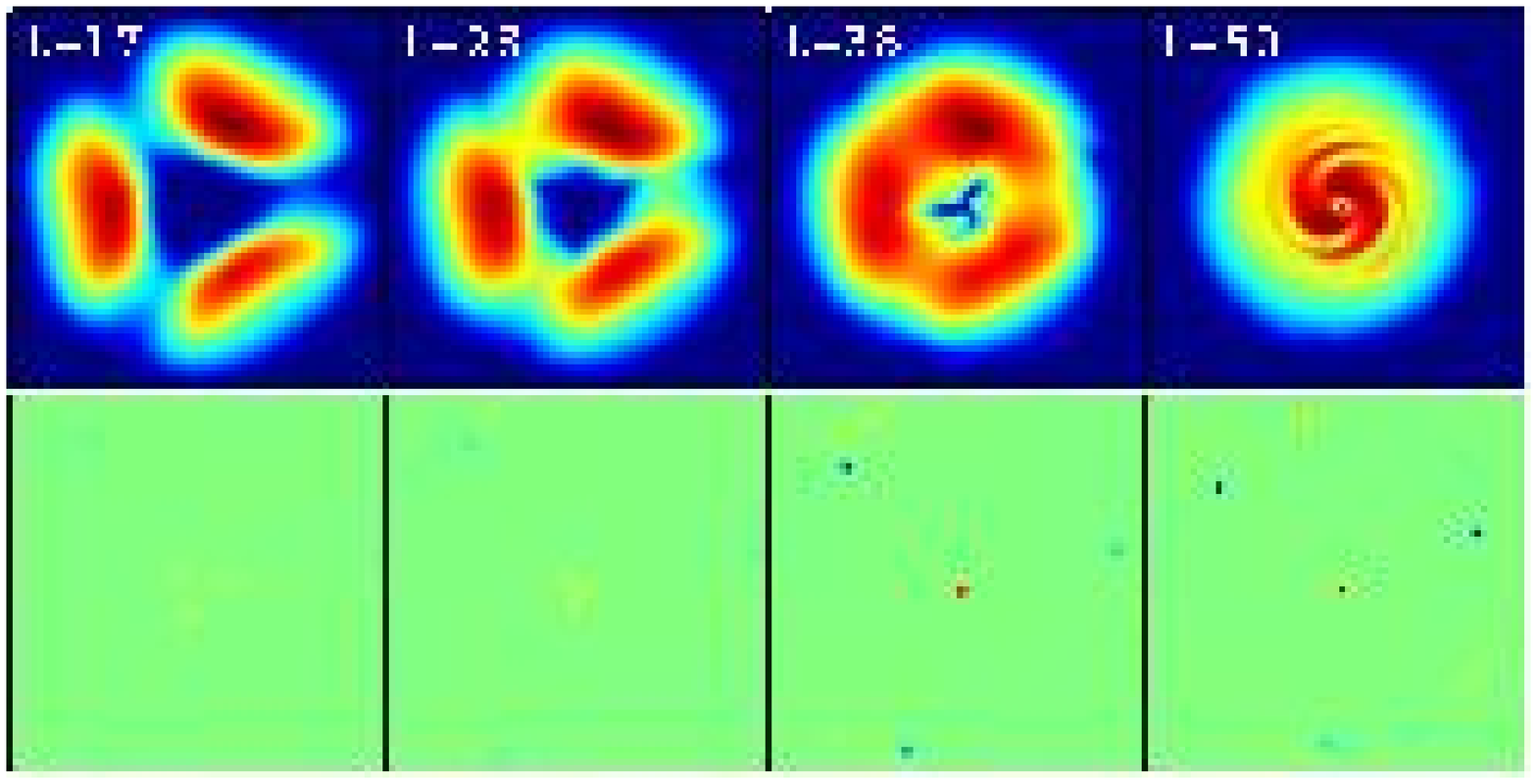}
\end{center}
\vskip-0.4cm
\caption{(Color online)
Evolution of the 2D condensate density (respective top series
of panels) and vorticity (respective bottom series of panels) for
three different ramp-down times of the optical potential barriers. 
From top to bottom, the three sequences correspond
to $t_b=0$~ms (first and
second row), $t_b=25$ ms (third and fourth row),
and $t_b=50$ ms (fifth and sixth row).
The times are indicated in the panels in ms
and the field of view is approximately 70 $\mu$m per side.
For the initial conditions in all cases, the different condensed fragments have relative
phases of $2\pi/3$, namely, $\phi_k=2\pi k /3$ ($k \in \{1,2,3\}$).
}
\label{bpa2d_allu}
\end{figure}

%%%%%%%%%%%%%%%%%%%%%%%%%%%%%%%%%%%%%%%%%%%%%%%%%%%%%%%%%%%%%%%%%%%%%%%%%%%%%%
\section{Numerics. \label{SEC:num}}
%%%%%%%%%%%%%%%%%%%%%%%%%%%%%%%%%%%%%%%%%%%%%%%%%%%%%%%%%%%%%%%%%%%%%%%%%%%%%%

%%%%%%%%%%%%%%%%%%%%%%%%%%%%%%%%%%%%%%%%%%%%%%%%%%%%%%%%%%%%%%%%%%%%%%%%%%%%%%
\subsection{Two-dimensional BECs.}
%%%%%%%%%%%%%%%%%%%%%%%%%%%%%%%%%%%%%%%%%%%%%%%%%%%%%%%%%%%%%%%%%%%%%%%%%%%%%%

For 
the 2D rendering of the experiment of 
Ref.~\cite{bpa_prl}, we restrict our system to the $(x,y)$
coordinates and we use the {\em same} chemical potential
as in Ref.~\cite{bpa_prl}.
First we explore the effect of 
the ramp-down time of the potential barrier on the formation of vortices through the merging
of the different fragments of the condensate. For this purpose
we use a linear ramp:
\begin{equation}
\alpha(t) = \max\left\{ \frac{\alpha_0}{t_b}(t_b-t), \,0\right\},
\label{alpha}
\end{equation}
where $\alpha_0\equiv\alpha(0)$ is the maximum barrier energy as defined
above and $t_b$ is the ramping time (in ms) of the barrier; 
note that a similar ramp was used in the experiments of Ref.~\cite{bpa_prl}.

We monitor the formation of vortices, as well as the overall vorticity
of the system, using various diagnostics. These are based on the
corresponding fluid velocity of the superfluid 
given by \cite{Jackson:prl:98},
\begin{equation}
{\mathbf v}_s=-\frac{i \hbar}{2m} \frac{\psi^*\nabla \psi - \psi\nabla \psi^*}{|\psi|^2},
\label{fluid_vel}
\end{equation}
where $(\cdot)^*$ stands for complex conjugation. The fluid vorticity
is then defined 
as $\mbox{\boldmath$\omega$}=\nabla\times{\mathbf v}_s$.
The results for the merger of the three BEC fragments
with relative phases $\phi_k=2\pi k/3$, $k \in \{1,2,3\}$, for different
merging times are depicted in Fig.~\ref{bpa2d_allu}.
As can be seen from the figure, the number of
vortex pairs nucleated by the merger is extremely
sensitive to the ramping time $t_b$. Shorter
ramping times give rise to an extremely rich vorticity pattern
as the fragments merge [see, for example, the top two rows in
Fig.~\ref{bpa2d_allu} corresponding to the instantaneous
($t_b=0$) removal of the barrier], including the appearance of structures resembling vortex streets.
However, for longer ramp times, i.e.~slower ramping, 
just a handful of vortices are nucleated. In fact, 
for $t_b>100$ ms (results not shown here), 
the only vortex that is nucleated is the central one.
It is evident that, independently of the ramping time,
the central vortex is always formed for this pair of
relative phases between the fragments. This vortex is
the consequence of the intrinsic vorticity present in
the initial condition where the three fragments have
been phase imprinted with a total of a $2\pi$ 
phase gain about the condensate center.

\begin{figure}[ht]
\begin{center}
\includegraphics[width=8.5cm]{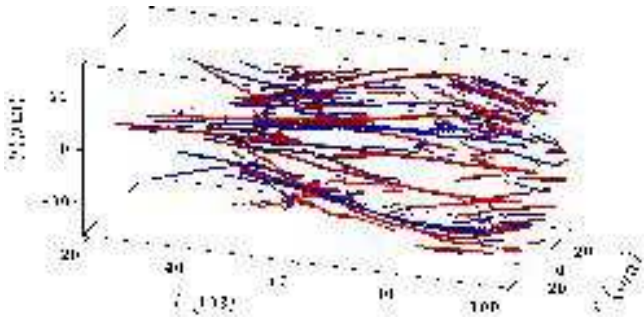}
\end{center}
\vskip-0.4cm
\caption{(Color online)
Evolution of the vortex structures in the 2D condensate density
for $t_b=25$ ms and $\phi_k=2\pi k/3$ (cf.~middle rows in
Fig.~\ref{bpa2d_allu}).
}
\label{filaments_fig}
\end{figure}

In Fig.~\ref{filaments_fig} we present a spatio-temporal
rendering of the vortex formation for the middle row
example of Fig.~\ref{bpa2d_allu} (i.e., phases given
by $\phi_k=2 \pi k/3$ and a ramp down time of $t_b=25$ ms).
In the figure we depict a space-time contour plot of the
vorticity where blue/red contours correspond to 
negatively/positively charged vortices, respectively.
The figure clearly shows 
the formation of pairs of vortices with opposite charge, some of which
self-annihilate at later times, while others oscillate
together with the cloud: upon formation, they 
expand to the rims of the cloud and are subsequently
reflected from its outskirts and contract anew, following
the density profile oscillations resulting from the merging process.

\begin{figure}[t]
\begin{center}
\includegraphics[width=8.5cm,height=3cm]{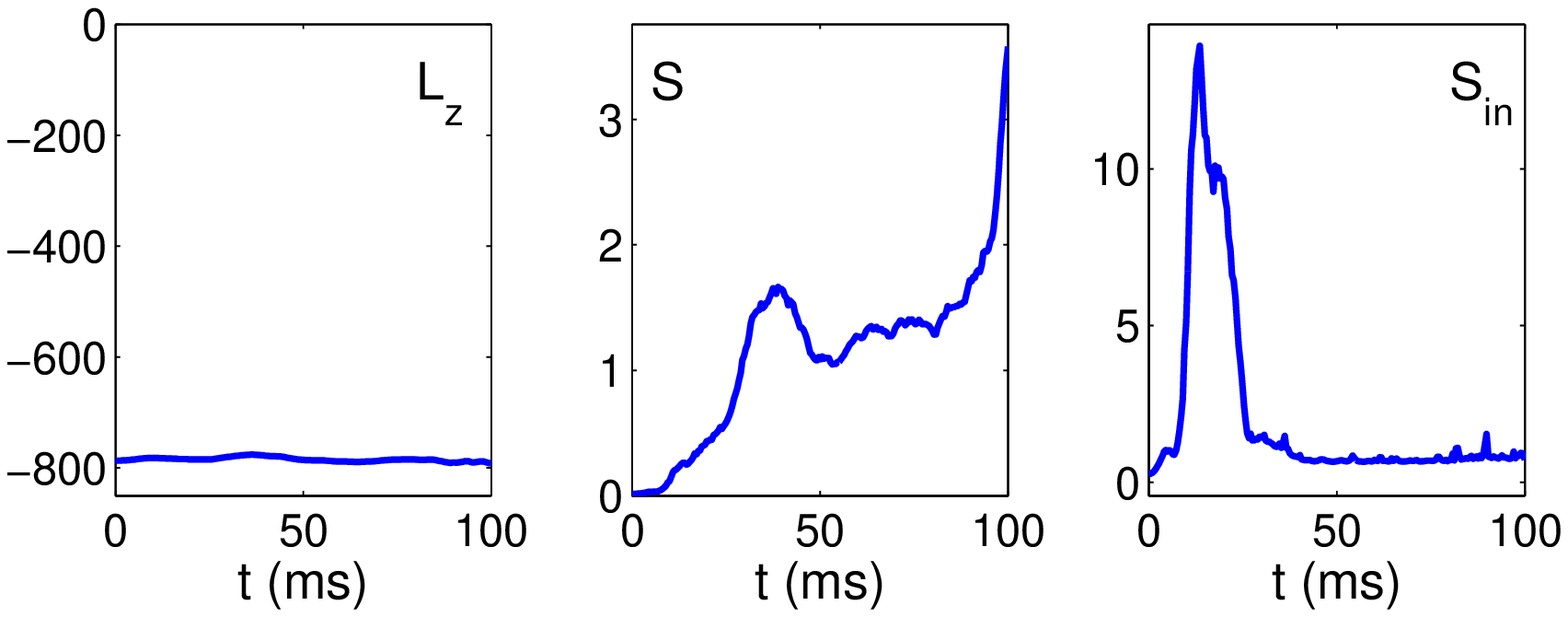}
\includegraphics[width=8.5cm,height=3cm]{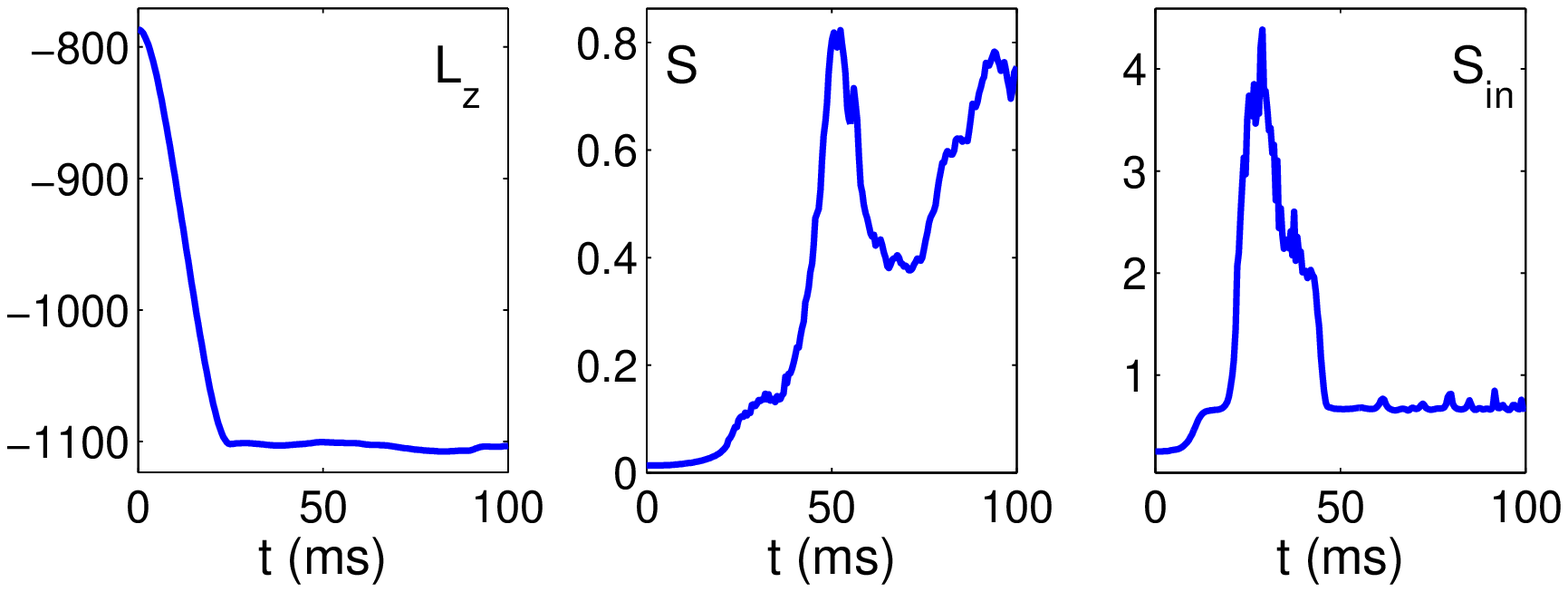}
\includegraphics[width=8.5cm,height=3cm]{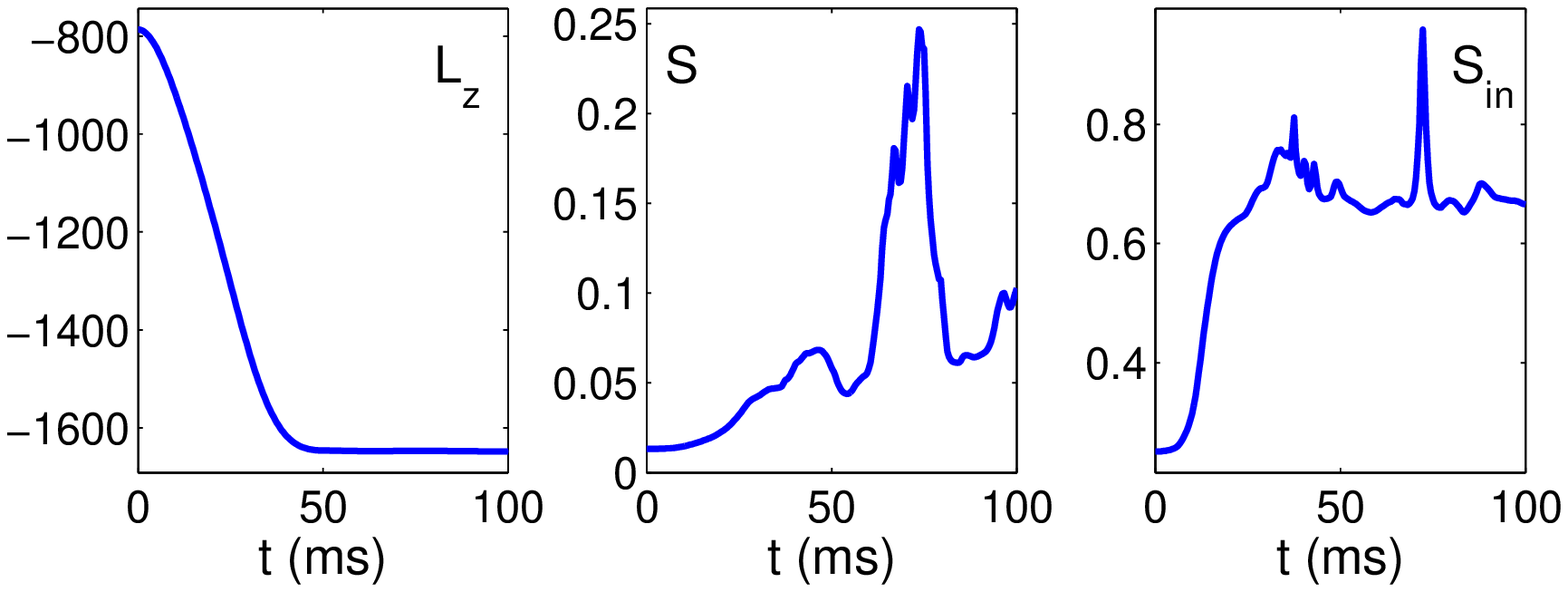}
\end{center}
\vskip-0.4cm
\caption{(Color online)
Vorticity indicators for the cases presented in Fig.~\ref{bpa2d_allu}
(namely, $t_b=0$ ms, $t_b=25$ ms, and $t_b=50$ ms, from top to bottom)
with $\phi_k=2\pi k/3$. The left panels correspond to the total
angular momentum [Eq.~(\ref{Lz_eq})] normalized by $\hbar$,
while the middle and right
panels correspond to the total fluid velocity
[Eq.~(\ref{S_eq})] for the
whole cloud (middle) and the central portion (see text) of the cloud (right).
}
\label{bpa2d_Lz}
\end{figure}

\begin{figure}[ht]
\begin{center}
\includegraphics[width=8.5cm]{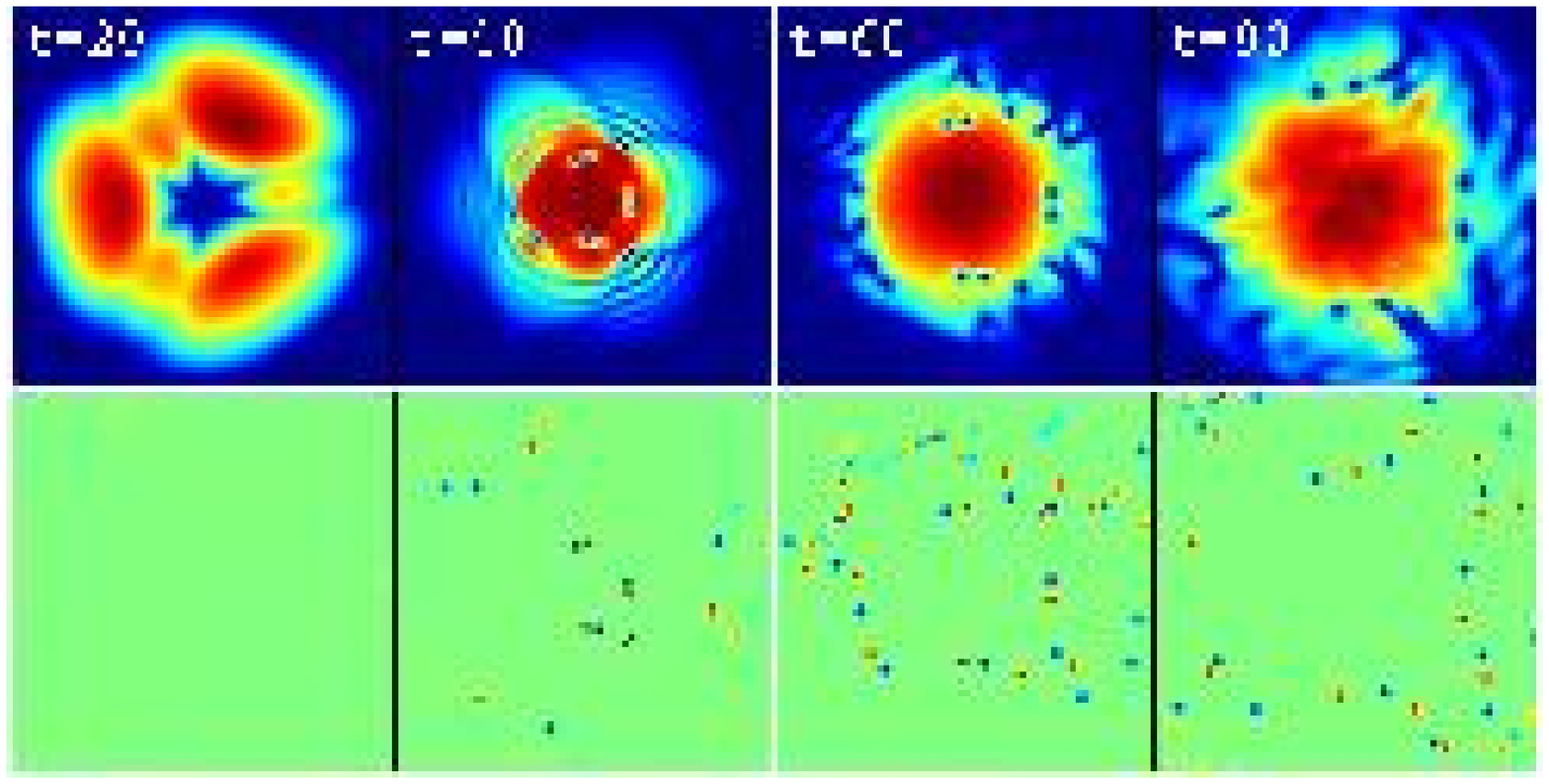}
\includegraphics[width=8.5cm]{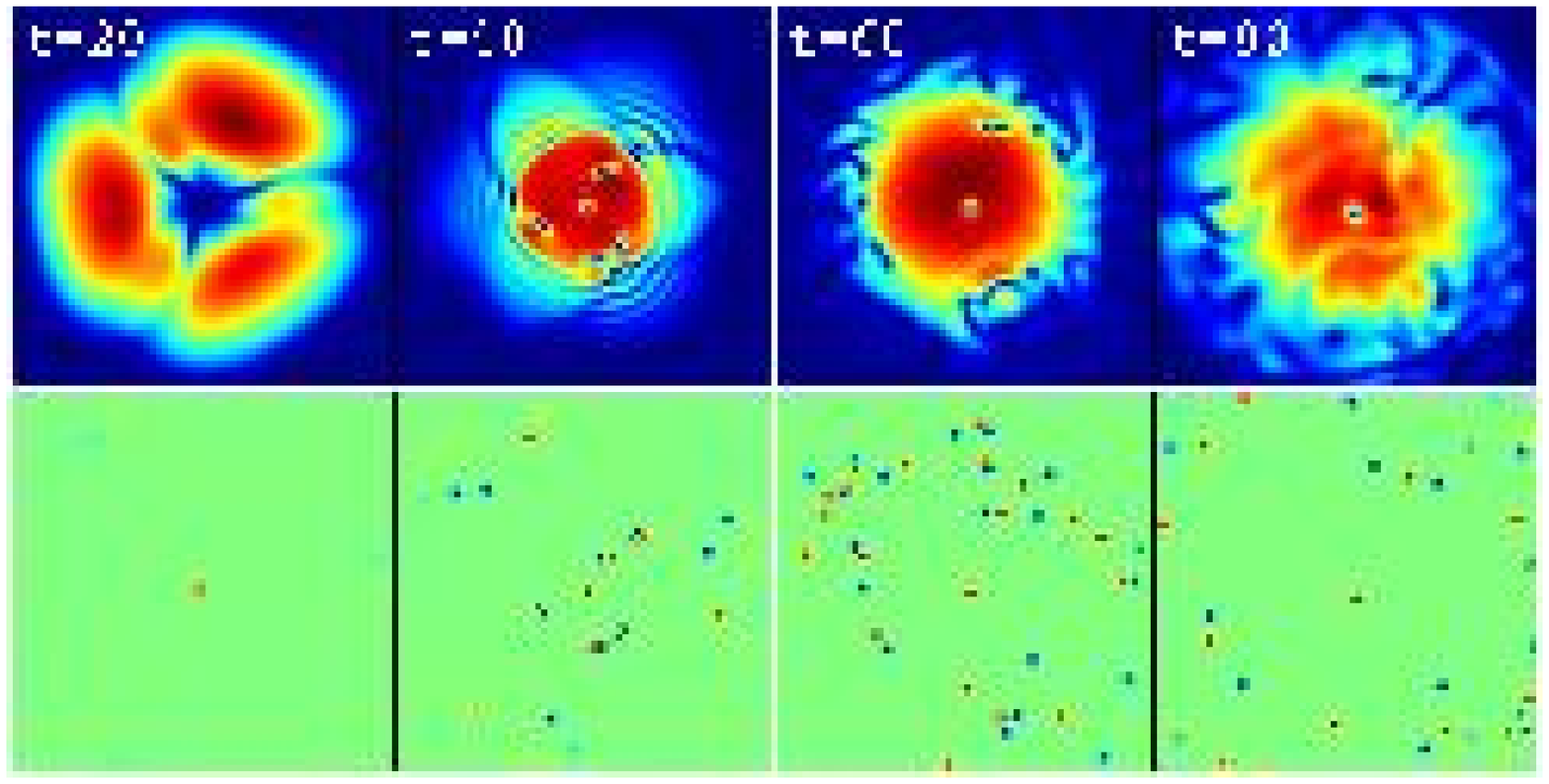}
\includegraphics[width=8.5cm]{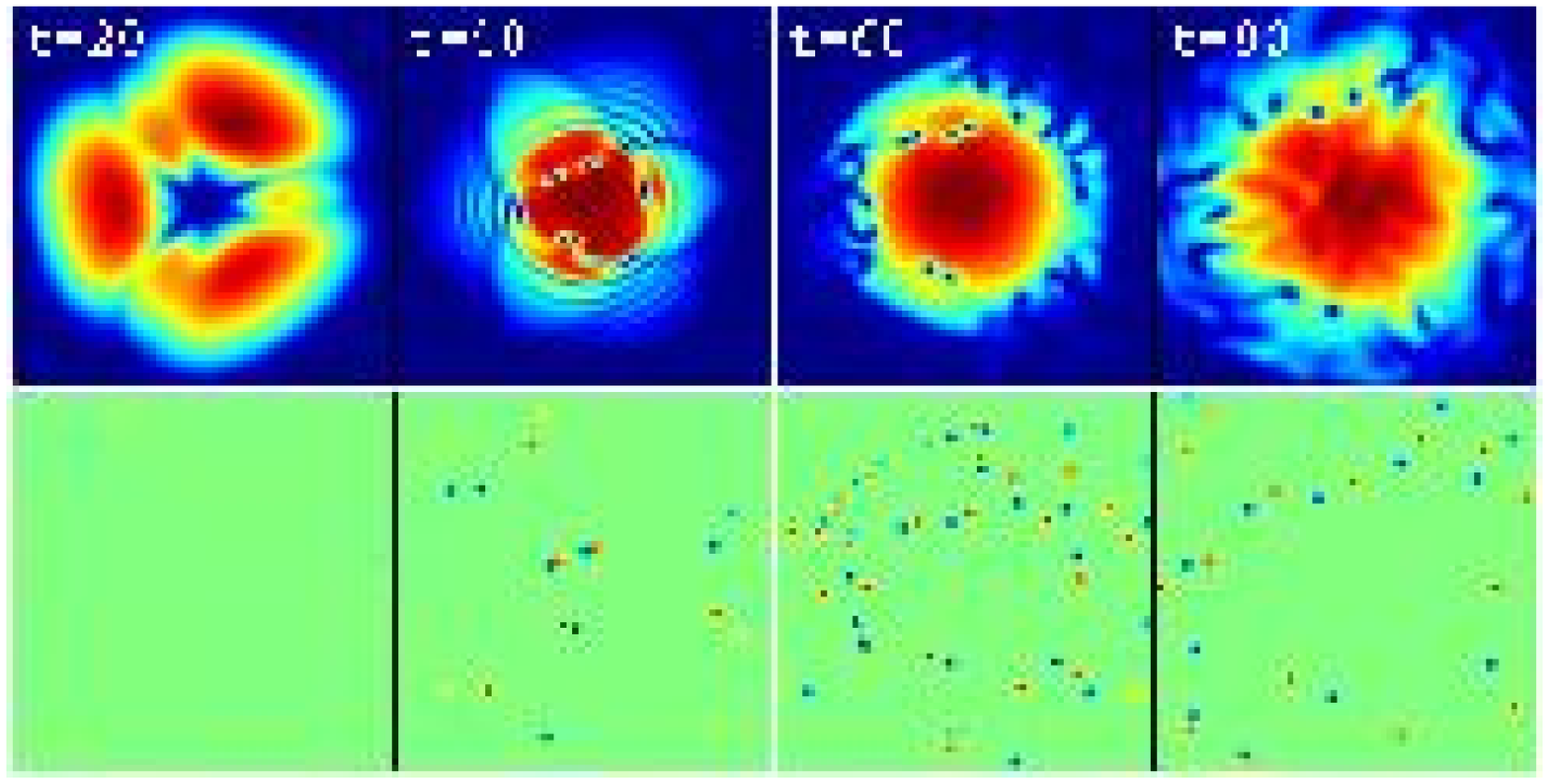}
\end{center}
\vskip-0.4cm
\caption{(Color online)
Same as in Fig.~\ref{bpa2d_allu} for
three different relative phases and a ramp down of $t_b=25$ ms.
The top two, middle two and bottom two series correspond,
respectively, to:
a) $\phi_k=0$,
b) $\phi_1=0$,  $\phi_2=2\pi/3$, and $\phi_3=4\pi/3$, and
c) $\phi_1=0$,  $\phi_2=\pi/3$, and $\phi_3=2\pi/3$.
}
\label{bpa2d_allu_phases}
\end{figure}

\begin{figure}[ht]
\begin{center}
\includegraphics[width=8.5cm,height=3cm]{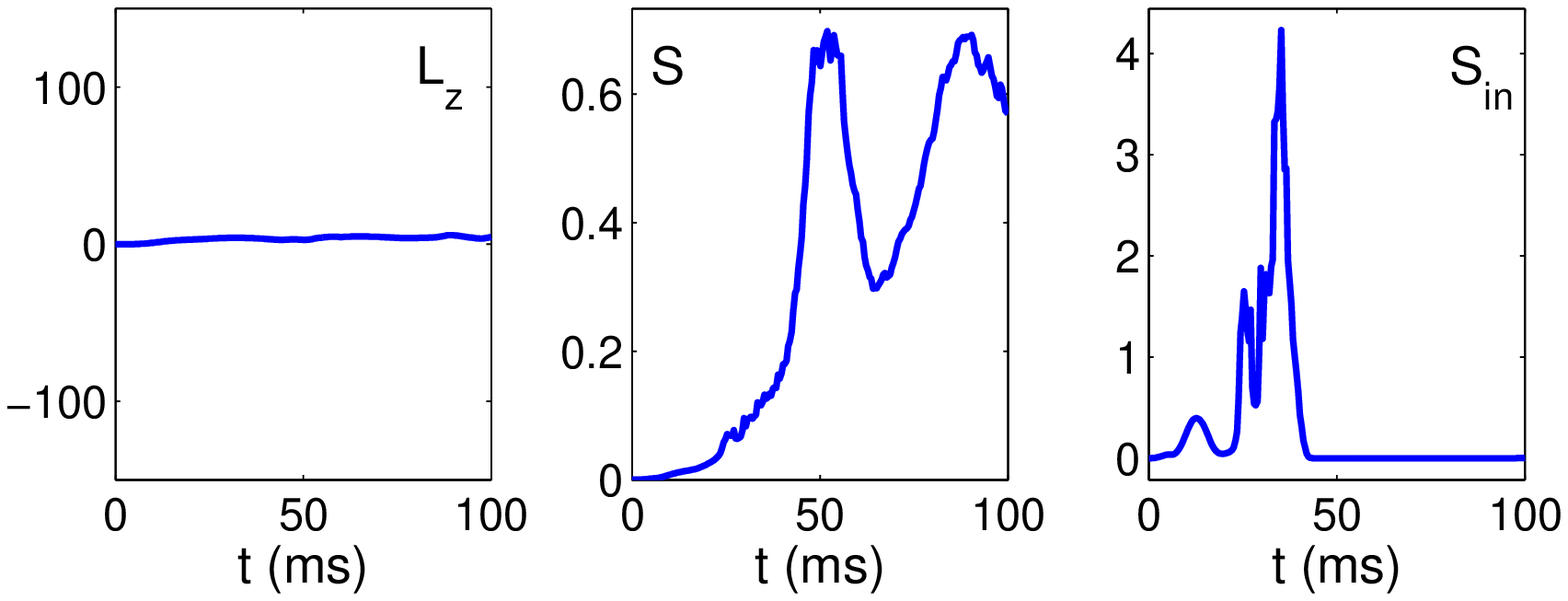}\\
\hskip-0.2cm
\includegraphics[width=8.65cm,height=3cm]{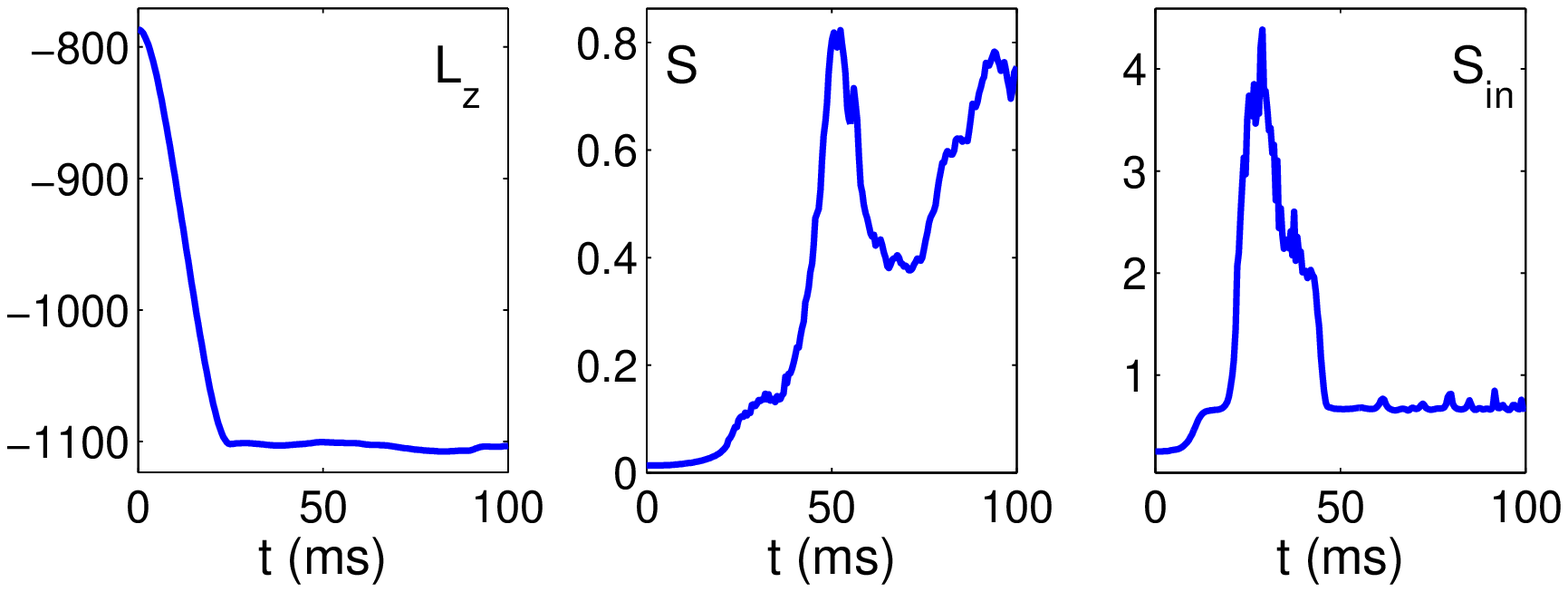}\\
\includegraphics[width=8.5cm,height=3cm]{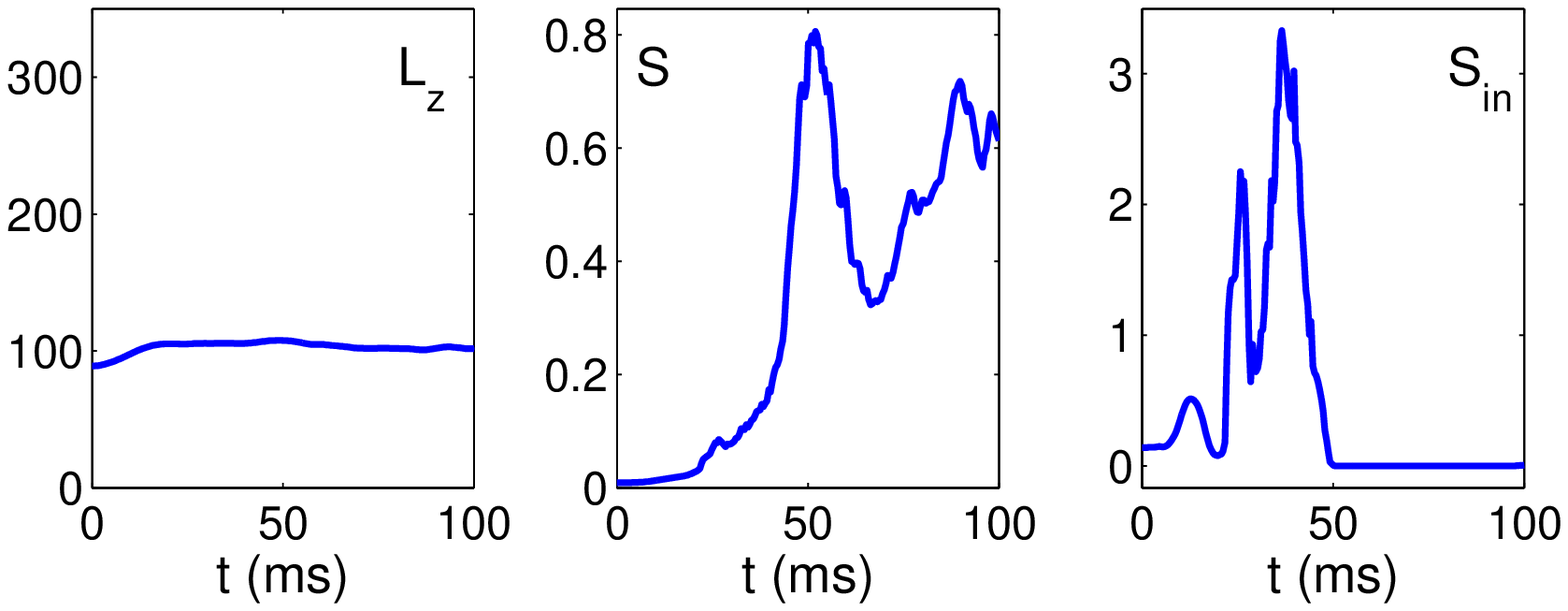}\\
\end{center}
\vskip-0.4cm
\caption{(Color online)
Vorticity indicators for the cases depicted in
Fig.~\ref{bpa2d_allu_phases} (the different panels
are presented in the same manner as in Fig.~\ref{bpa2d_Lz}).
}
\label{bpa2d_Lz_phases}
\end{figure}

In order to measure the vorticity generated
during the merger of the condensate fragments at
any given time, we compute the expectation value
of the $z$-component of the angular momentum of the BEC
\begin{equation}
L_z = i\hbar\, \int \psi^*\, \partial_\theta\psi \,d{\bf r}
\equiv \left(\psi, (\hat{\bf r} \times \hat{\bf p})_z \psi \right),
\label{Lz_eq}
\end{equation}
where $\hat{\bf r}$ and $\hat{\bf p}$ 
denote the corresponding
position and momentum operators, the subscript $z$ indicates the
component of the corresponding cross-product and $(\cdot,\cdot)$ indicates
the complex inner product defined from $C \times C \rightarrow C$.
Using Eq.~(\ref{Lz_eq}) and Eq.~(\ref{gpe}), 
we can evaluate 
the time derivative of this expectation value:  
\begin{eqnarray}
\frac{d L_z}{dt}= \frac{\alpha(t)}{\hbar} \int |\psi|^2
(\hat{\bf r} \times \hat{\bf p})_z V_L({\bf r}) d {\bf r}.
\label{lz2}
\end{eqnarray}
In obtaining this result, we have assumed the isotropy of
$V_{\rm MT}({\bf r})$ in the $(x,y)$ plane. This result also has
some important consequences including the fact that during the
ramp down of the optical barrier, we should not expect
the angular momentum of the BEC to be conserved, while we should expect
such a conservation to appear once $\alpha(t)=0$ 
or, more generally, when the full potential is azimuthally isotropic in the $(x,y)$ plane
(cf.~Figs.~\ref{bpa2d_Lz} and \ref{bpa2d_Lz_phases}). 
In the left column of Fig.~\ref{bpa2d_Lz} we depict the $z$-component of the angular momentum.
The middle column of the same figure shows a quantity
that we refer to as total fluid velocity, defined as:
\begin{equation}
S = \frac{1}{V} \int |{\mathbf v}_s| \,d{\bf r},
\label{S_eq}
\end{equation}
where $V=\int d {\bf r}$ is the total volume of integration.
Finally, the right column of Fig.~\ref{bpa2d_Lz} shows the same quantity, as defined
in Eq.~(\ref{S_eq}), but only for the central portion of the cloud.
The central portion of the cloud was defined as a square,
centered at the center of the magnetic trap, with a side
equal to 10\% of the integration domain. This area
corresponds approximately to the void area between the
three initial fragments (see top-center panel in Fig.~\ref{pot}).

The three diagnostics defined above are shown in Fig.~\ref{bpa2d_Lz}
for the same cases presented in Fig.~\ref{bpa2d_allu}.
It is interesting to note how
the presence of a time-dependent component in the potential
generates angular momentum in the system. As can be
observed in the figure, the total angular momentum increases
(in absolute value) through the duration
of the barrier ramp-down and then settles to a constant value that
is larger (in absolute value) for slower ramps.
Also, it is worth noting that the initial value of the 
angular momentum is different from zero [$L_z(t=0)\approx -800$]
due to the intrinsic vorticity carried by the out-of-phase
fragments (see also discussion below).
For our initial condition, most of the total angular momentum is
seeded in the weak overlapping region between the fragments where the
phase gradient is large.
What we can observe about the second and third diagnostics by
comparing (the second row of) Fig.~\ref{bpa2d_allu} and
Fig.~\ref{filaments_fig}
with (the second row of)
Fig.~\ref{bpa2d_Lz} is roughly the following: the integrated
(throughout the cloud) velocity appears to peak when the
filamentation in the pattern of Fig.~\ref{filaments_fig}
is maximal, e.g., around times of 50 and 90
ms. On the other hand, the same diagnostic integrated within
the central core of the cloud peaks substantially earlier when the
vortices are formed through the collision of the fragments
around the end time of the ramp (i.e., around 25 ms).
Subsequently, and as the vortices are advected away from the core,
the latter quantity decreases.

\begin{figure}[ht]
\begin{center}
\includegraphics[width=6.0cm]{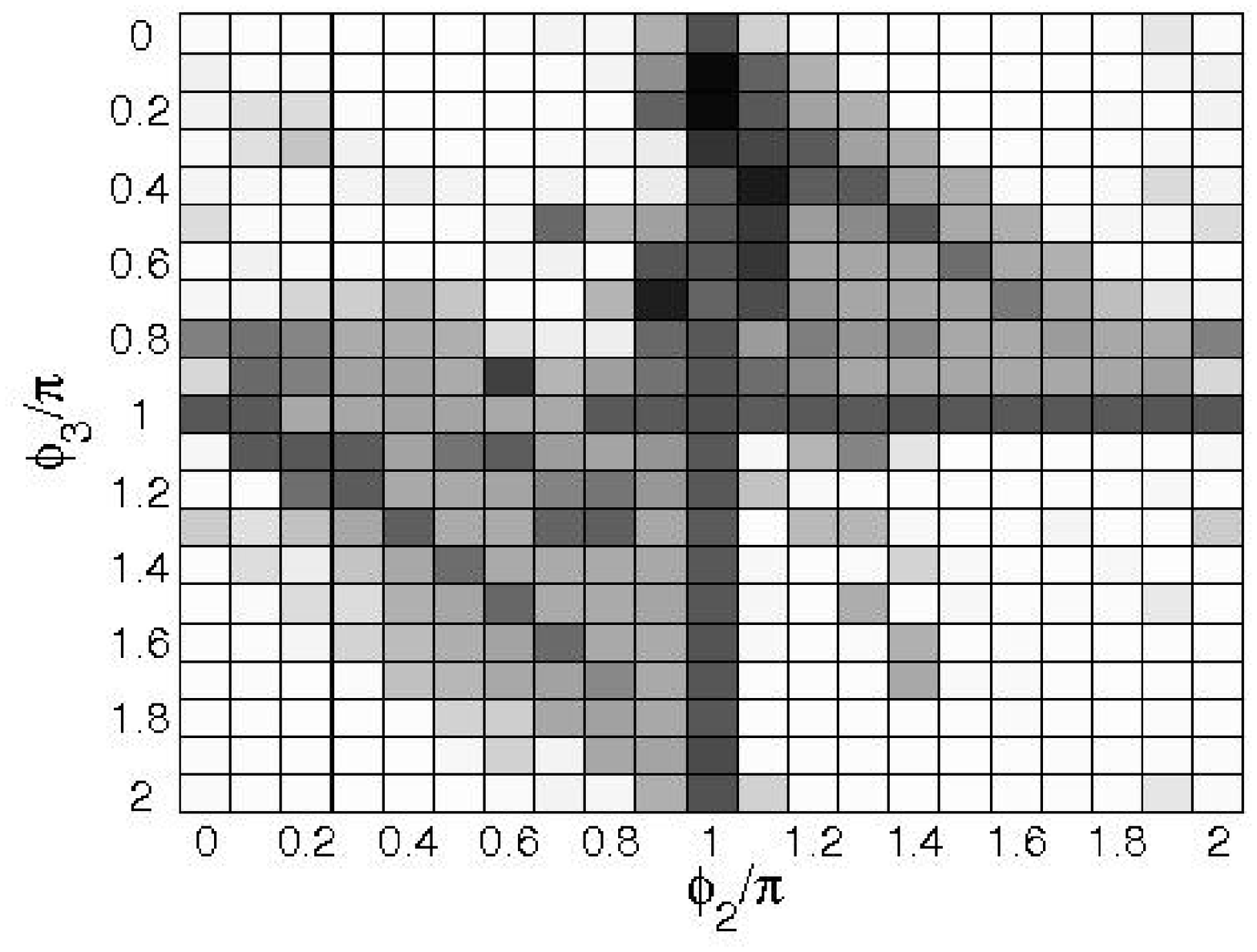}\\[2.0ex]
\includegraphics[width=6.0cm]{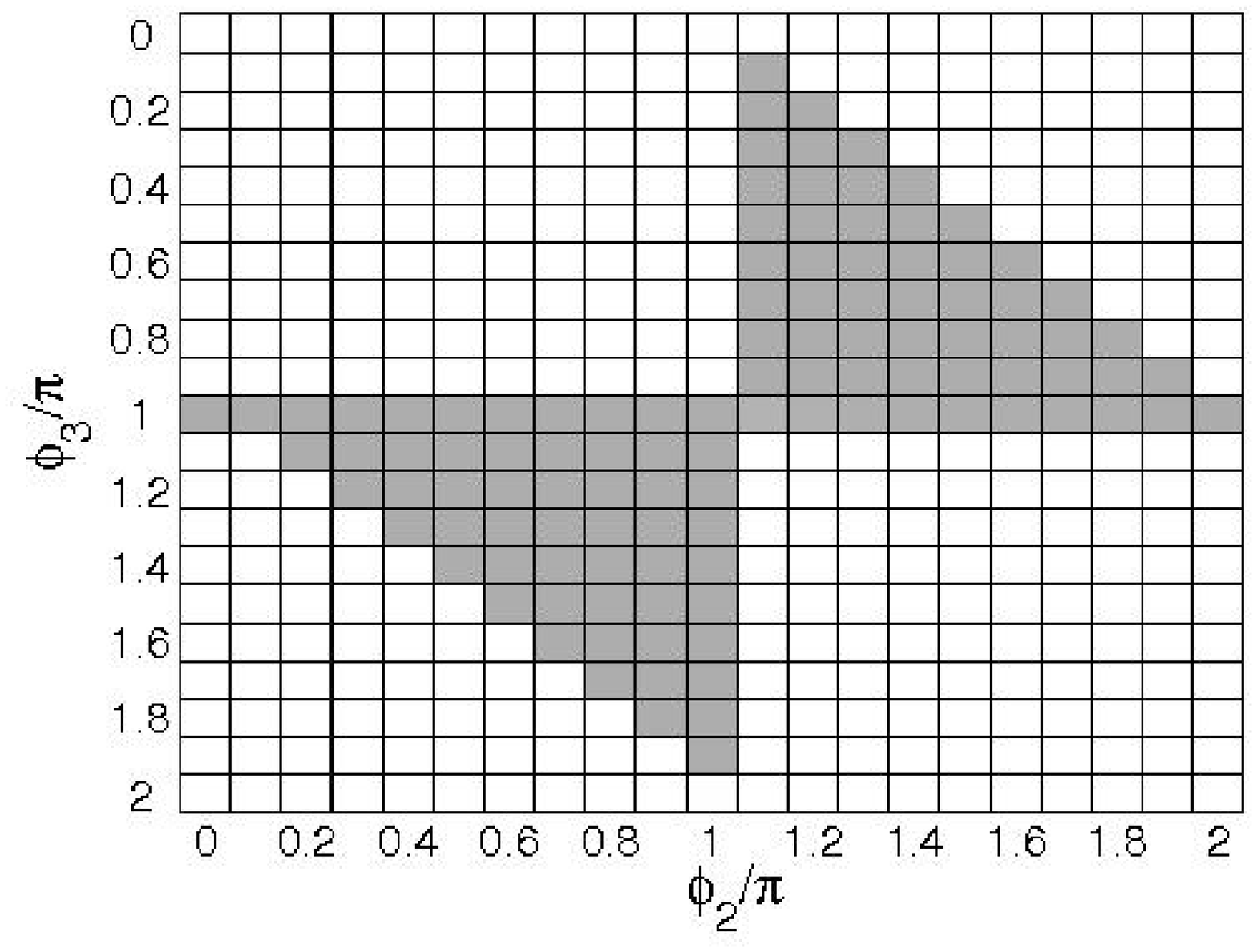}\\[2.0ex]
\includegraphics[width=6.0cm]{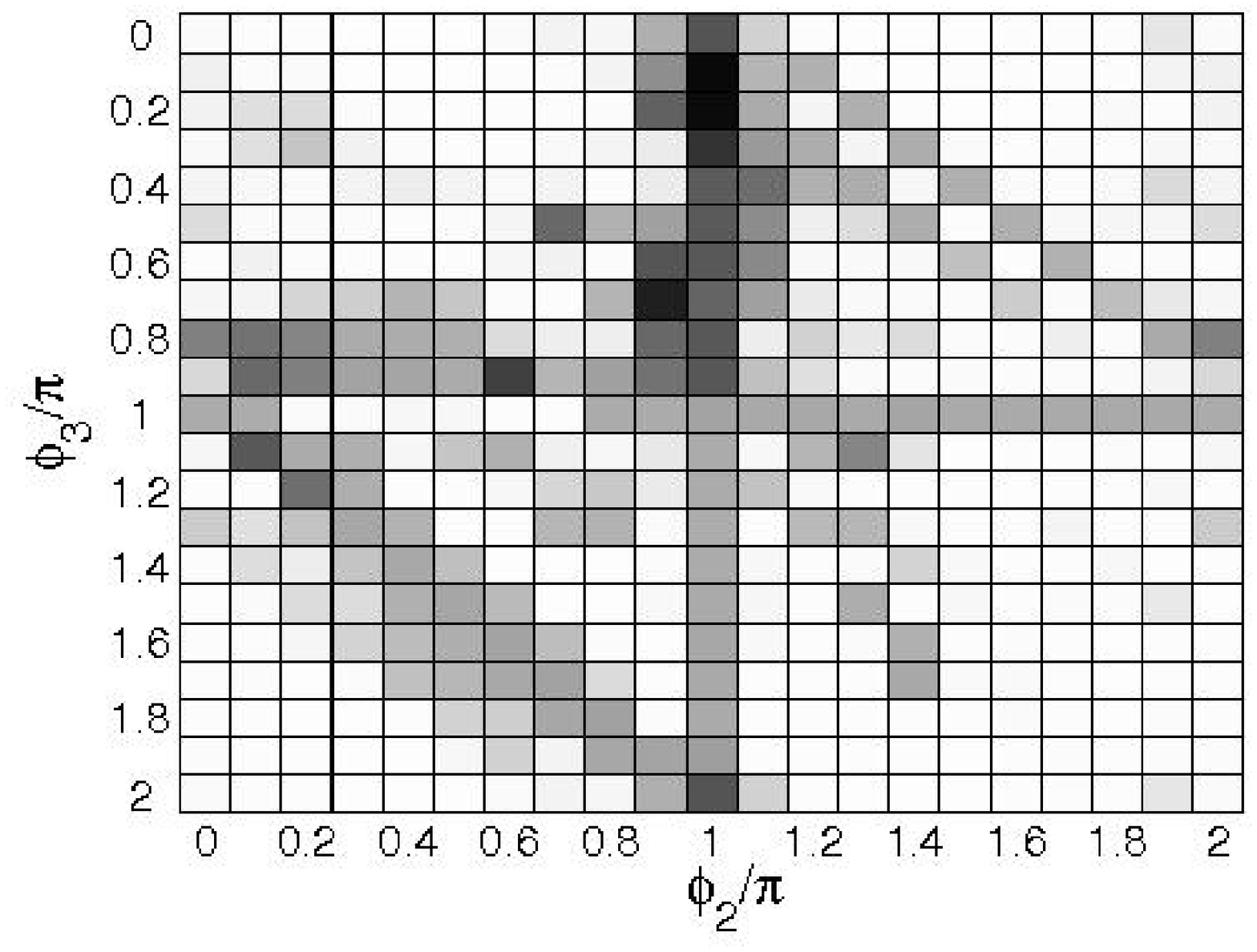}
\end{center}
\vskip-0.4cm
\caption{
Phase diagram for vortex formation at the trap center region
as a function of the relative phases of the merging fragments
(where $\phi_1$ is assigned the value of 0).
Top: Ramp down time of $t_b=0$ ms (i.e., instantaneous removal) of
the barrier. The different shades of gray indicate the amount of
vorticity contained in the central region after 100 ms.
White, light gray, gray, black correspond, respectively,
to 0, 1, 2, and 3 vortices.
Middle: Same as above for a ramp down time of $t_b=50$ ms.
This case only produces no vortices (white) or one
vortex (gray).
Bottom: difference between top and middle diagrams.
This corresponds to the vortices formed by the collision
of the fragments and not by the intrinsic vorticity of the
initial configuration (which depends on the relative phases
of the different fragments).
}
\label{diagram}
\end{figure}

\def\figwt{3.4cm}
\def\figw{3.5cm}

\begin{figure*}[t]
\begin{center}
\includegraphics[width=\figw]{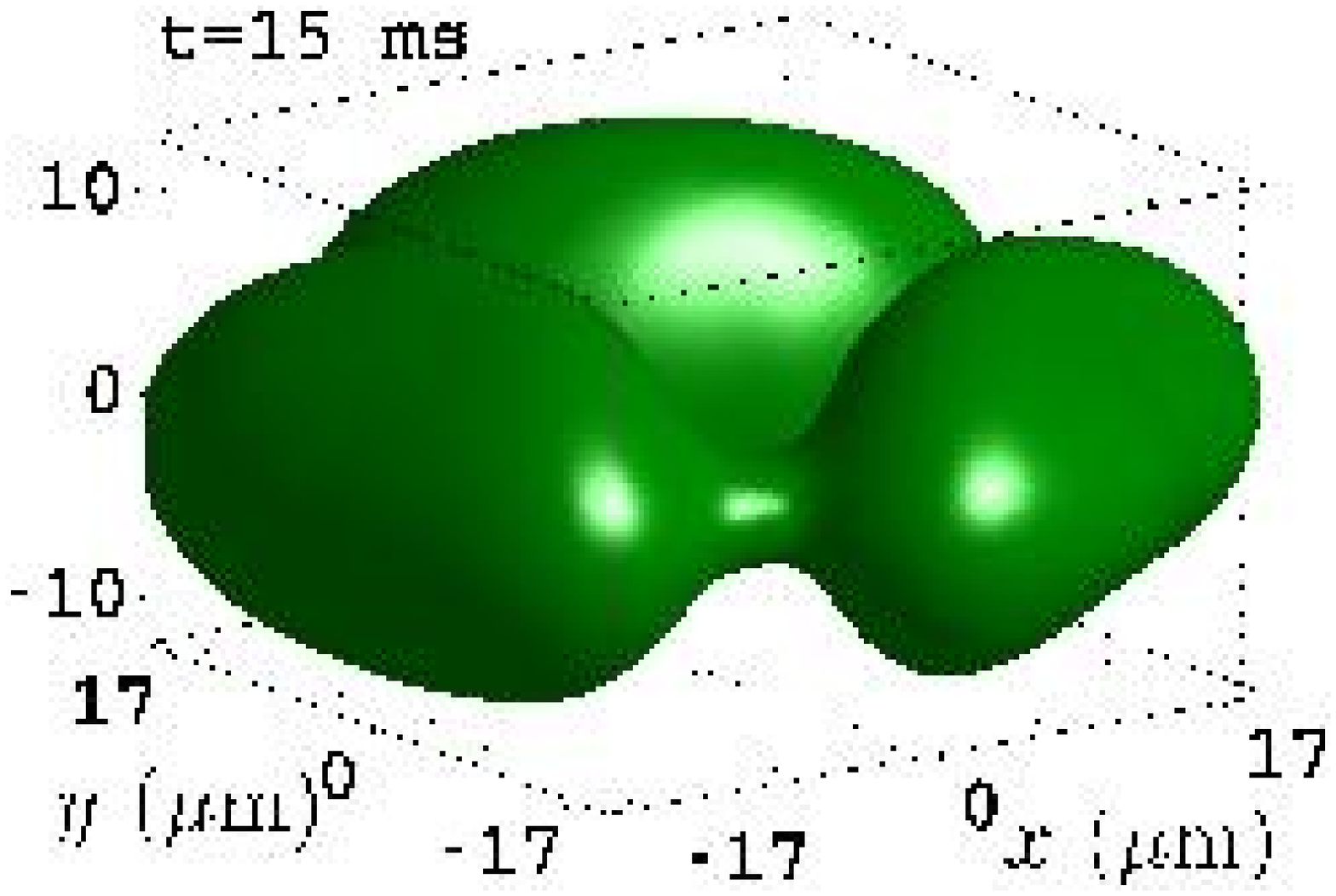}
\includegraphics[width=\figw]{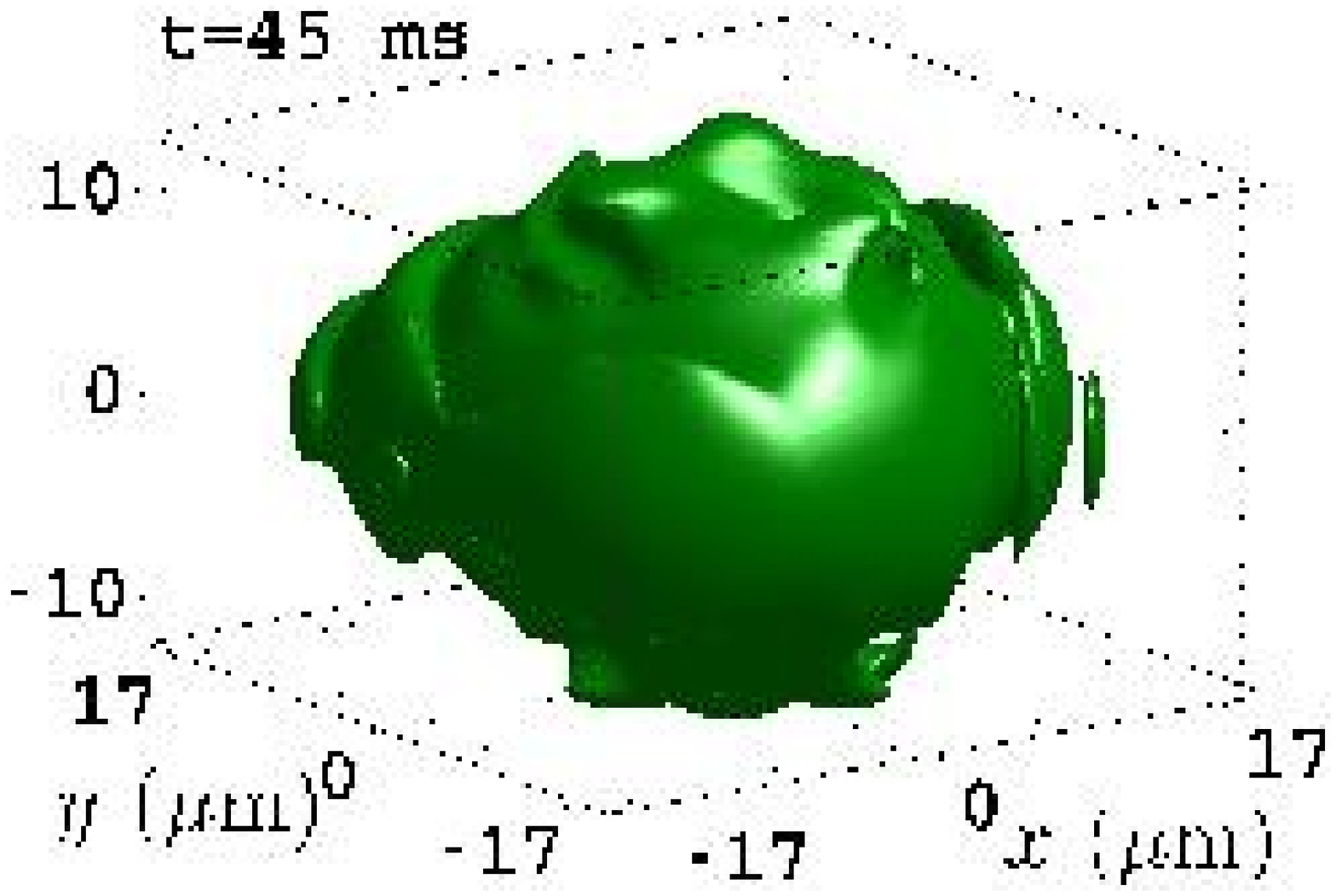}
\includegraphics[width=\figw]{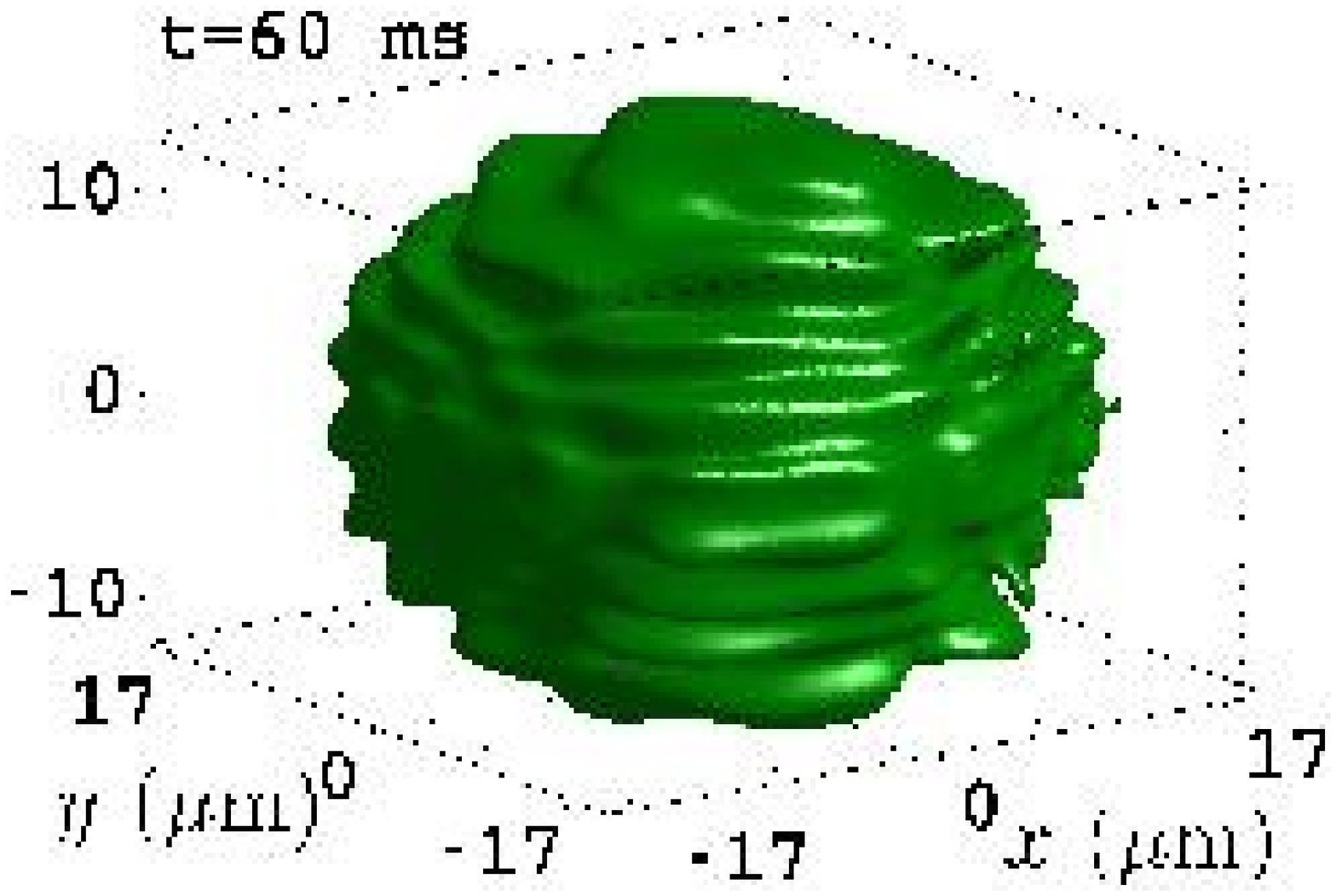}
\includegraphics[width=\figw]{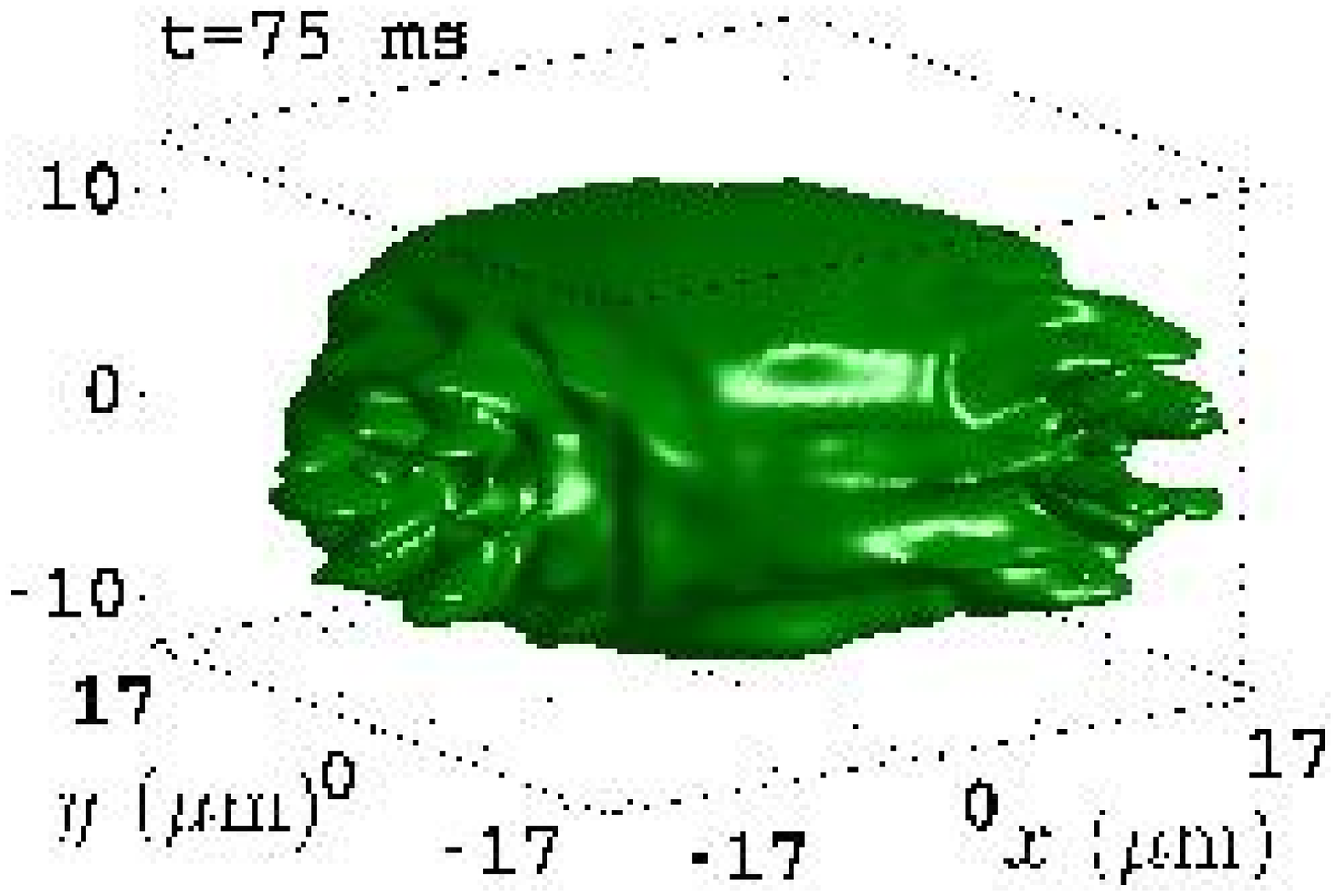}
\includegraphics[width=\figw]{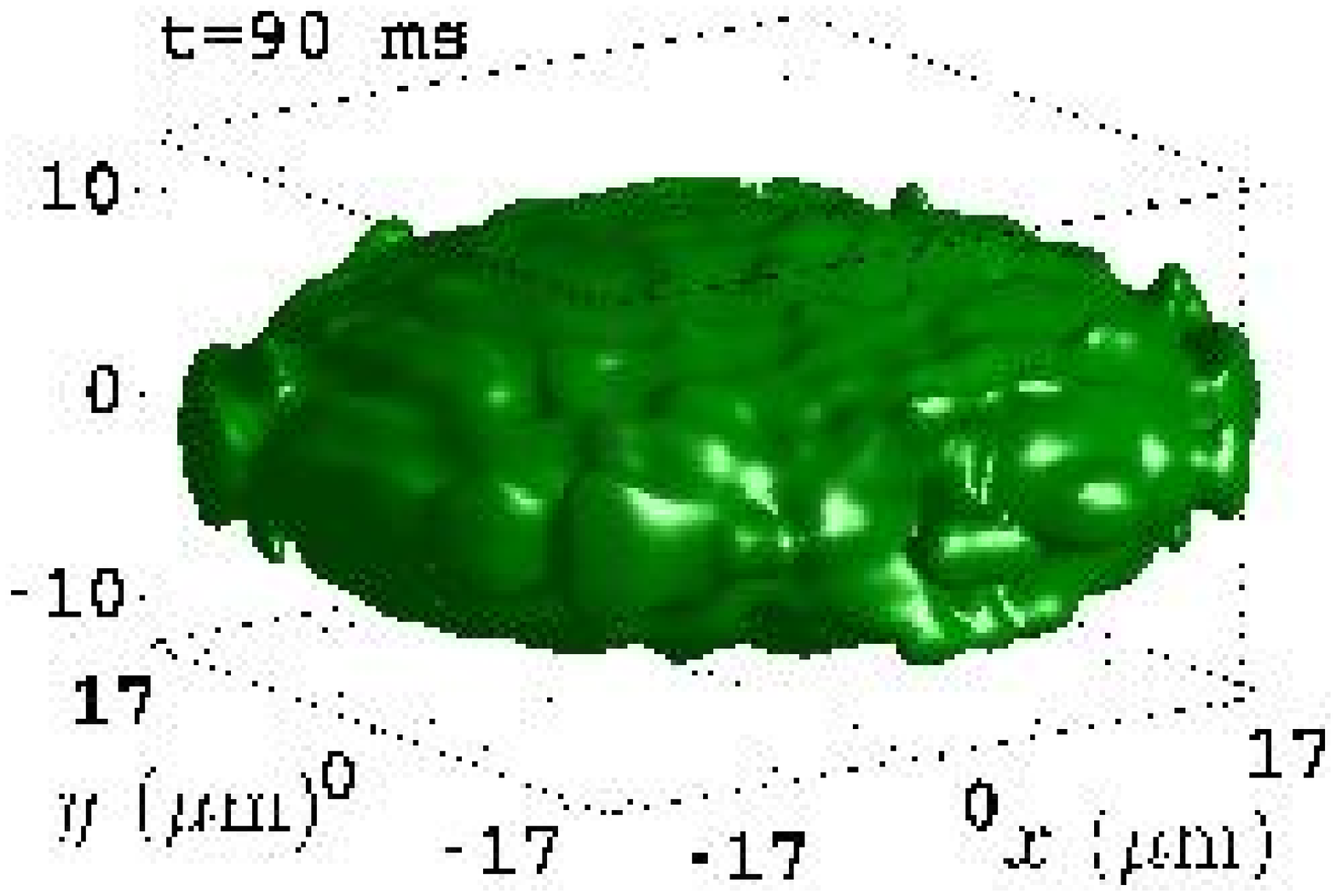}
\\[2.0ex]
\includegraphics[width=\figw]{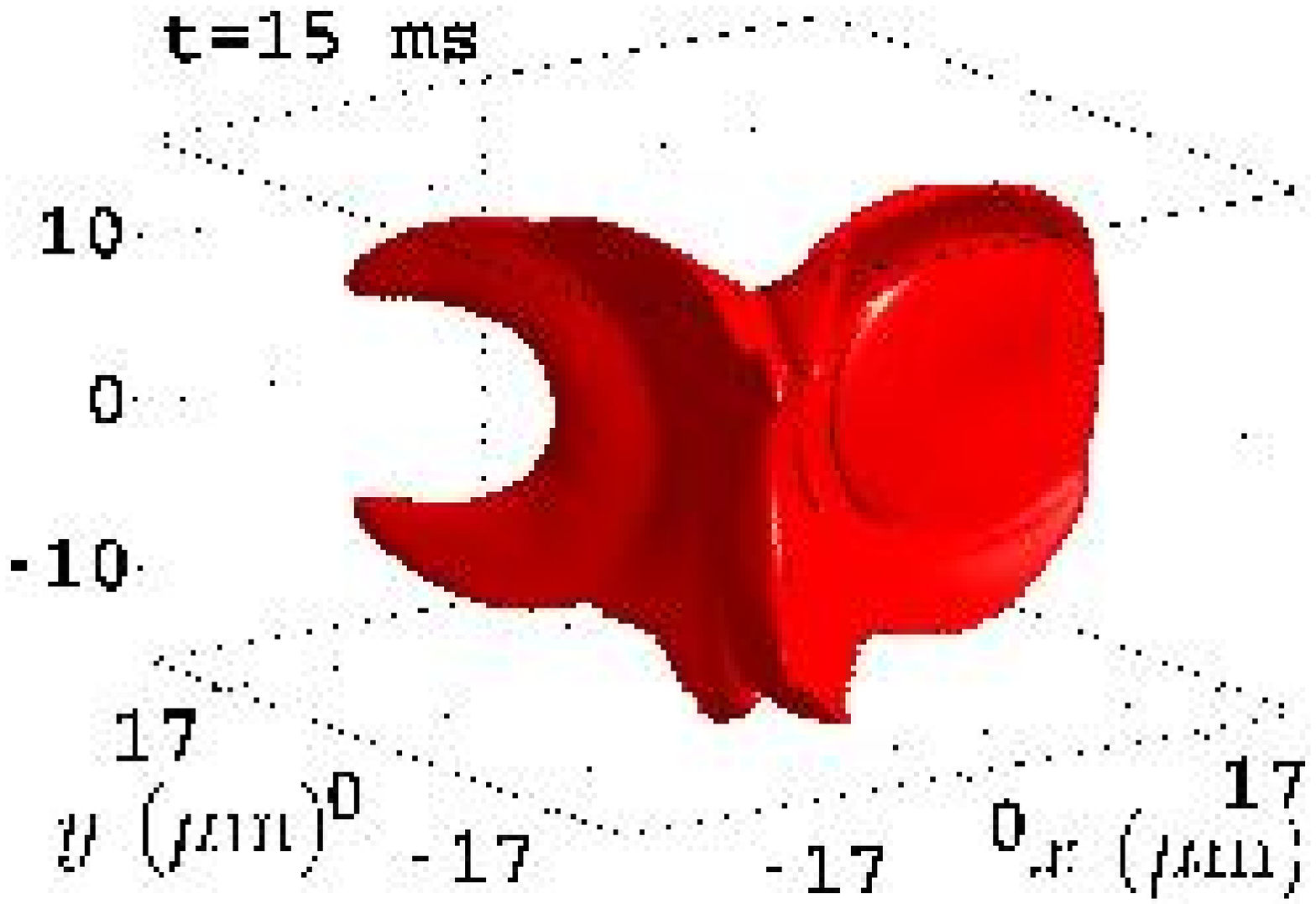}
\includegraphics[width=\figw]{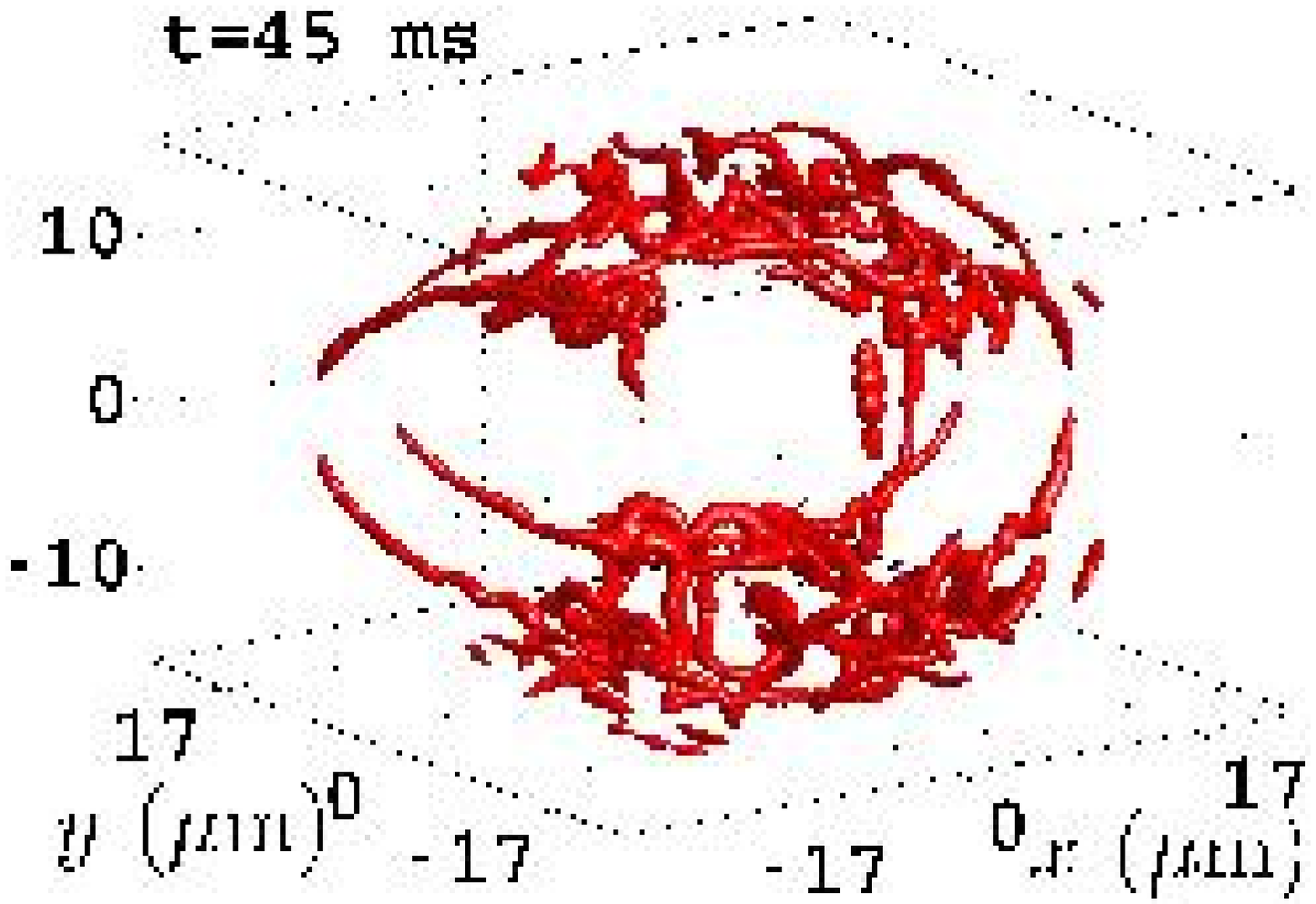}
\includegraphics[width=\figw]{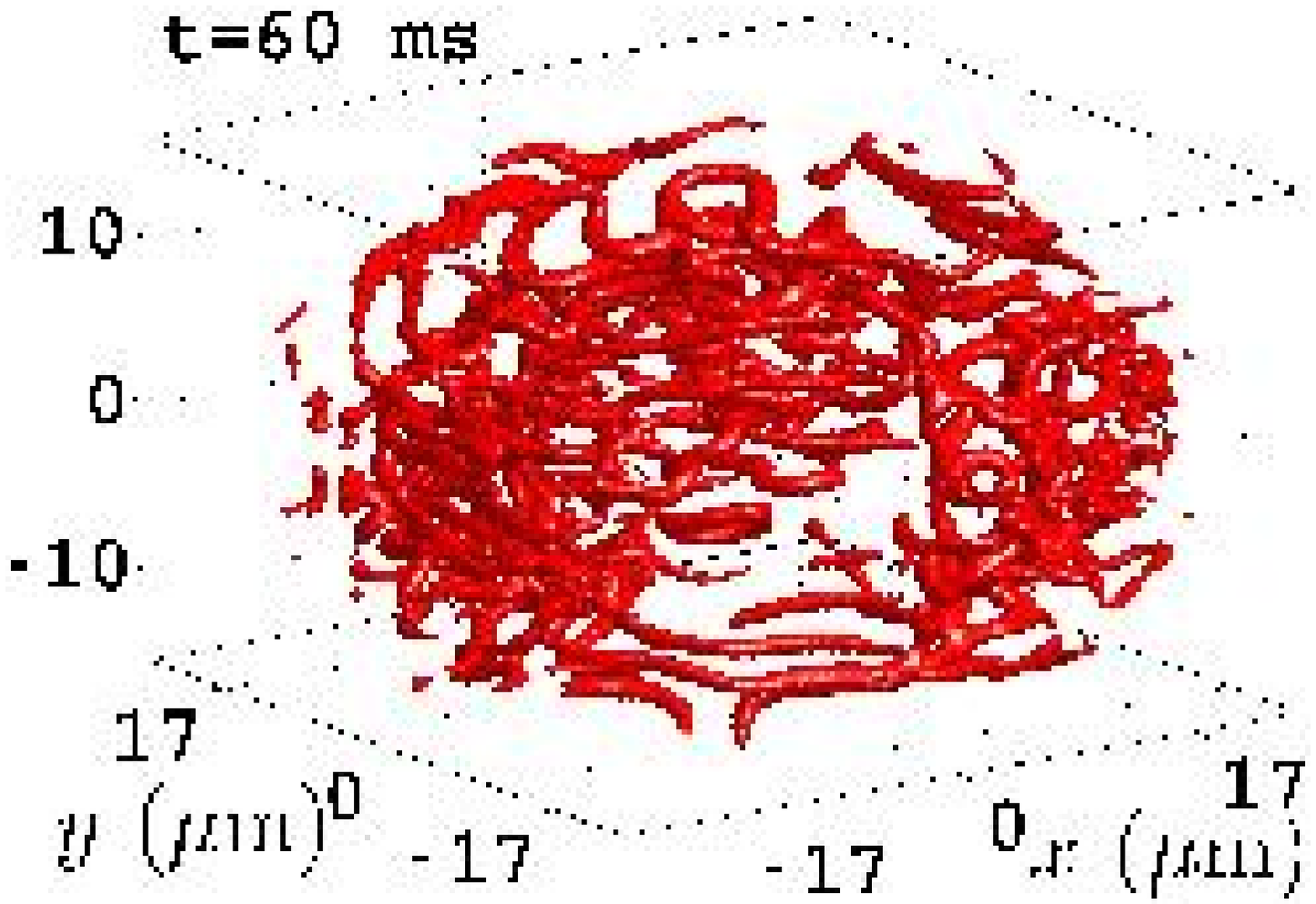}
\includegraphics[width=\figw]{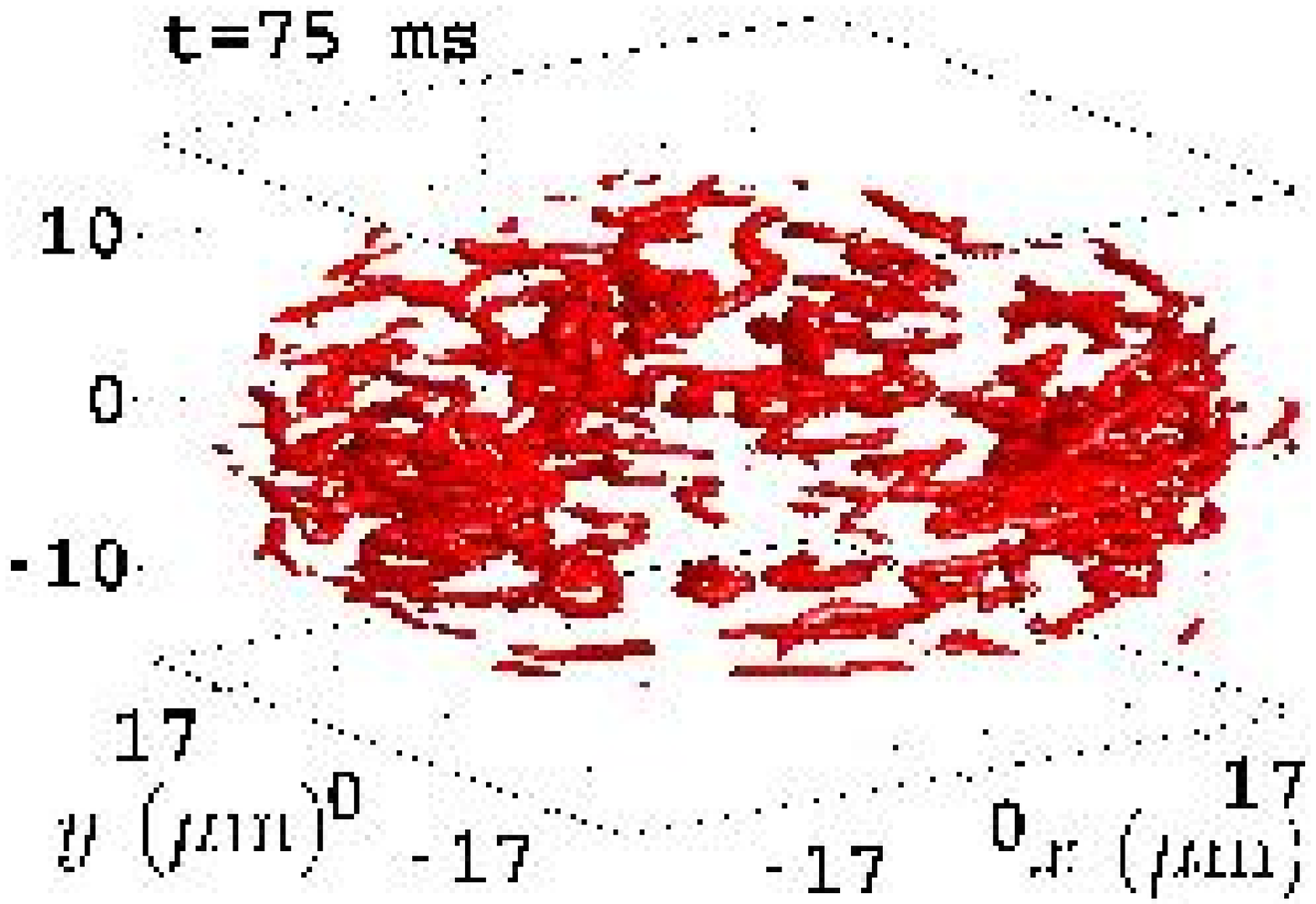}
\includegraphics[width=\figw]{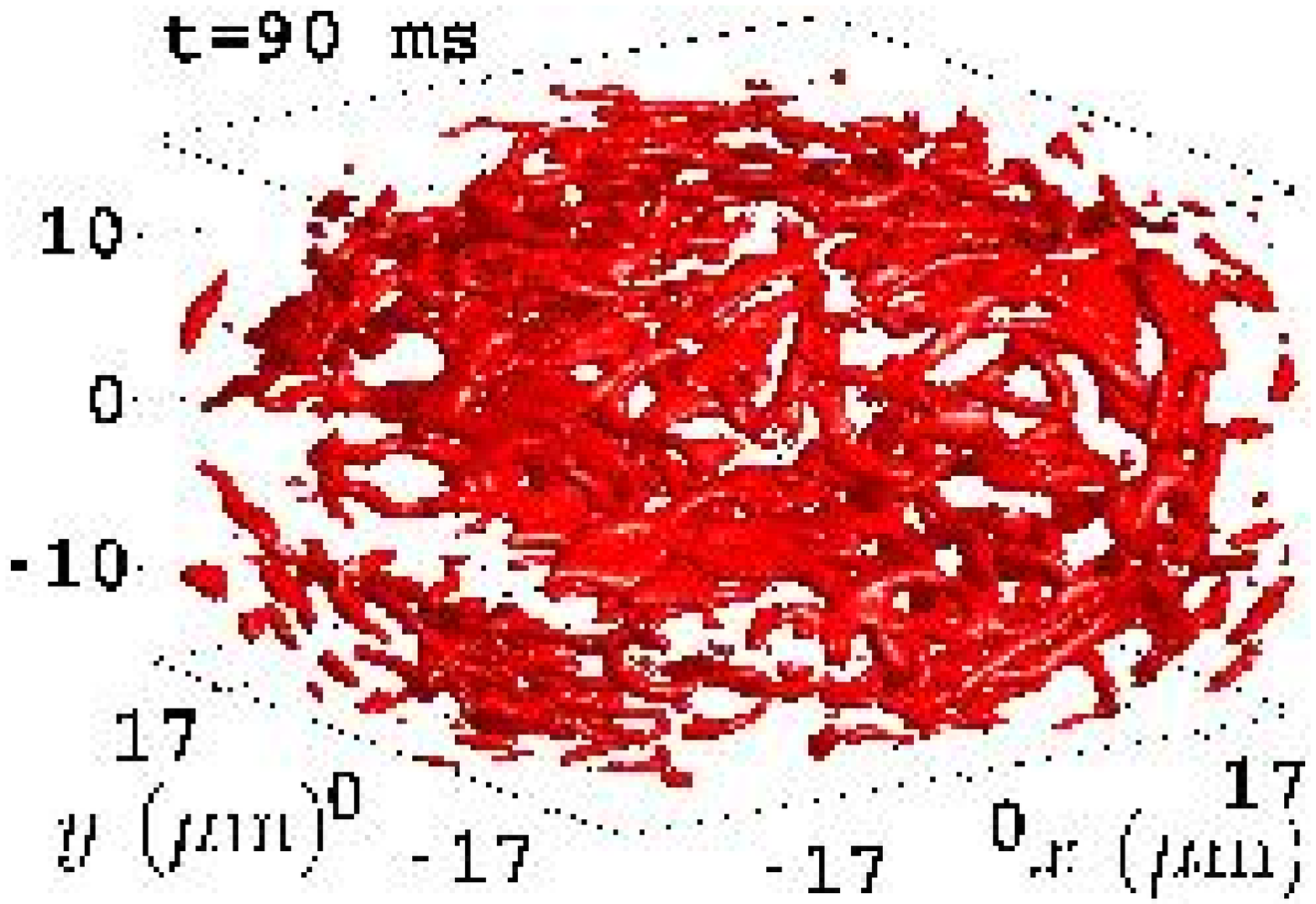}
\\[2.0ex]
\includegraphics[width=\figwt]{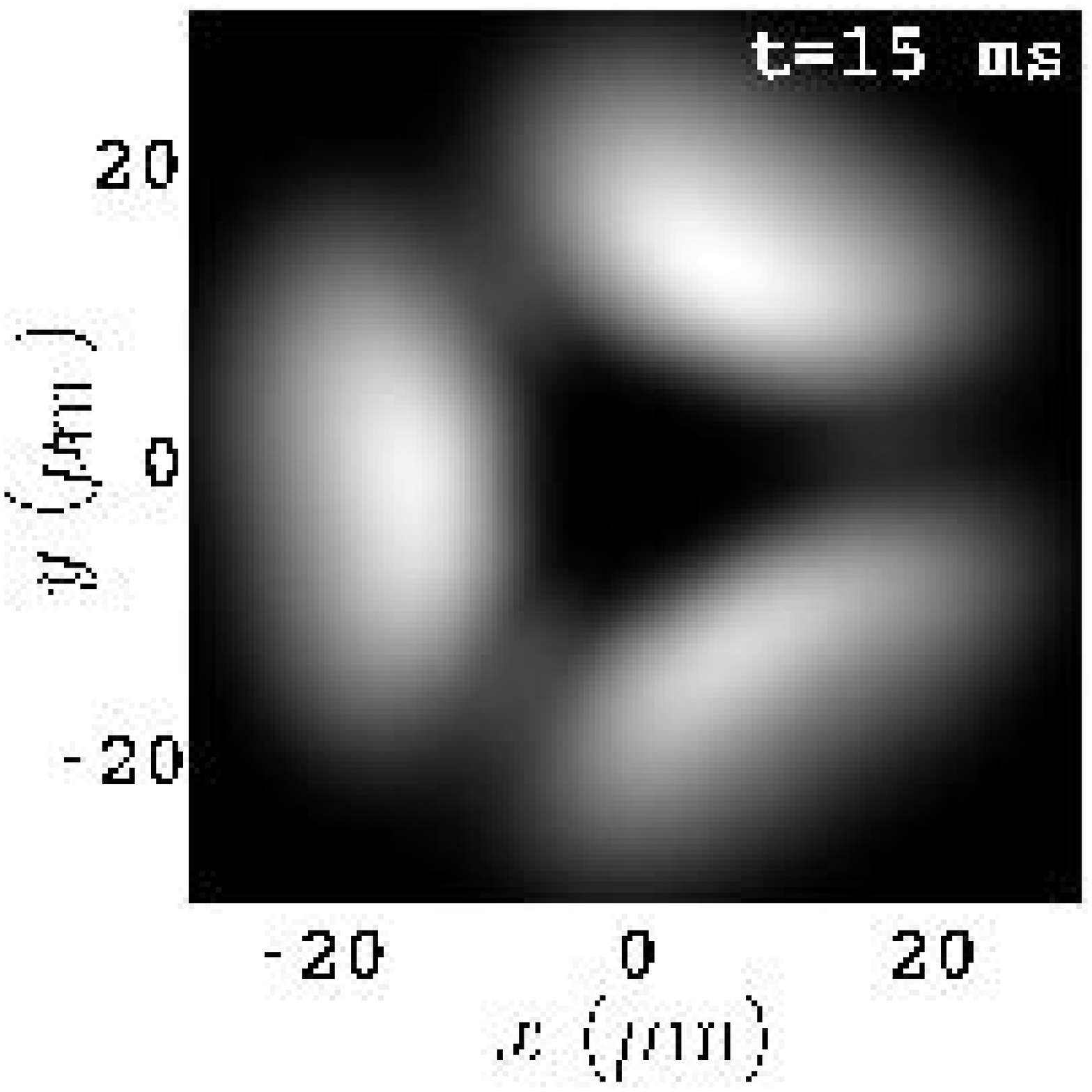}~~
\includegraphics[width=\figwt]{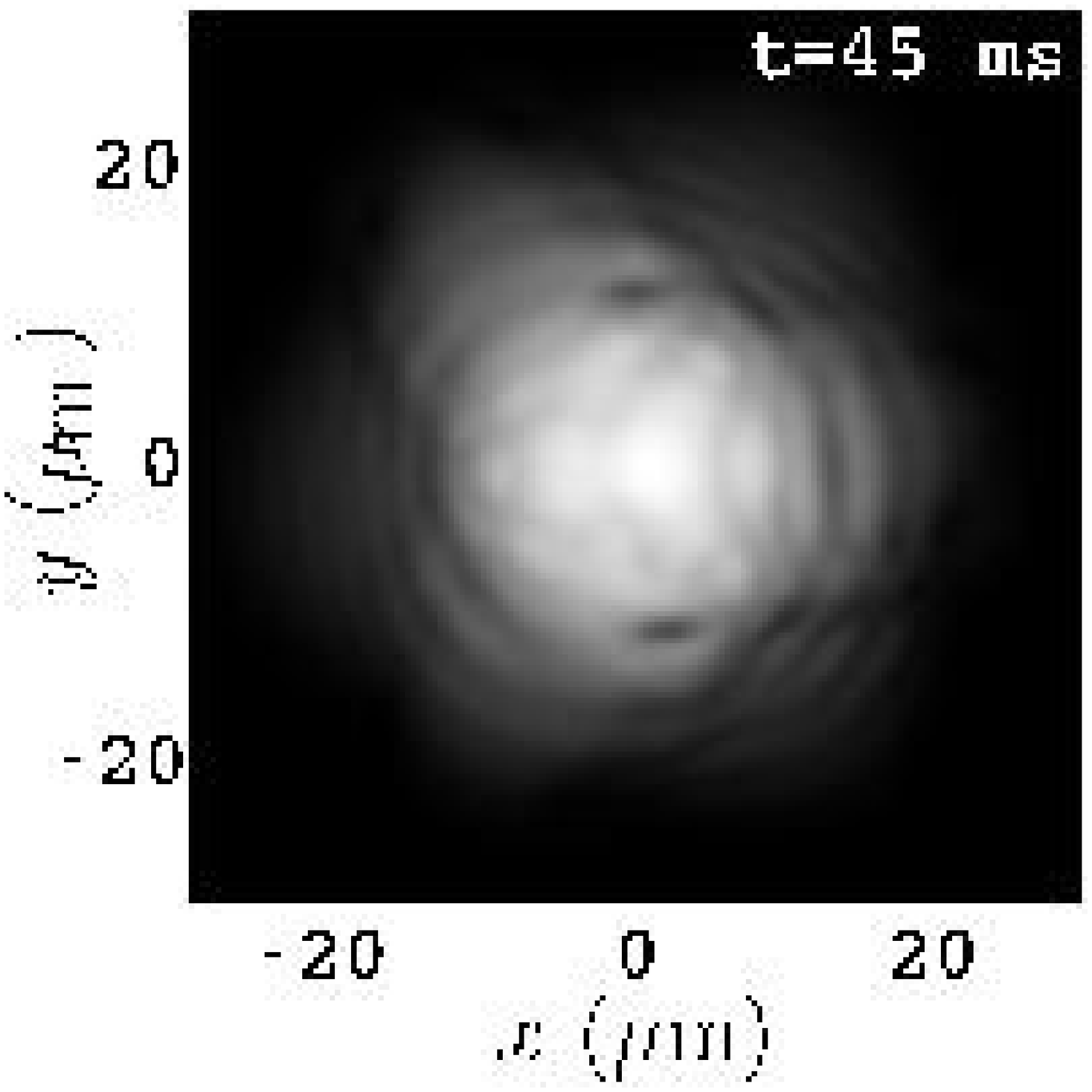}~~
\includegraphics[width=\figwt]{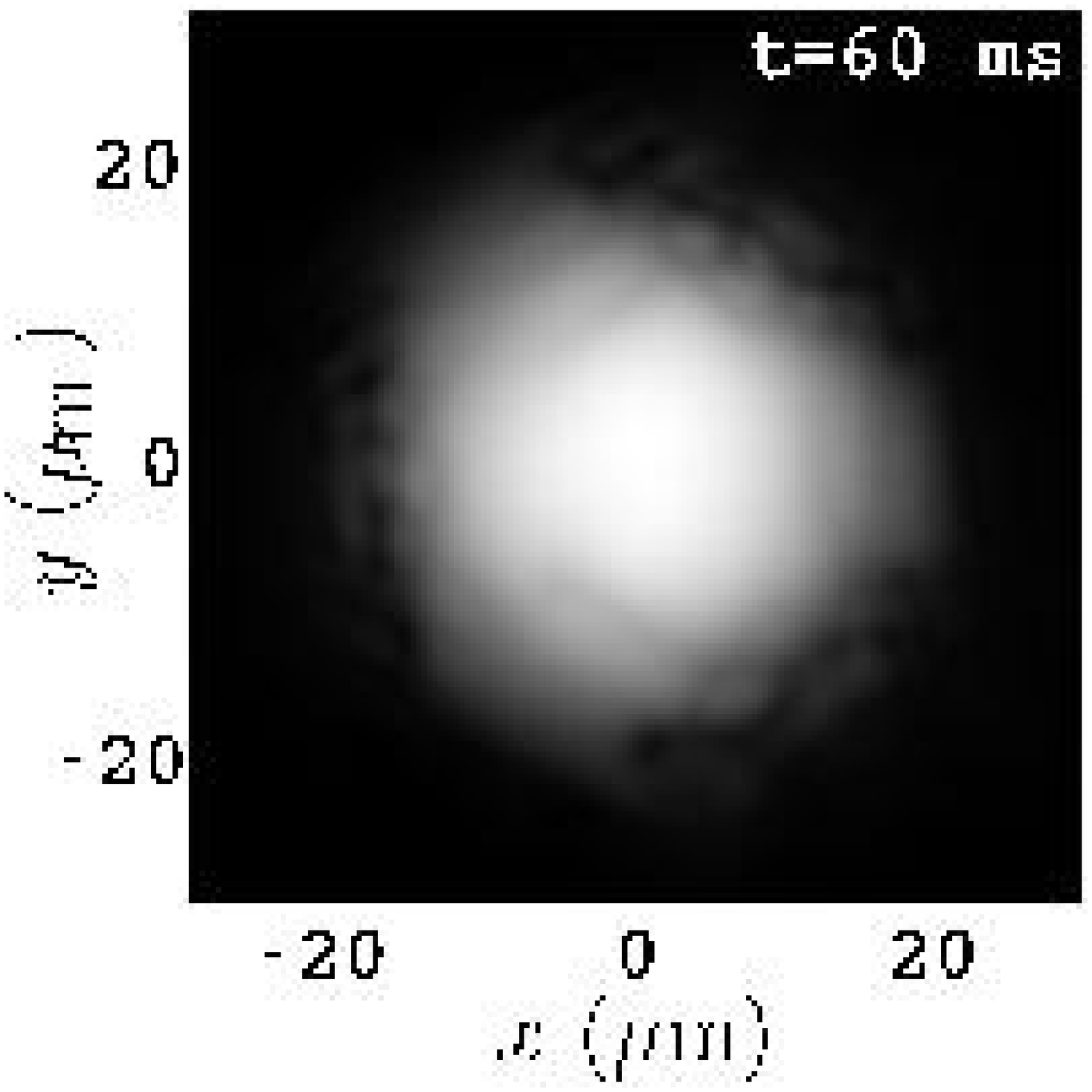}~~
\includegraphics[width=\figwt]{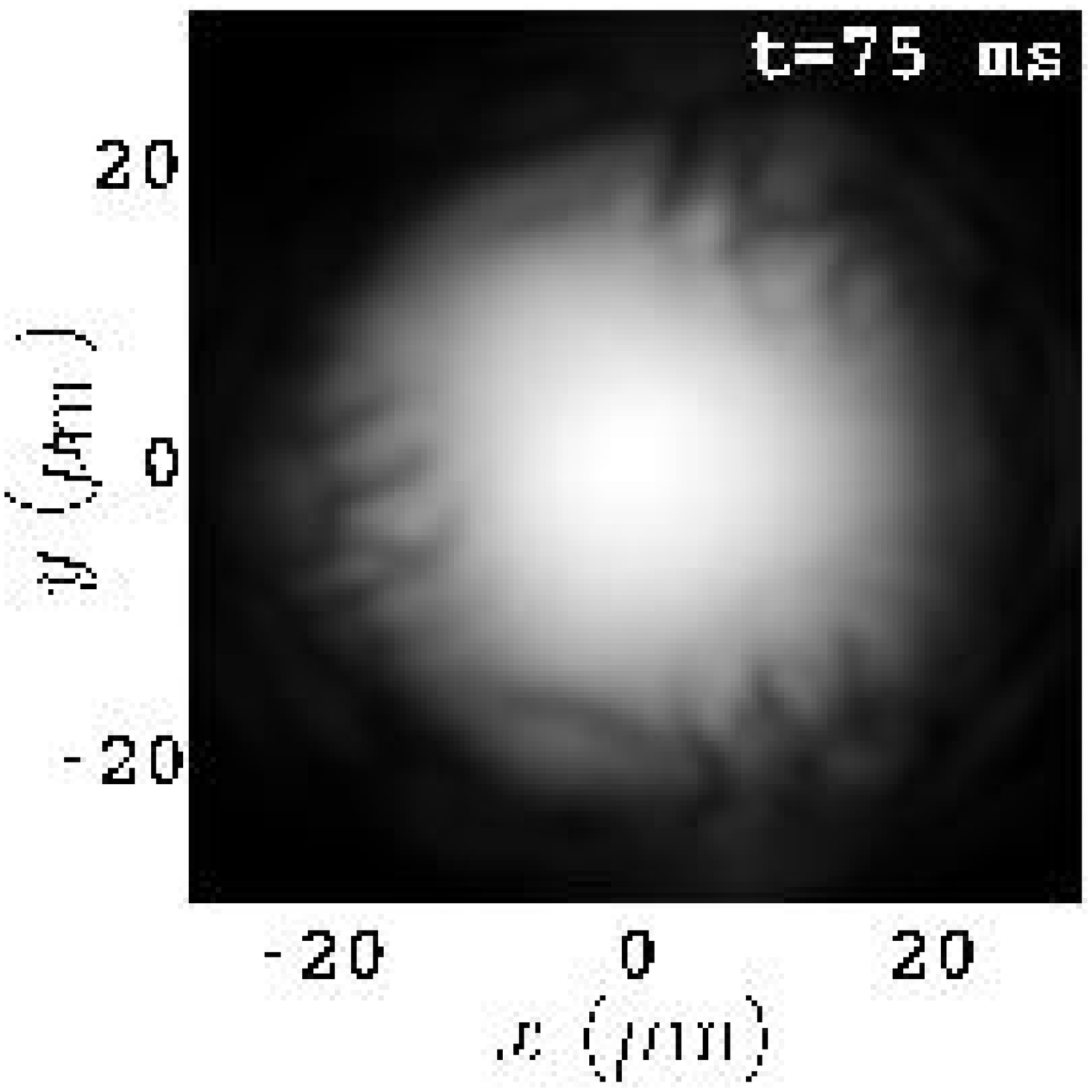}~~
\includegraphics[width=\figwt]{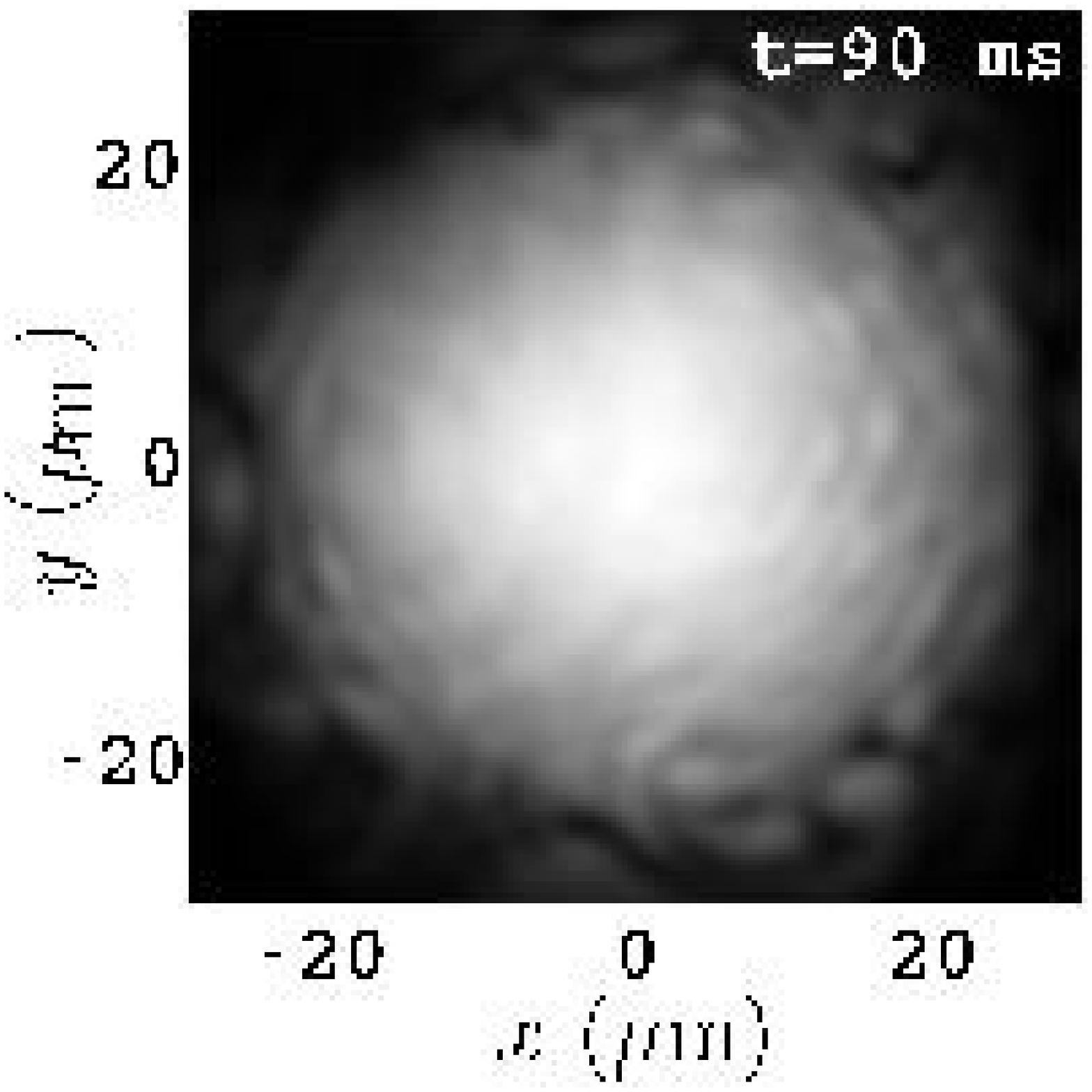}~~
\end{center}
\vskip-0.4cm
\caption{(Color online)
Vortex formation by the merging of three-dimensional BECs.
Top row: contour surfaces of constant atomic density.
Middle row: contour surfaces of the corresponding absolute
value of the vorticity.
Bottom row: $z$-projection of the density distribution (i.e., column density along $z$)
as it would be observed in the laboratory.
The snapshots are taken at the indicated times (in ms)
for an initial phase distribution corresponding to
$\phi_k=0$ and a ramping down time of 25 ms. For the contour-surface images, 
the axis on the left side of each plot represents the vertical ($z$) direction, expressed in units of microns.
}
\label{3Da}
\end{figure*}

We now turn to the examination of the effect of the relative
initial phases between the different fragments.
As an example, we show in Fig.~\ref{bpa2d_allu_phases} 
the behavior of the cloud density and its vorticity 
for three different phase combinations for a fixed ramp-down 
time of $t_b=25$~ms. As can be noticed, 
the complexity of the vorticity field is similar for the three
cases shown in the figure. However, the only phase combination
of the ones shown here that produces a vortex at the center of the cloud corresponds
to $\phi_k=2\pi k/3$, which can be understood by the 
intrinsic vorticity already present in the initial
configuration (see explanation below).
In Fig.~\ref{bpa2d_Lz_phases} we depict the vorticity
indicators for the three phase combinations of
Fig.~\ref{bpa2d_allu_phases}. As 
can be seen in the figure, the angular momentum $L_z$ for $\phi_k=0$ (top-left
panel) and $\phi_k=\pi k/3$ (bottom-left panel)
suffers almost no change since the initial configuration
does not carry any intrinsic vorticity and thus, its
interaction with the ramping down barrier does not produce
angular momentum. However, for $\phi_k=2\pi k/3$ (middle-left panel),
as explained before, the system does gain angular momentum
during the barrier removal.
The total fluid velocity indicators (middle and right 
columns in Fig.~\ref{bpa2d_Lz_phases} behave similarly 
for the three phase combination with the notable 
difference that for the $\phi_k=2\pi k/3$ (middle-left panel) 
case, its final value is different from zero since the 
fragments produce a vortex at the center of the trap (due
to the intrinsic vorticity carried by the initial condition).

In order to follow in more detail the formation of vortices
in the central portion of the cloud as a function of
the relative initial phases of the different fragments, we perform
systematic simulations for a large set of relative initial phases.
The results are presented in 
Fig.~\ref{diagram}, where we show the existence of
vortices as a function of $(\phi_2,\phi_3)$ for $\phi_1=0$.
Darker shades correspond to the presence of more vortices.
The top panel of Fig.~\ref{diagram} corresponds to an immediate
release of the barrier (ramp down time of $t_b=0$ ms) where
the presence of 0, 1, 2 or 3 vortices 
(white, light gray, gray, and black respectively) can be observed for different
phase combinations. The middle panel depicts the same
diagram, but for a ramping-down time of $t_b=50$ ms. It is
clear that, for this relatively slow ramp down, the formation
of a vortex in the central region is exclusively determined
by the vorticity of the initial configuration. Namely,
if any of the relative phases is larger than $\pi$, the
initial configuration resembles more that of a {\it discrete vortex} 
(see Refs.~\cite{panos} and \cite{desyatnikov,our} for reviews). 
In particular, each fragment can be thought of a ``unit'' and
the whole configuration corresponds to a discrete vortex with three
units with a net vorticity different from zero.
The middle panel of the figure clearly reveals that {\it only}
within an arc of the second (and essentially symmetrically
of the fourth) quadrant of the plane of the phases $(\phi_2-\pi,\phi_3-\pi)$,
a single vortex will form at the center of the configuration.
It is interesting that discrete
vortex-like configurations consisting of three fragments have
been considered in the context of 
nonlinear optics in Ref.~\cite{sukh}
and even in genuinely discrete systems as, e.g., in Ref.~\cite{kouk}.
In the bottom panel of Fig.~\ref{diagram} we depict the
difference between the top and middle panels to show
the amount of vortices that are created exclusively by the
fragment collision and not by the intrinsic vorticity
of the initial configuration.
It is important to mention that some of the vortices counted
in the top panel come from vortices that are created {\em outside}
the central region but that migrate towards the center as
time progresses.

\begin{figure*}[htb]
\begin{center}
\includegraphics[width=\figw]{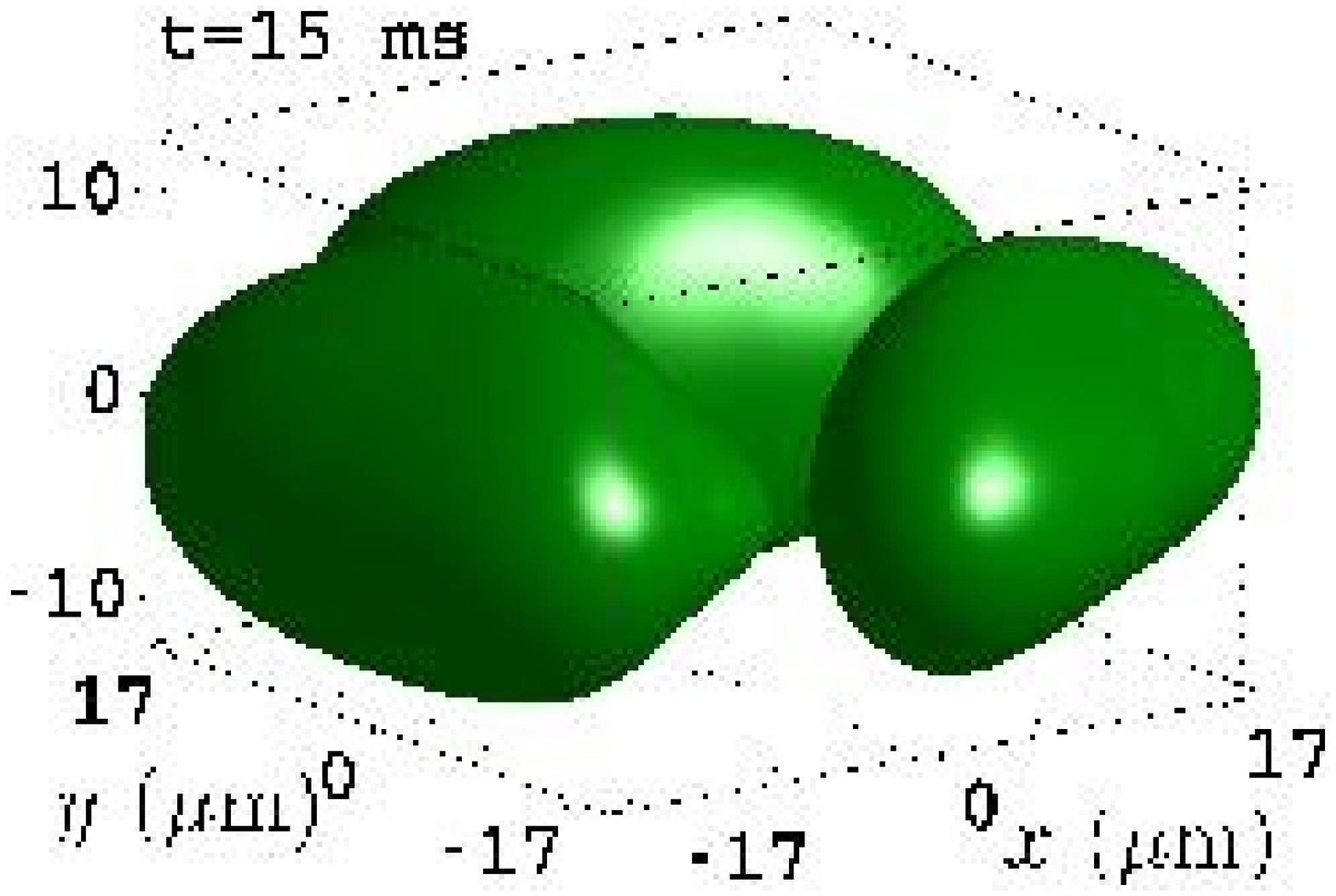}
\includegraphics[width=\figw]{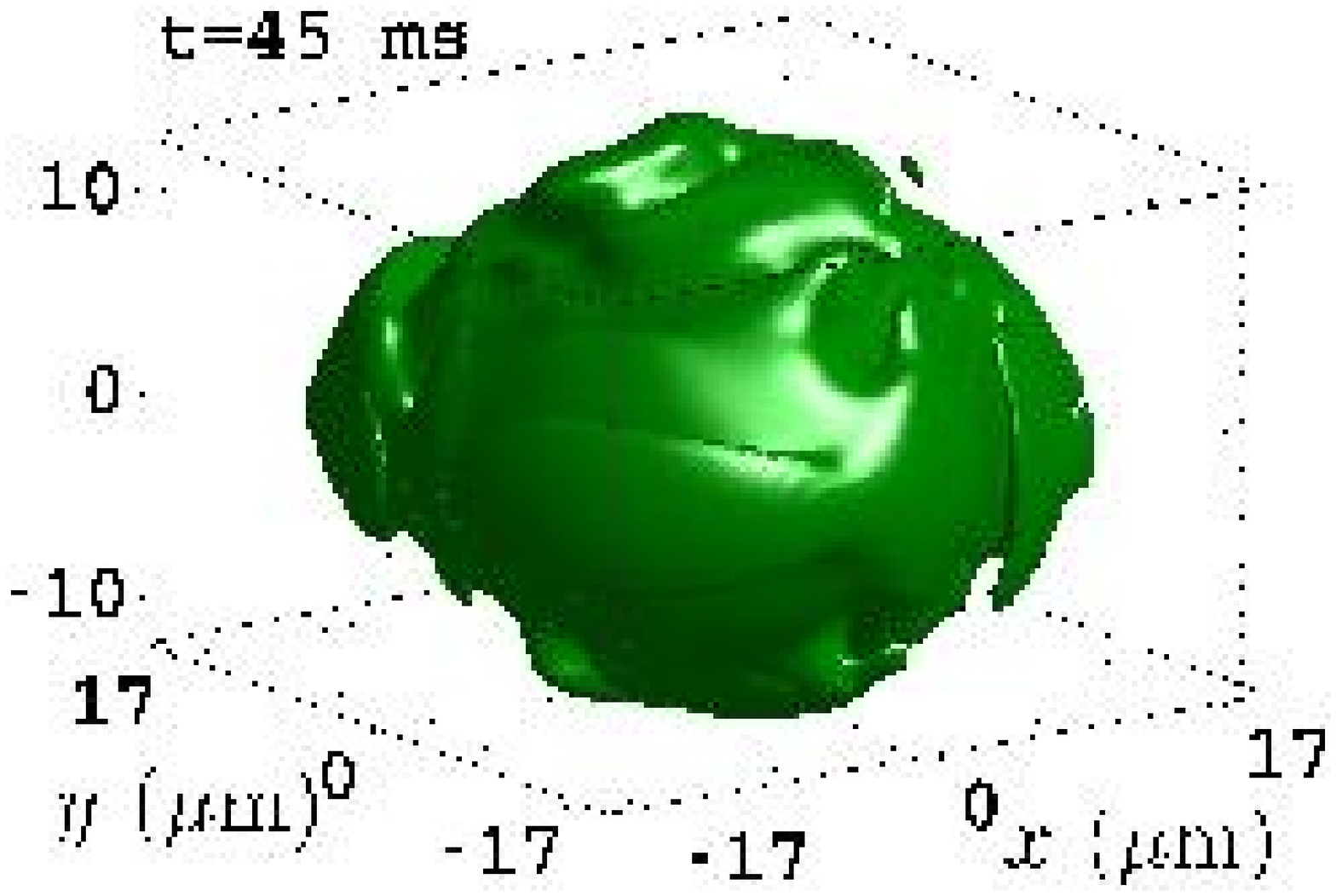}
\includegraphics[width=\figw]{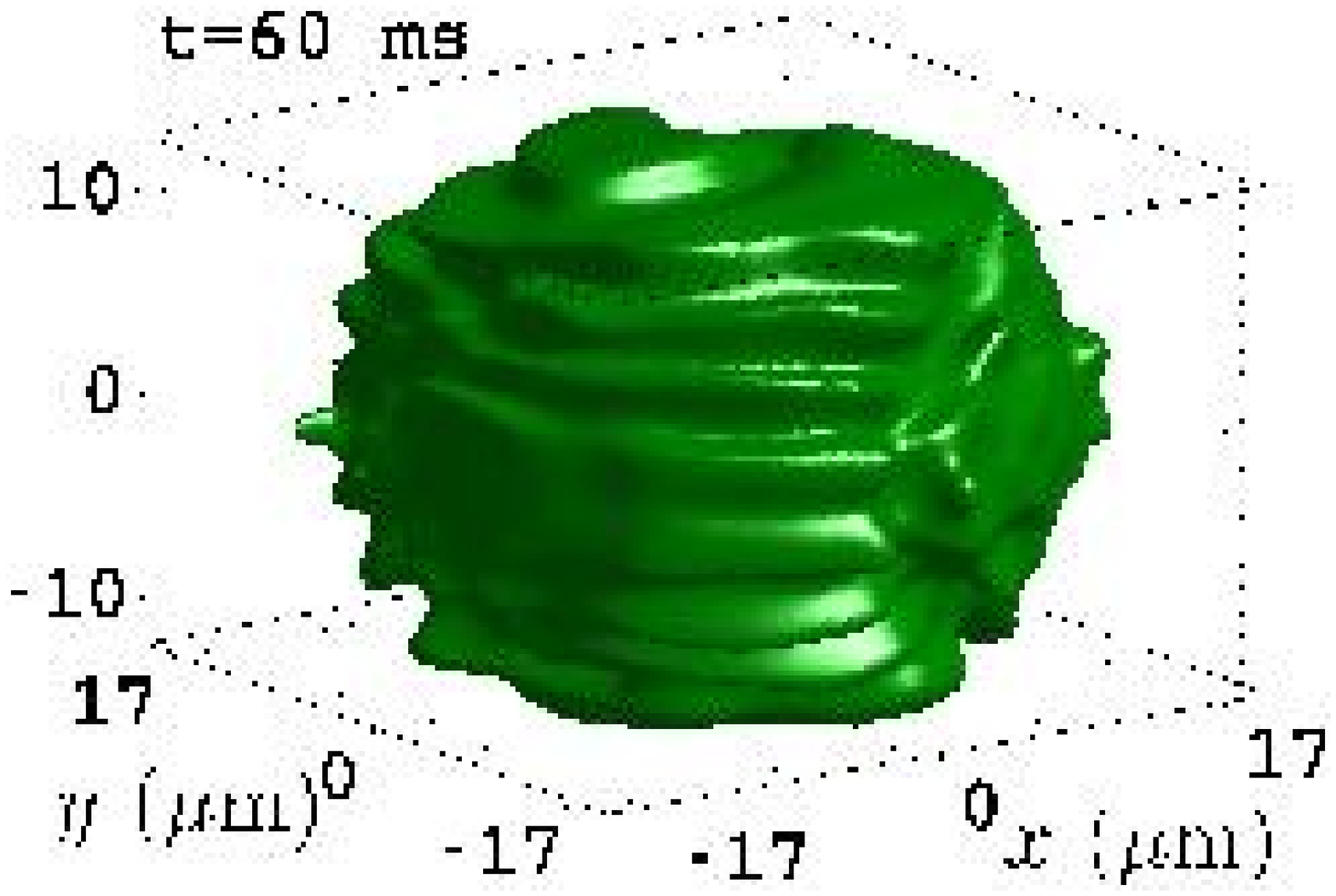}
\includegraphics[width=\figw]{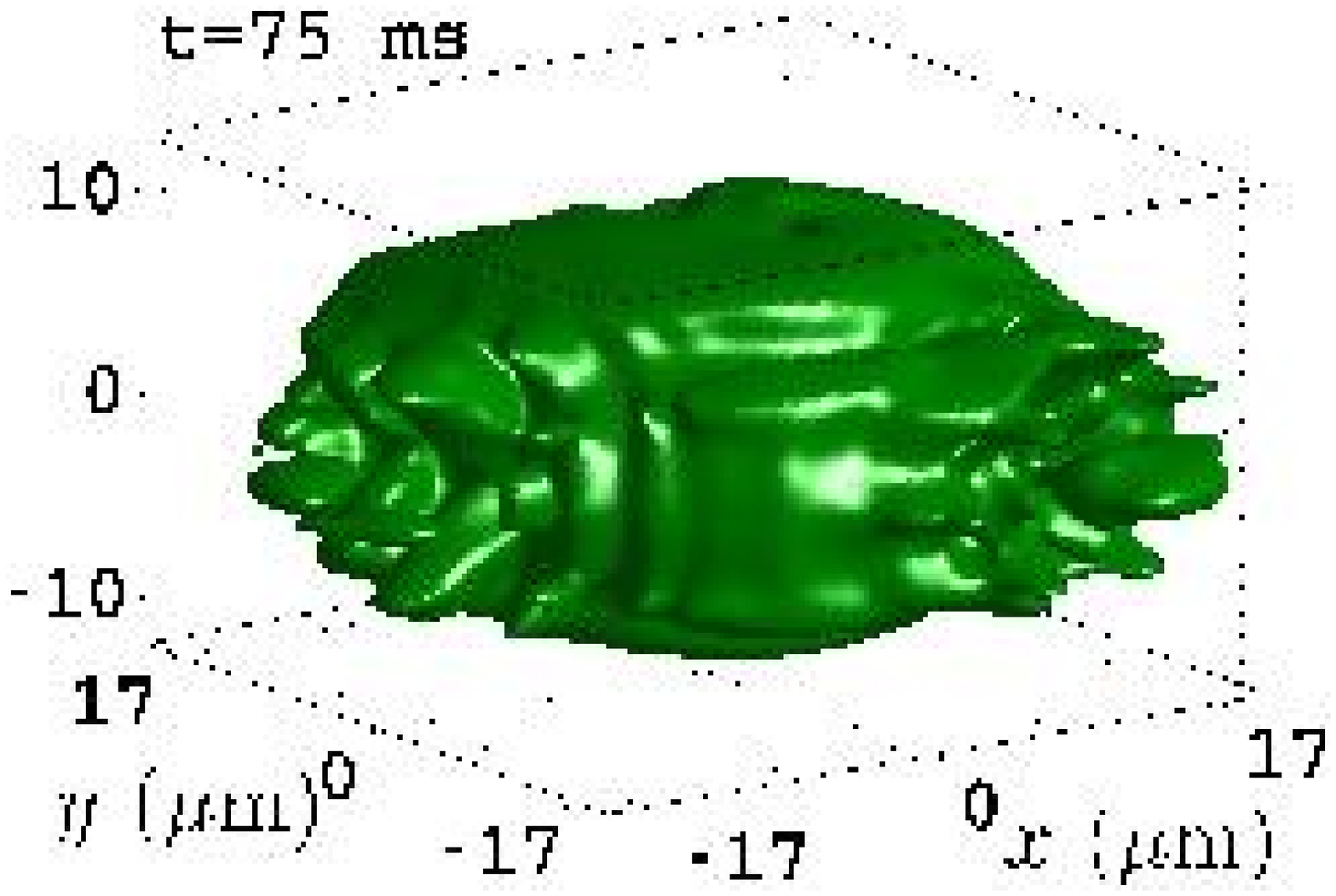}
\includegraphics[width=\figw]{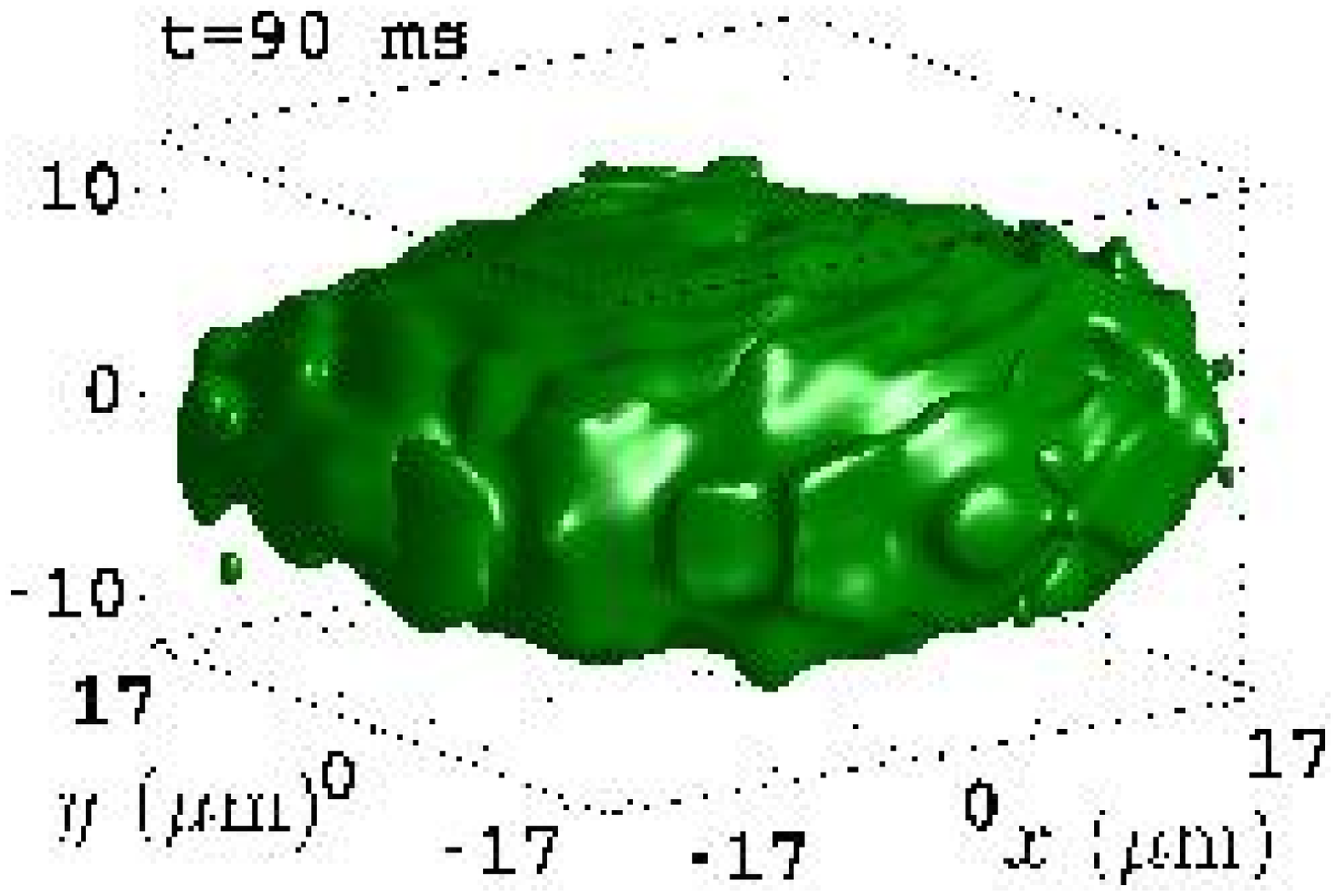}
\\[2.0ex]
\includegraphics[width=\figw]{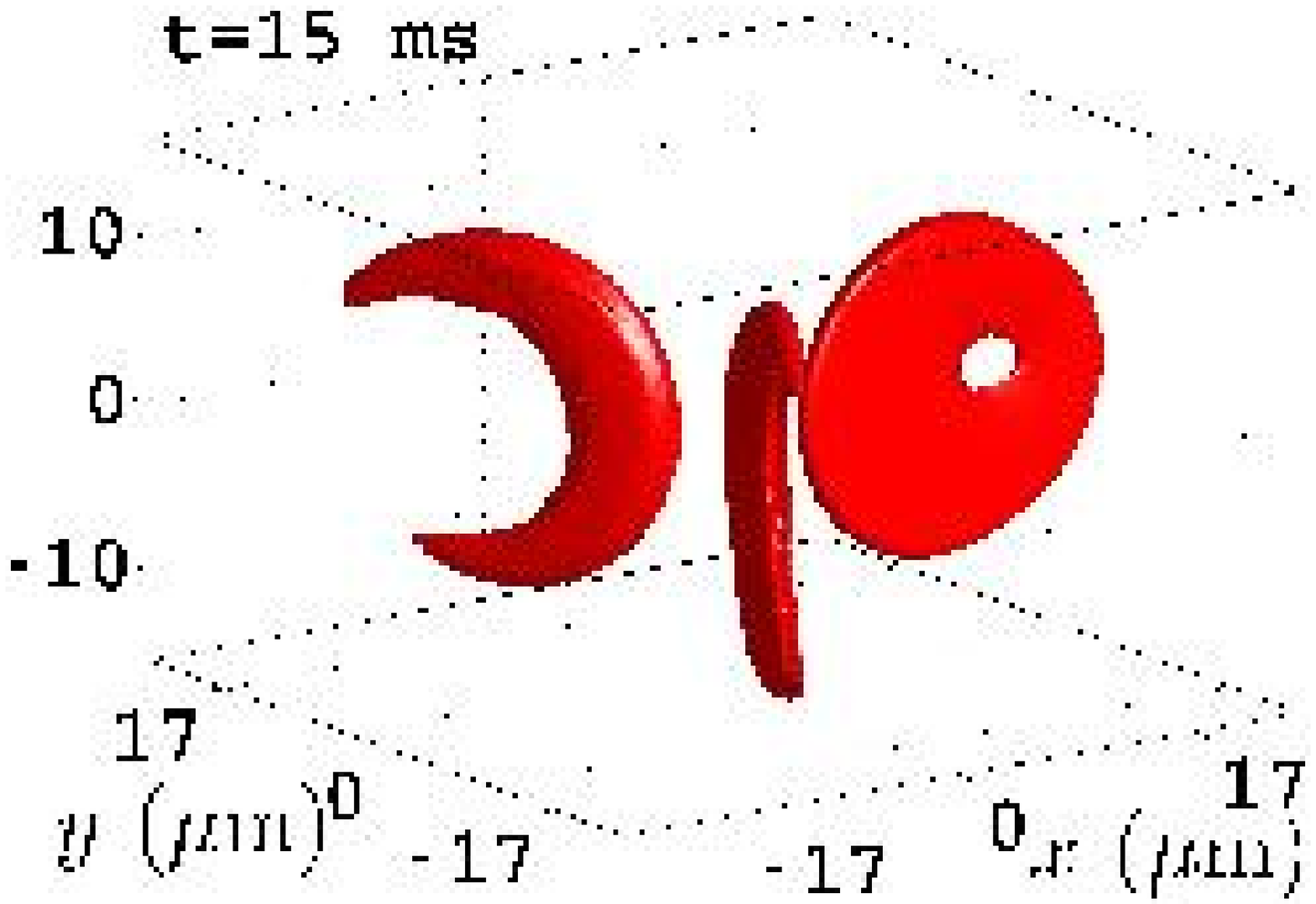}
\includegraphics[width=\figw]{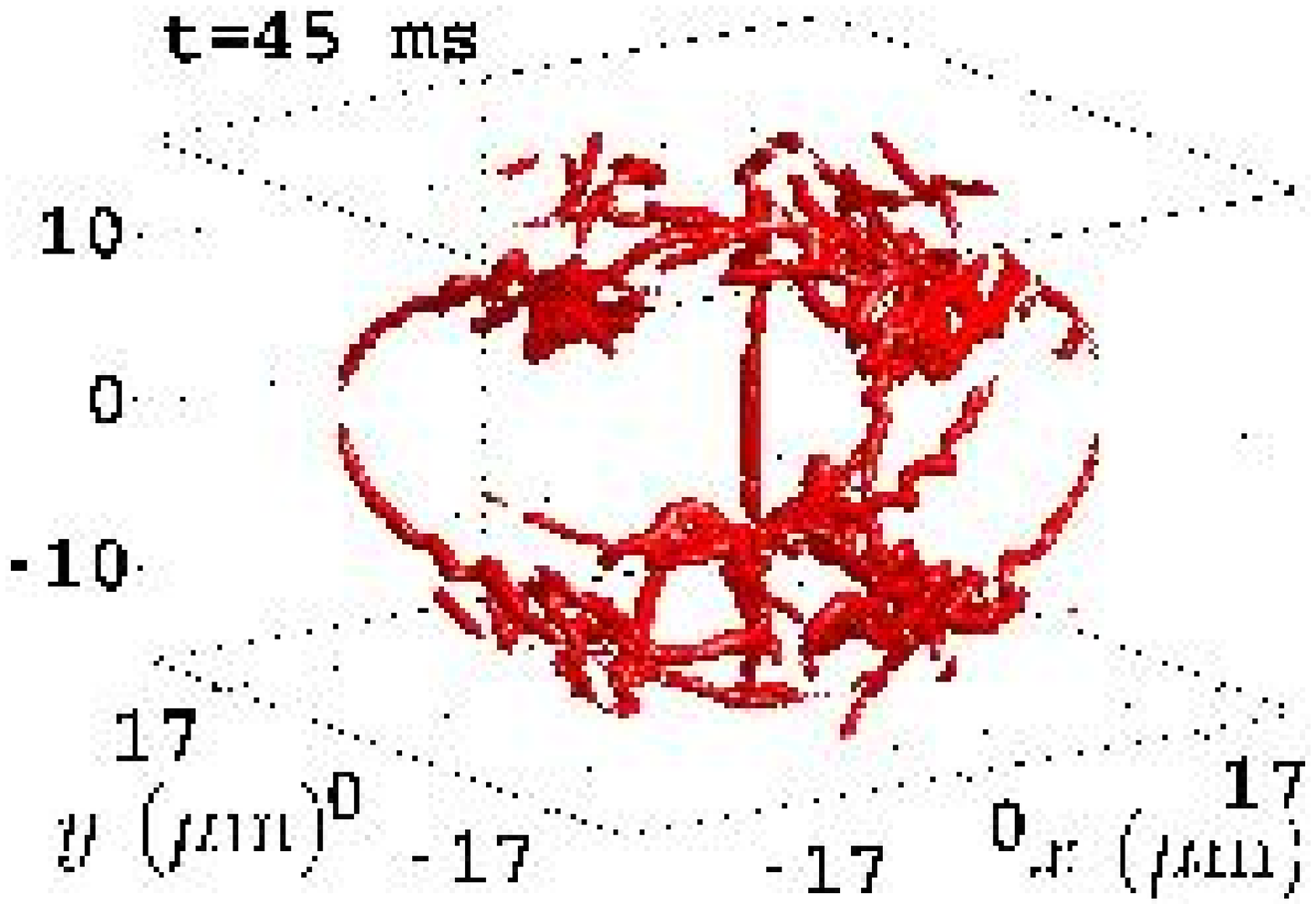}
\includegraphics[width=\figw]{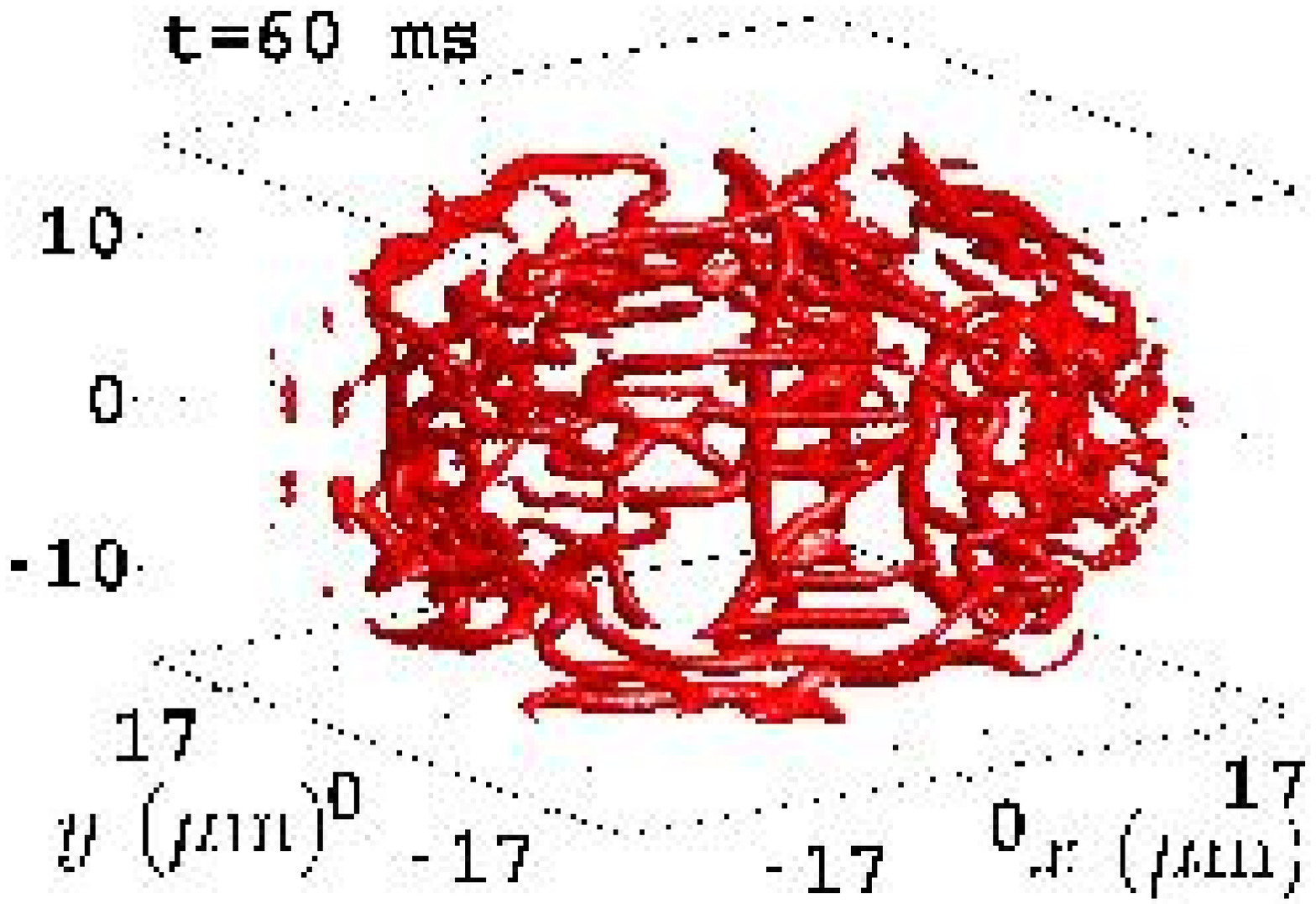}
\includegraphics[width=\figw]{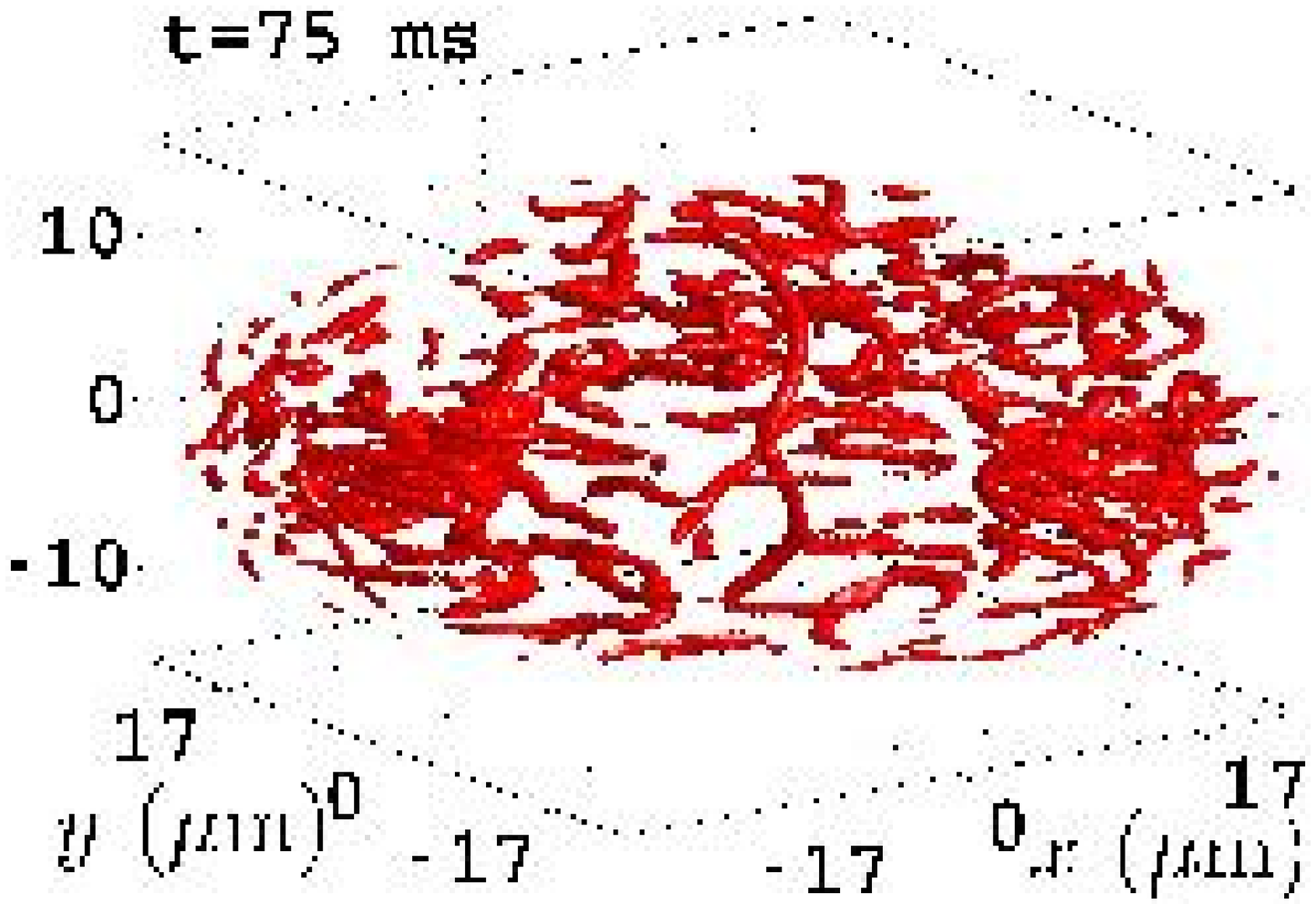}
\includegraphics[width=\figw]{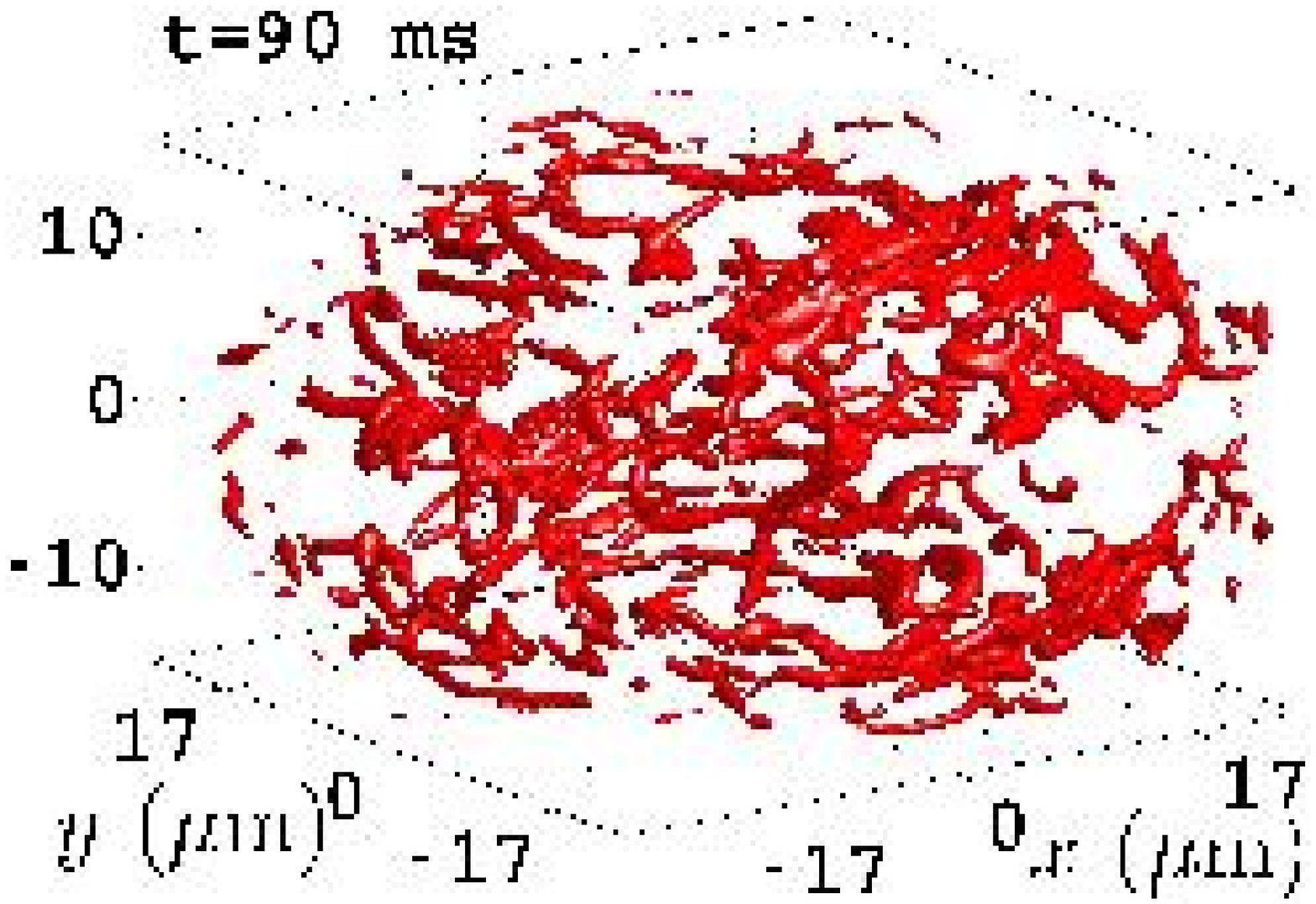}
\\[2.0ex]
\includegraphics[width=\figwt]{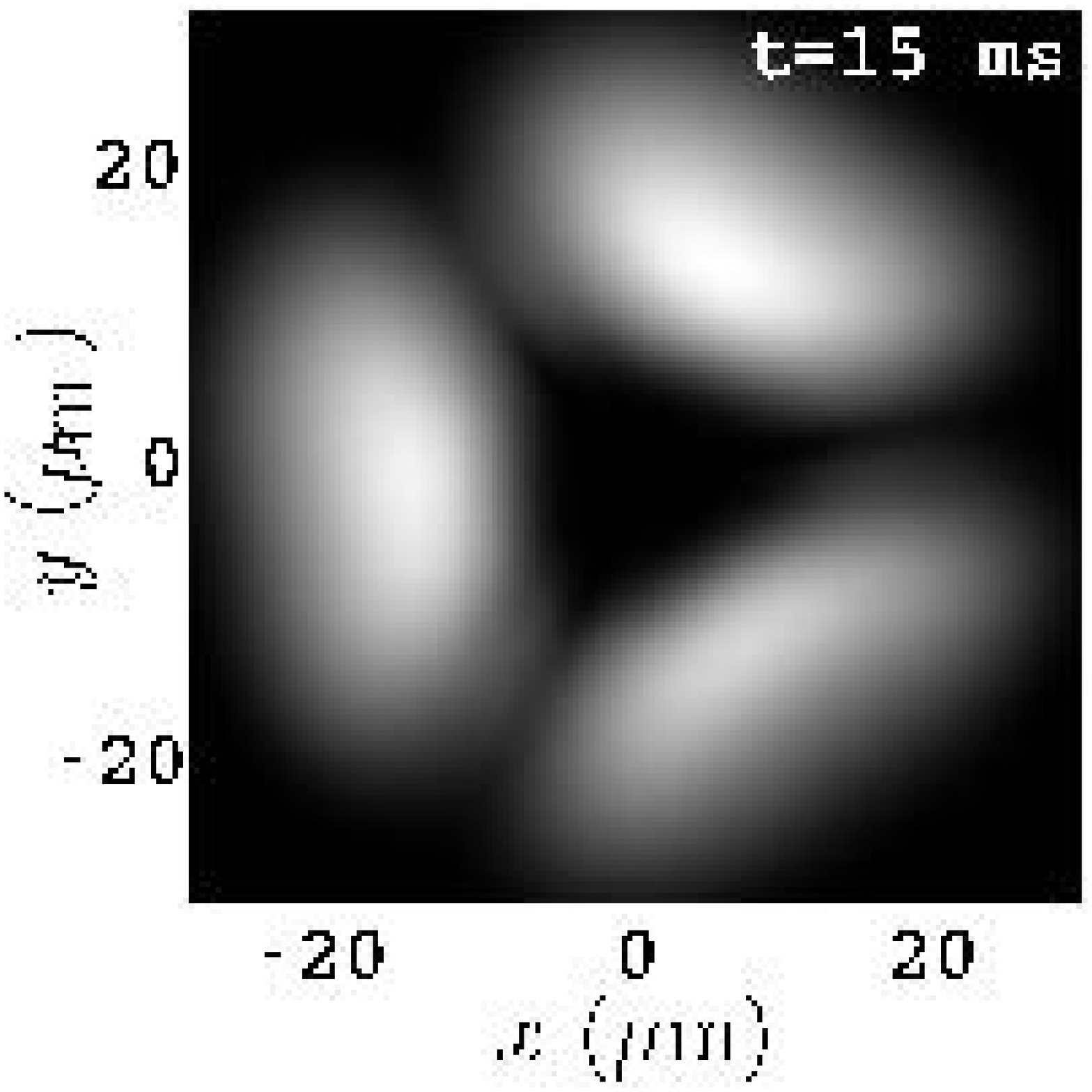}~~
\includegraphics[width=\figwt]{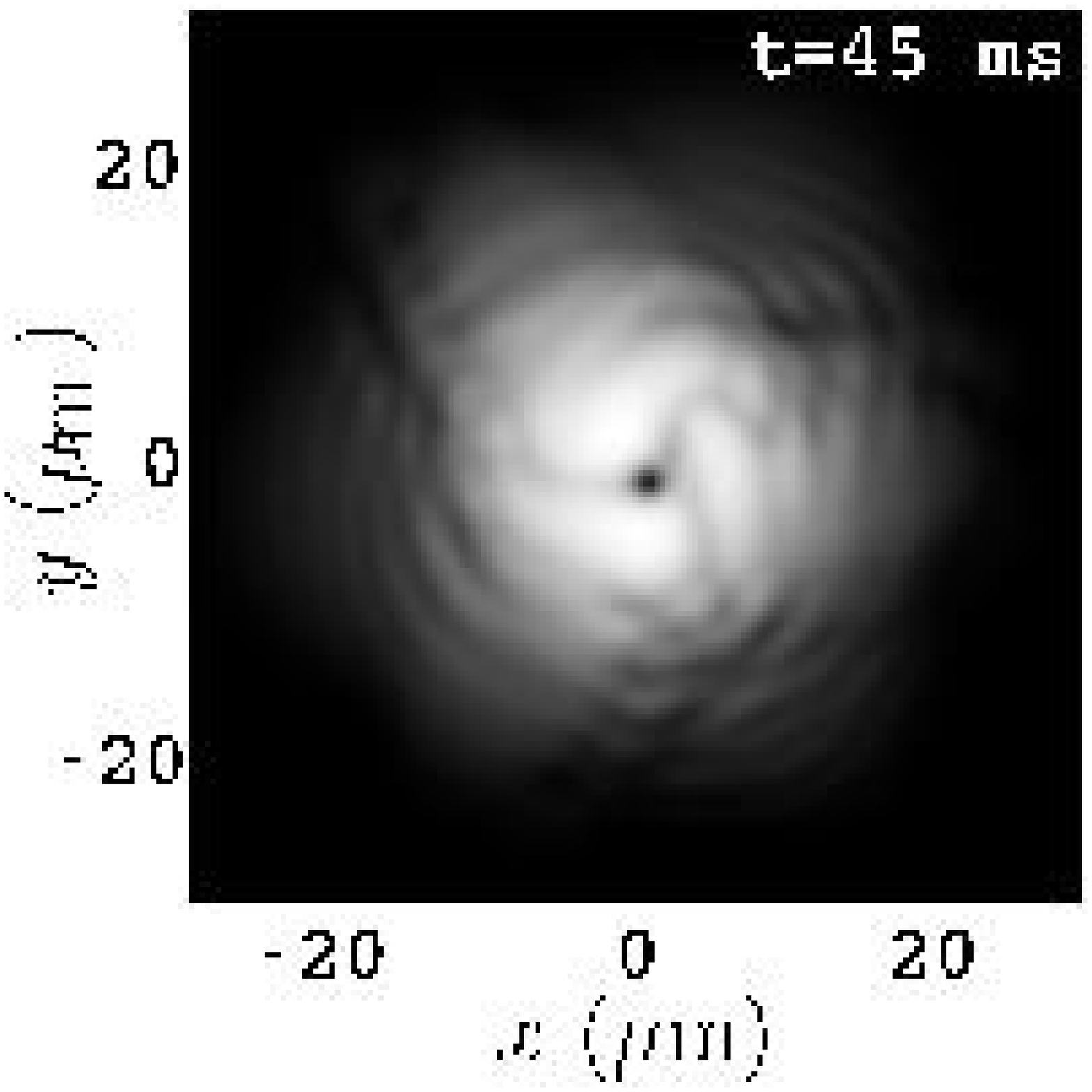}~~
\includegraphics[width=\figwt]{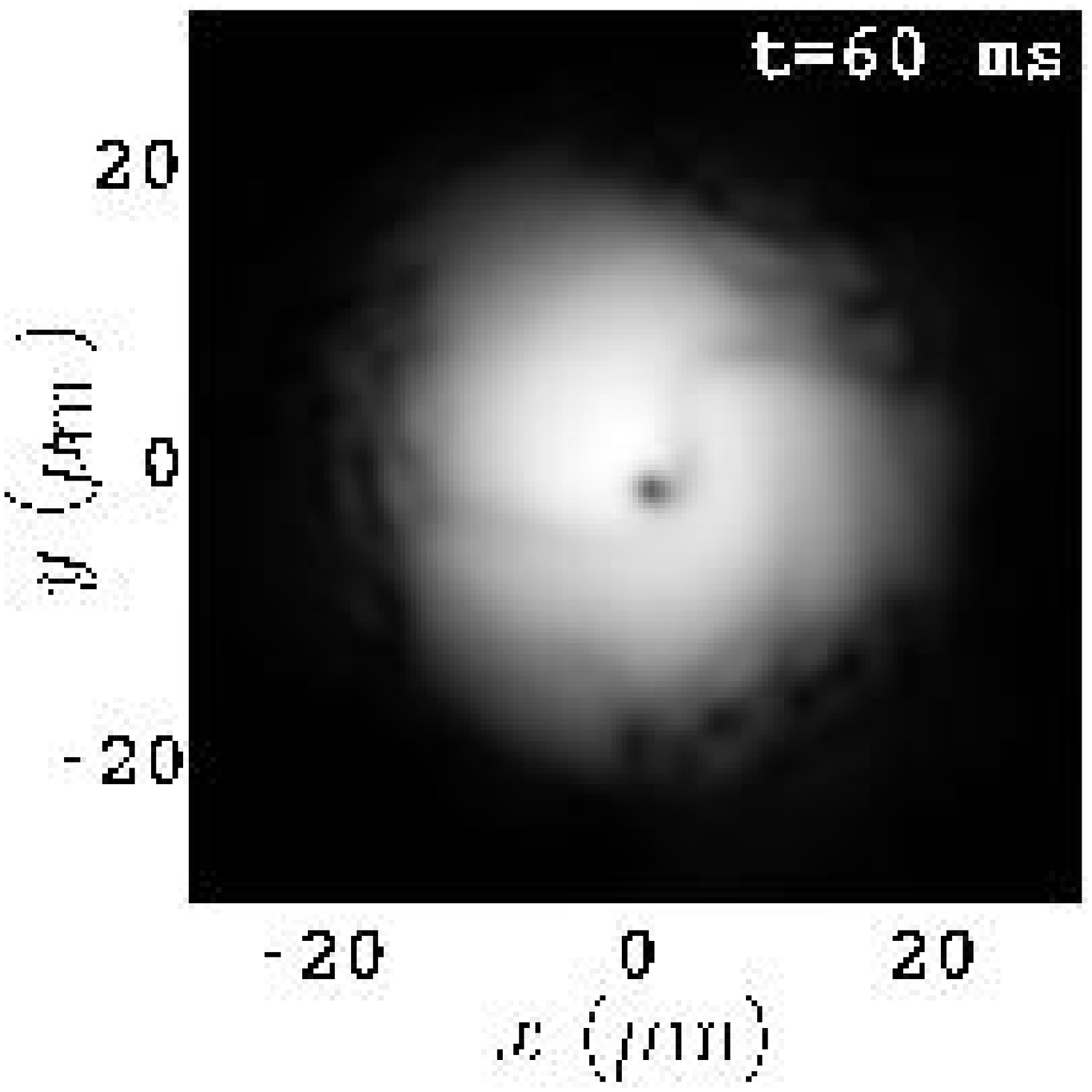}~~
\includegraphics[width=\figwt]{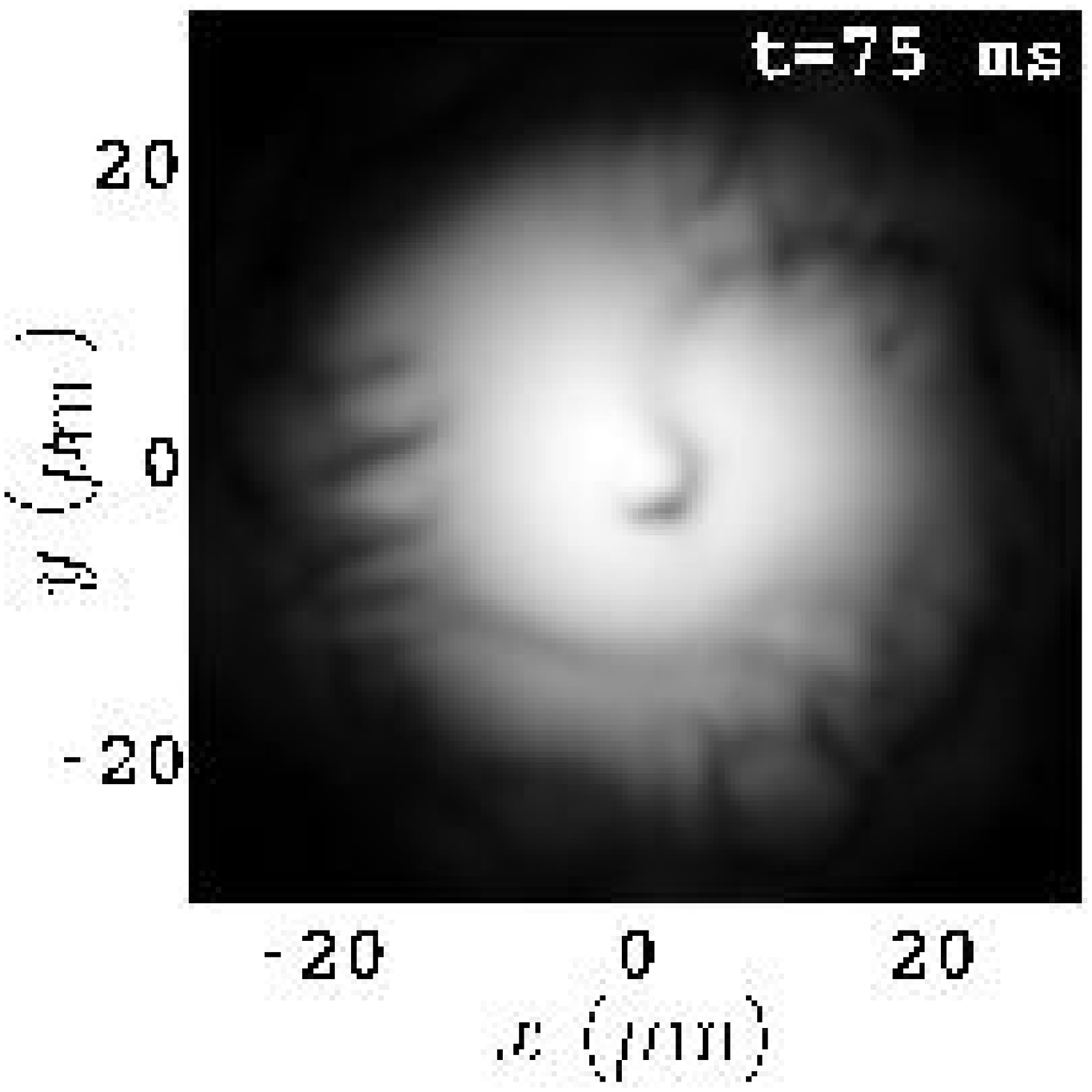}~~
\includegraphics[width=\figwt]{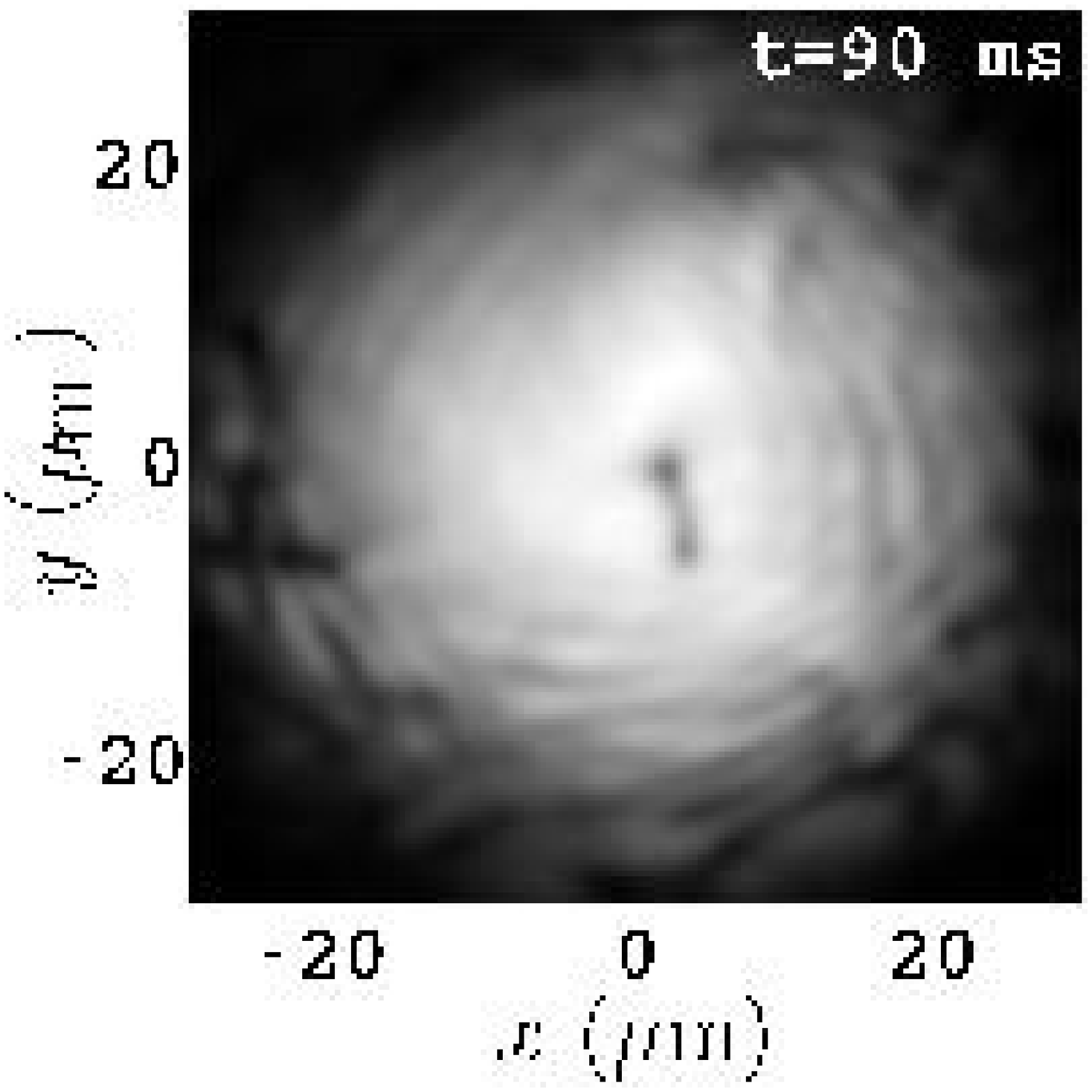}~~
\\[4.0ex]
\end{center}
\vskip-0.4cm
\caption{(Color online)
Same as in Fig.~\ref{3Da} for an initial phase distribution
corresponding to $\phi_k=k2\pi/3$.
}
\label{3Db}
\end{figure*}
\begin{figure*}[htb]
\begin{center}
\includegraphics[width=\figw]{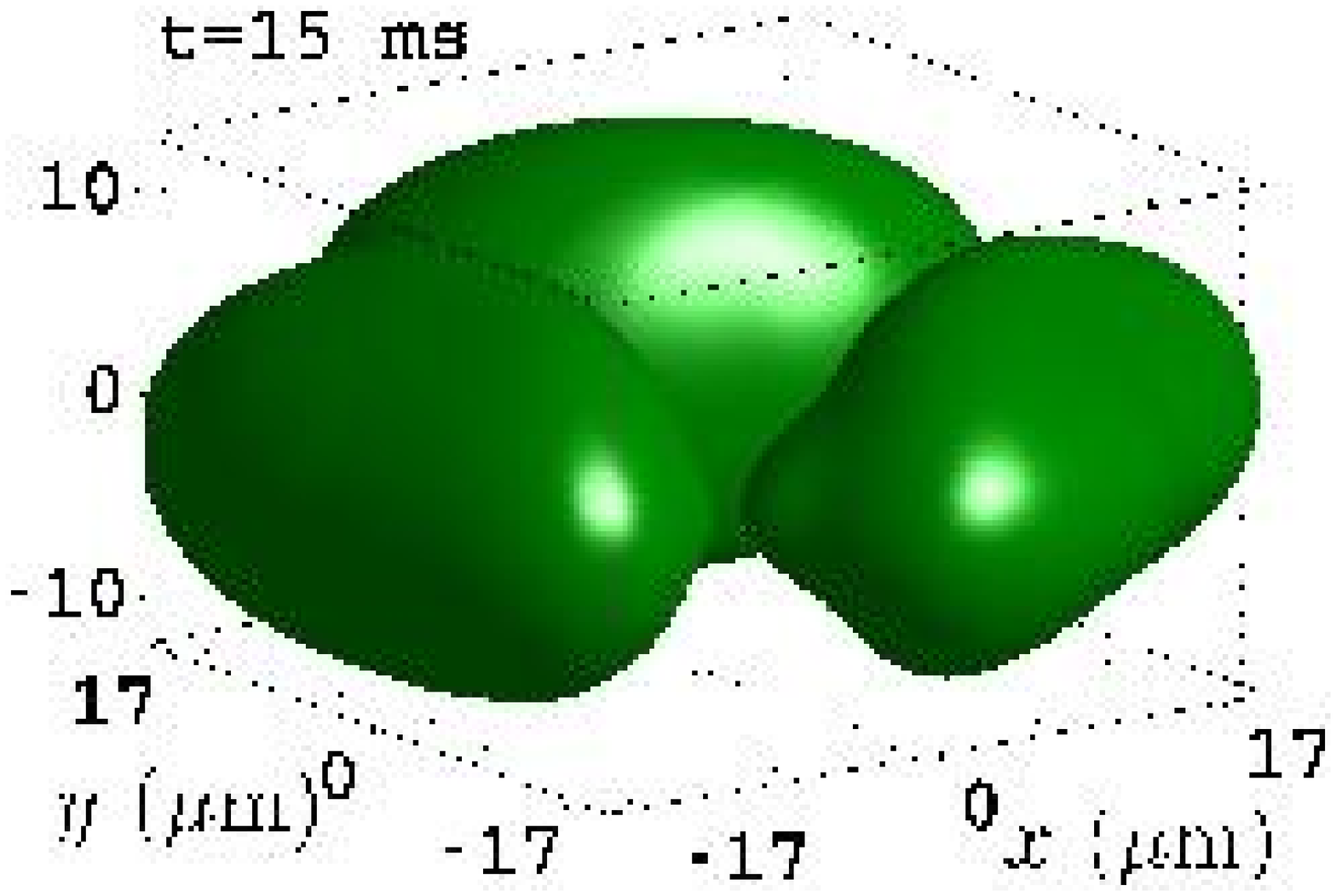}
\includegraphics[width=\figw]{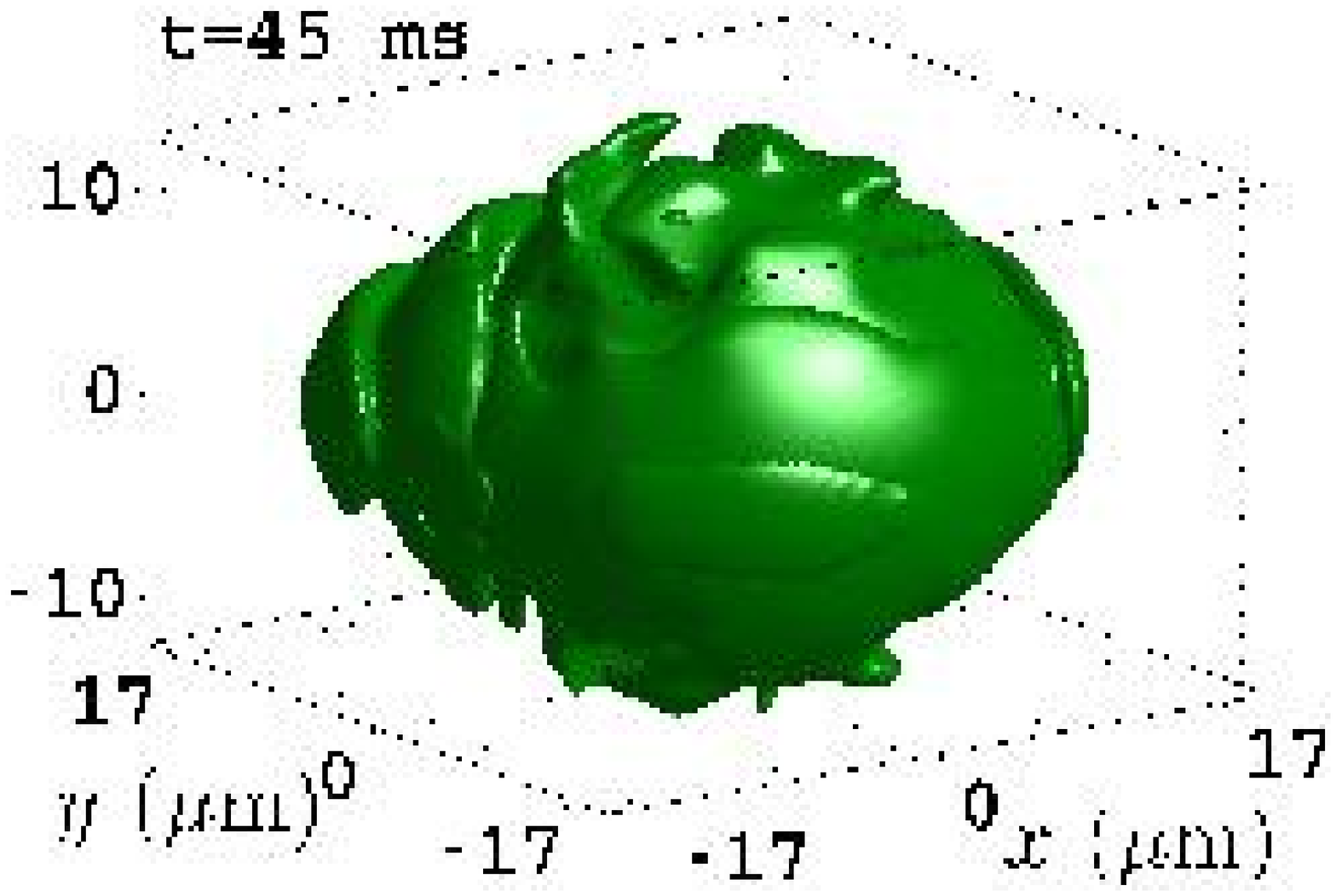}
\includegraphics[width=\figw]{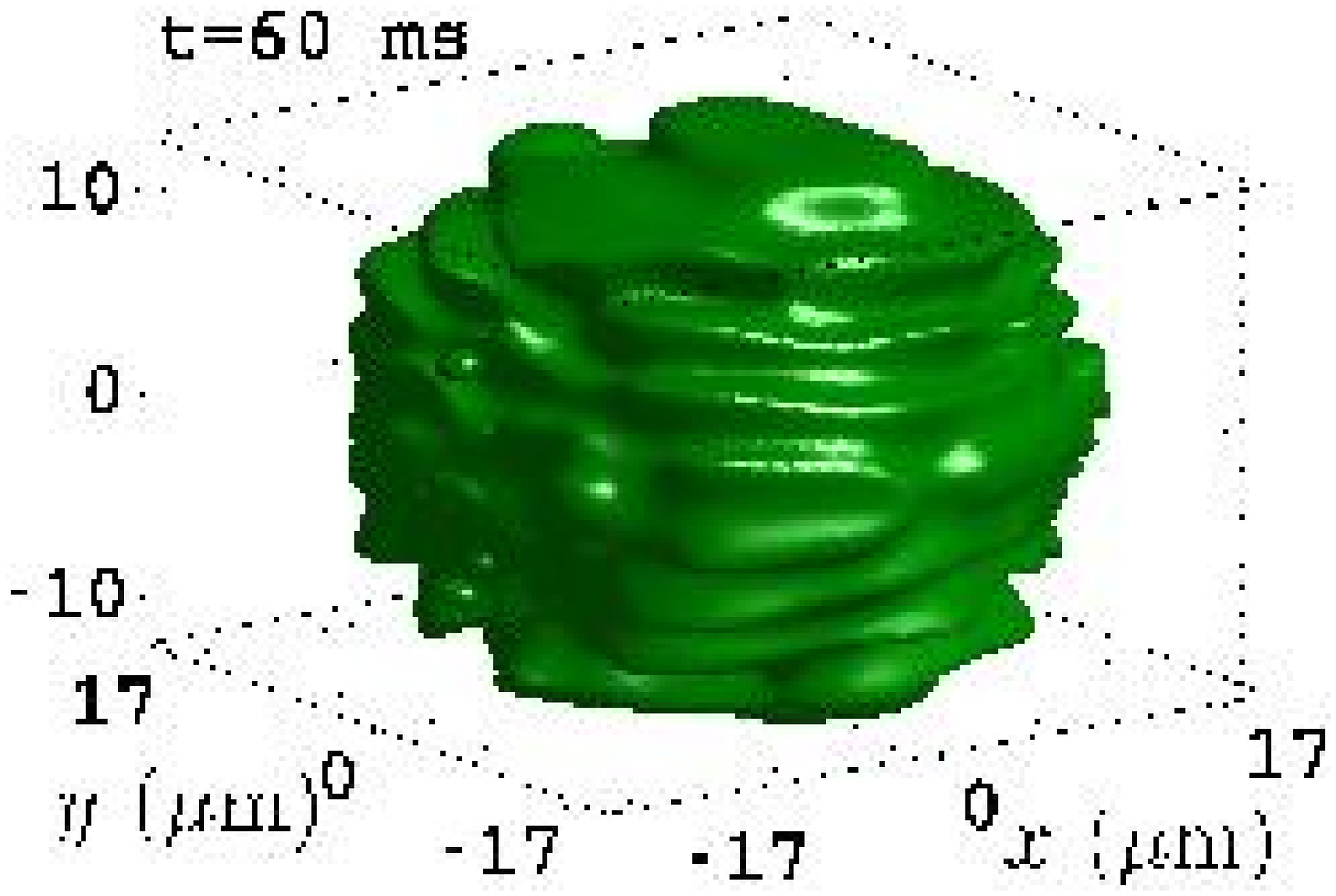}
\includegraphics[width=\figw]{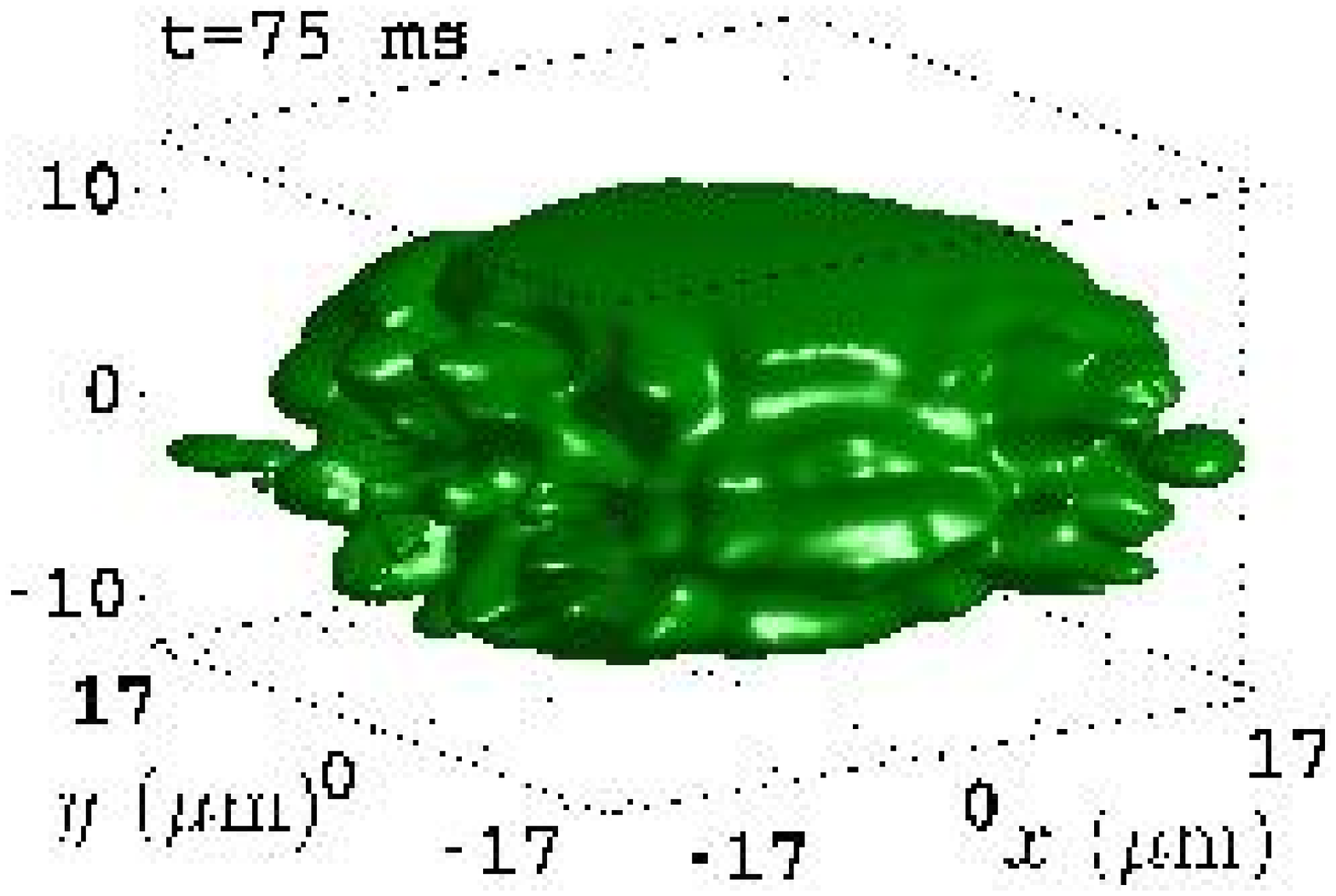}
\includegraphics[width=\figw]{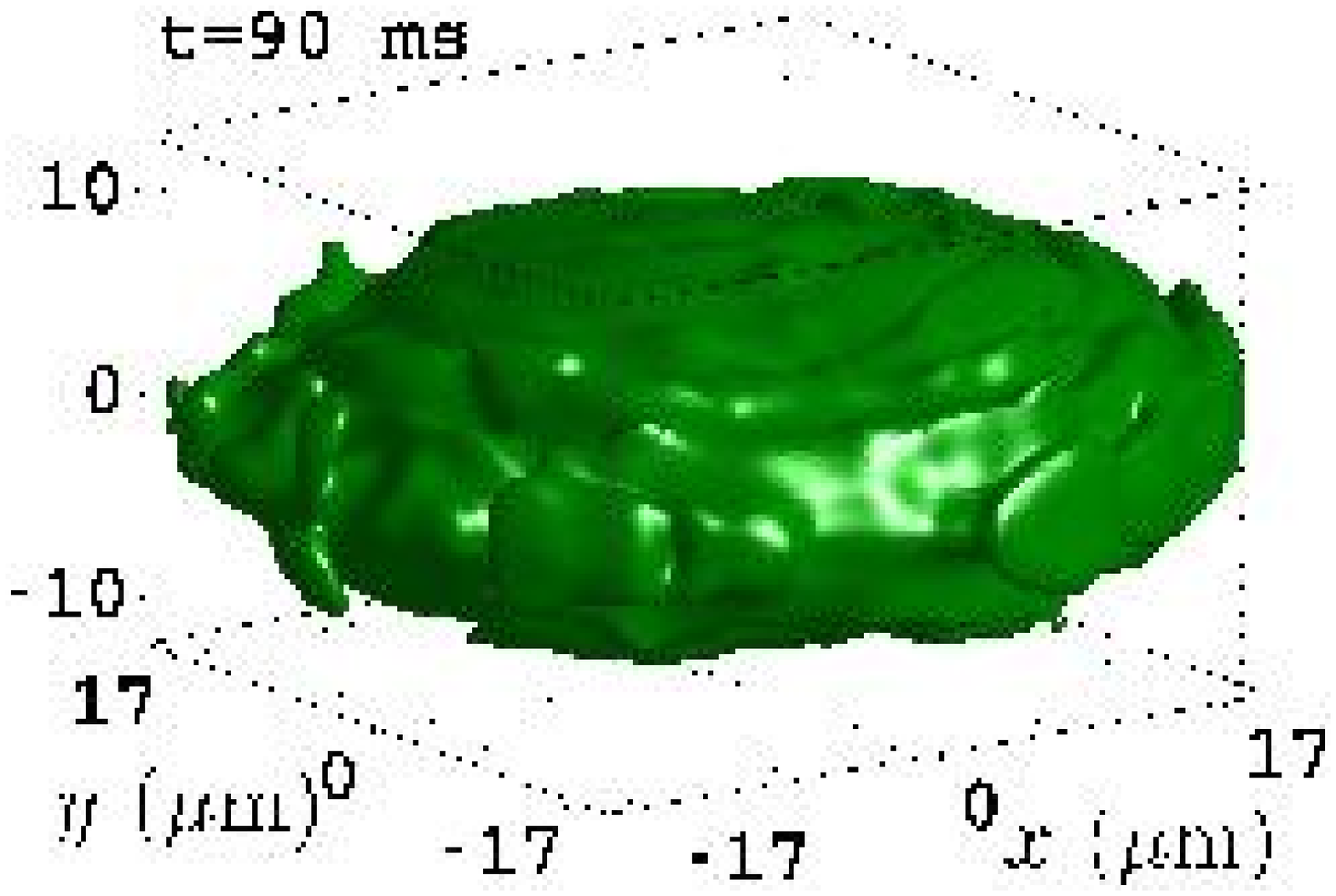}
\\[2.0ex]
\includegraphics[width=\figw]{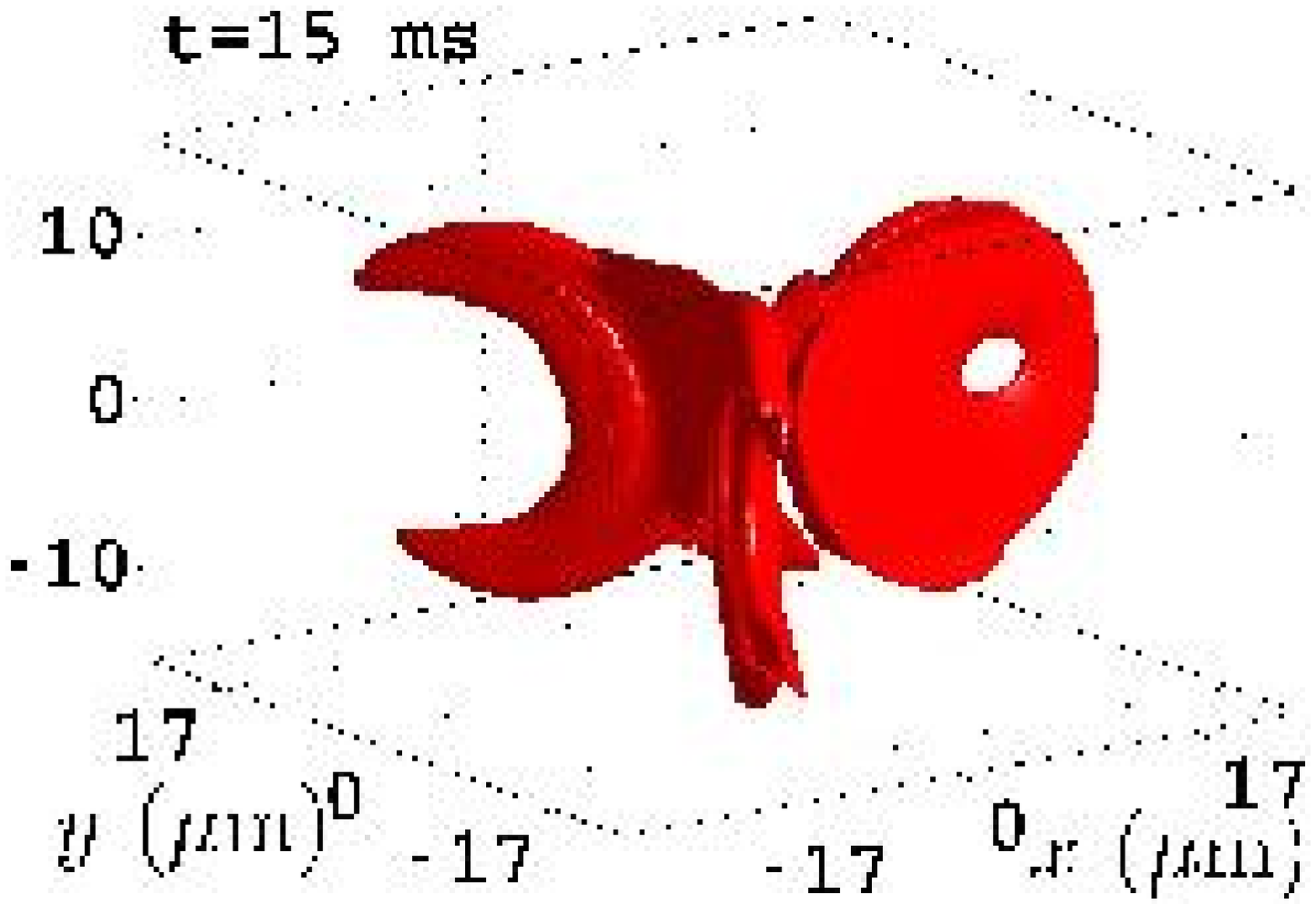}
\includegraphics[width=\figw]{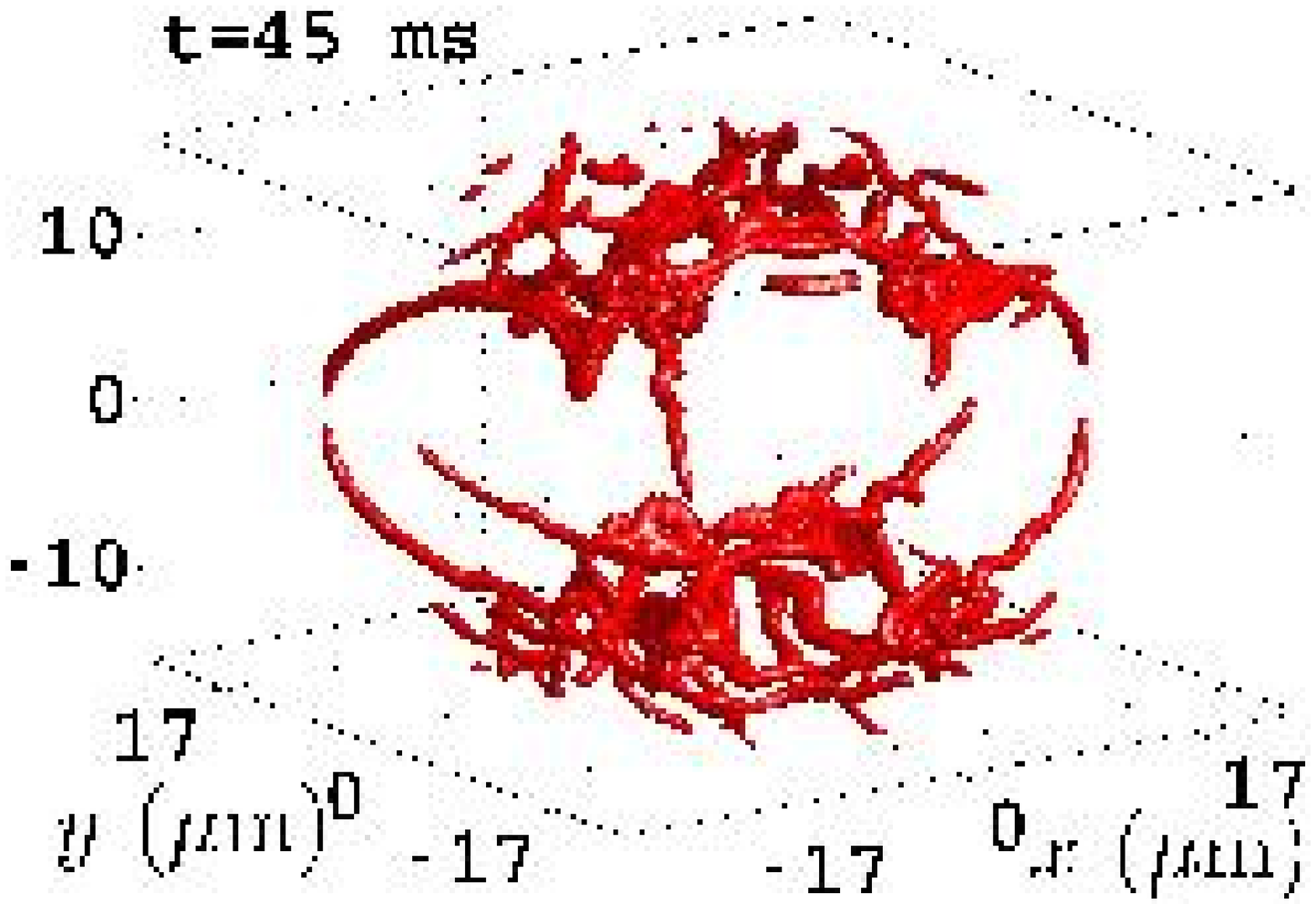}
\includegraphics[width=\figw]{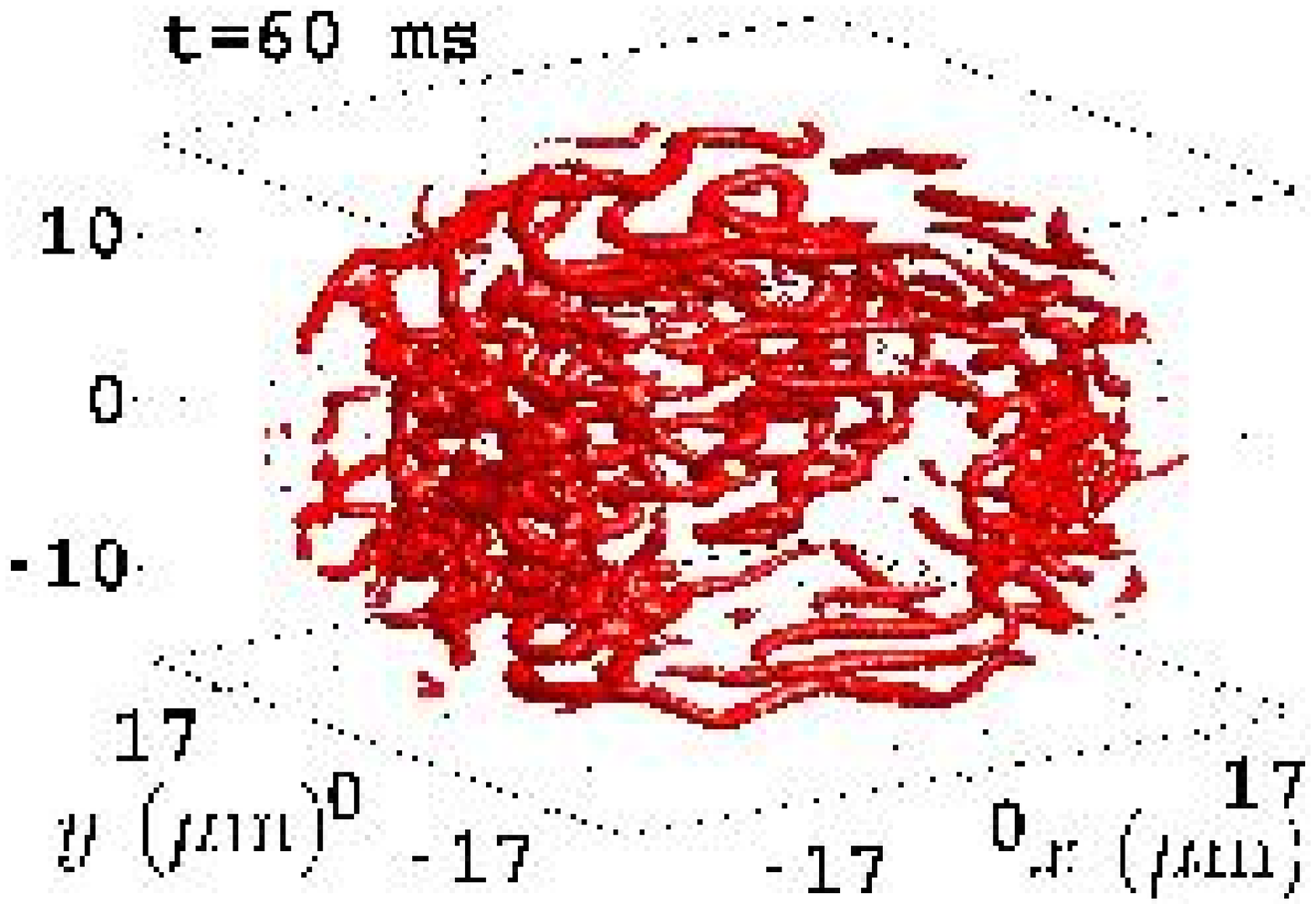}
\includegraphics[width=\figw]{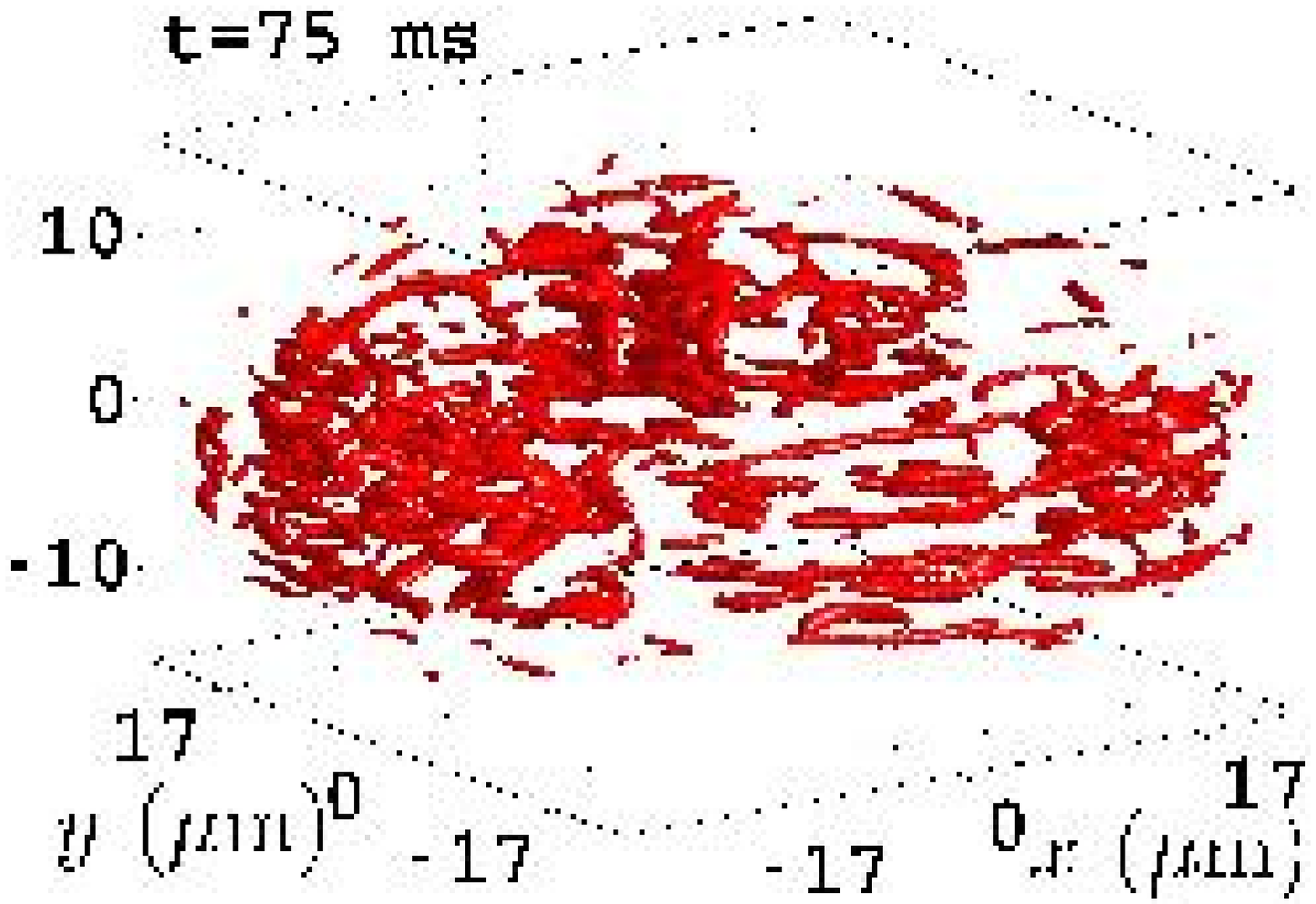}
\includegraphics[width=\figw]{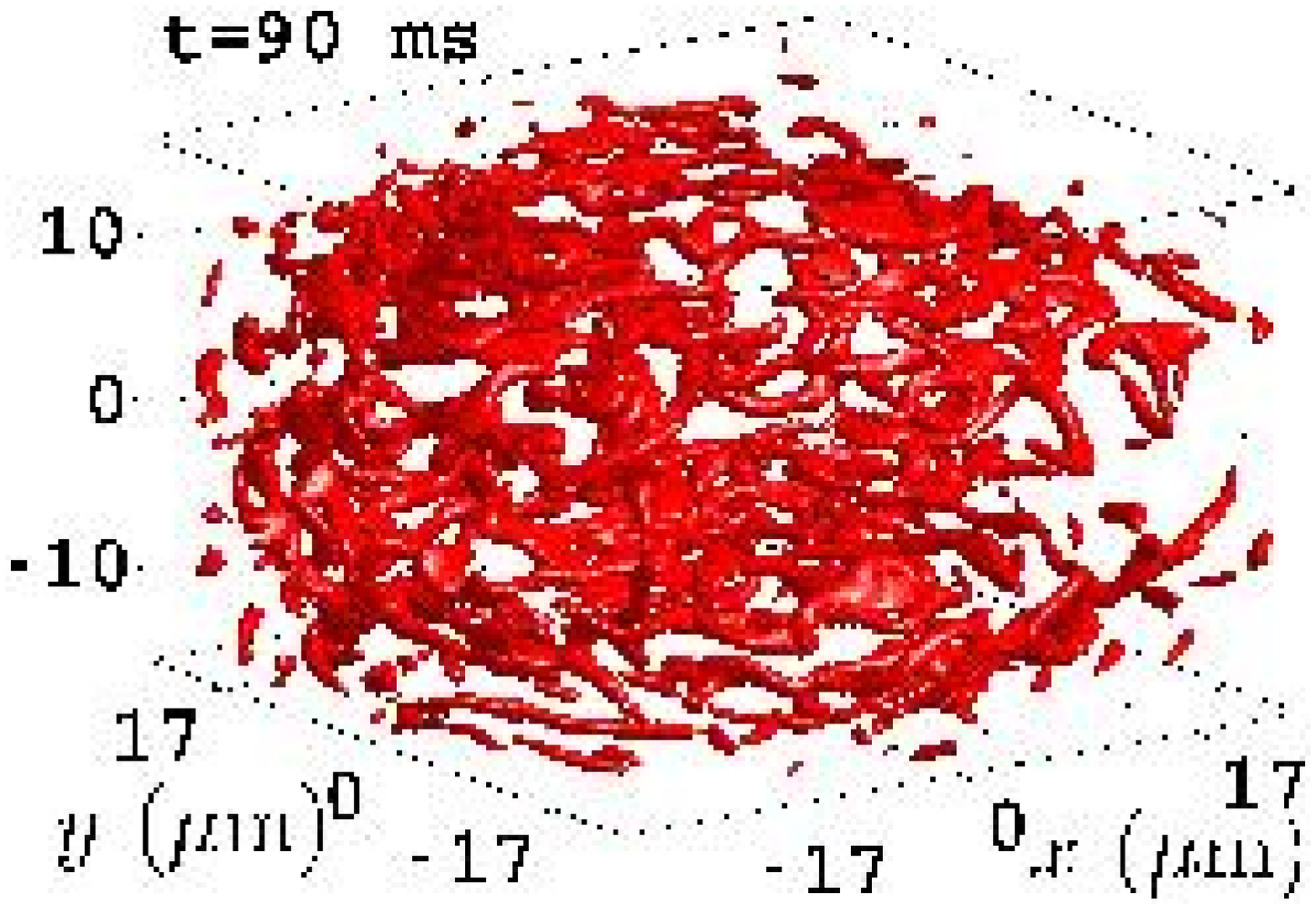}
\\[2.0ex]
\includegraphics[width=\figwt]{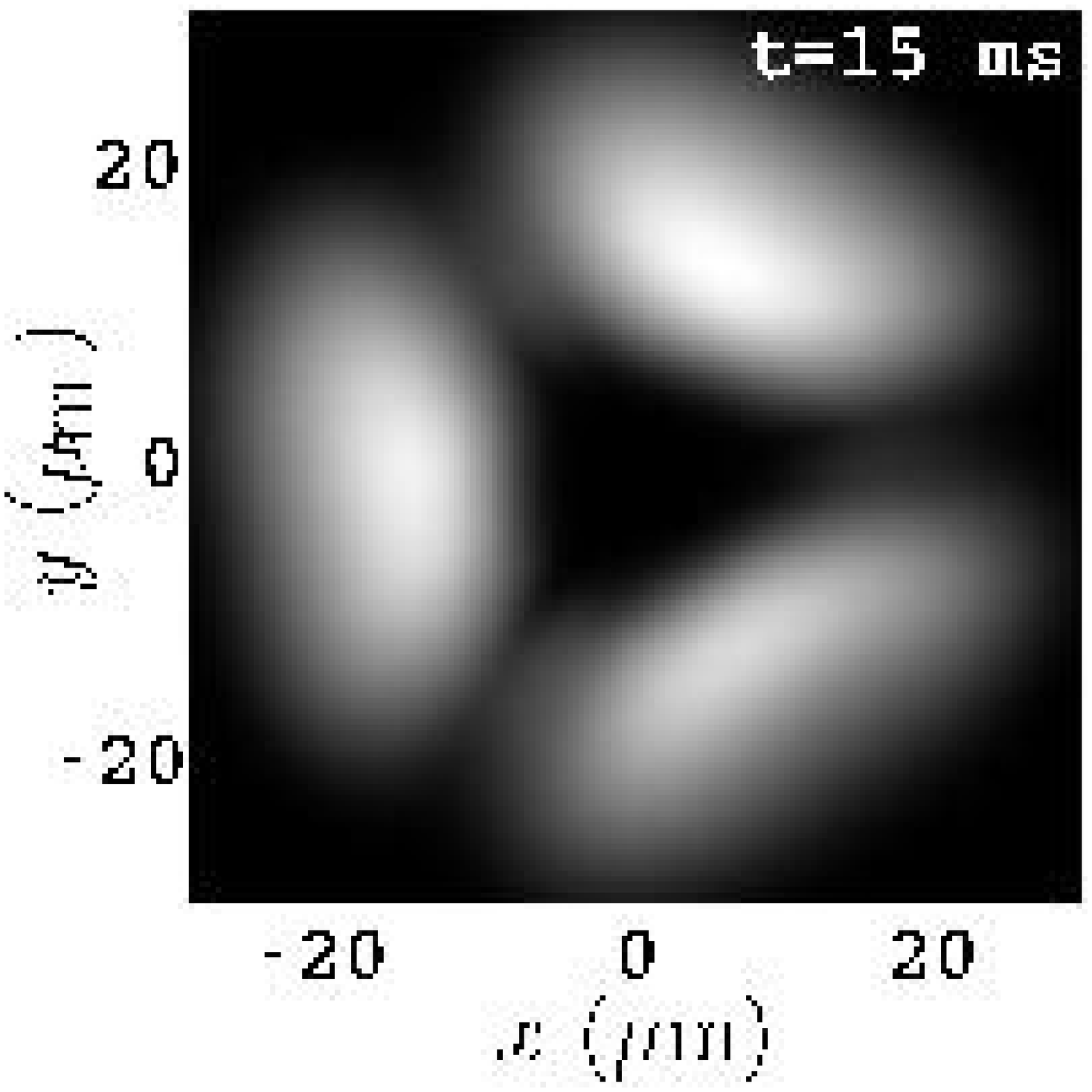}~~
\includegraphics[width=\figwt]{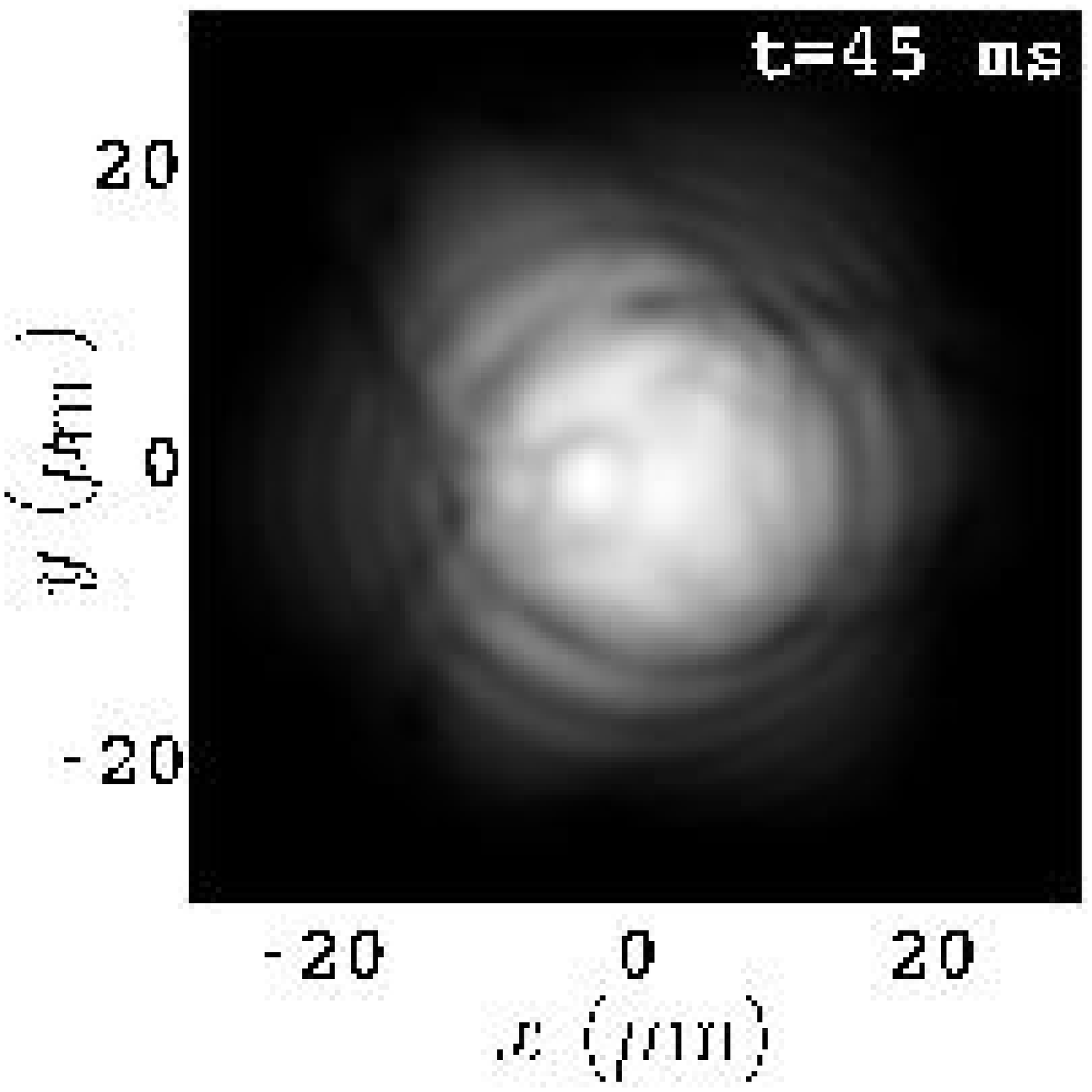}~~
\includegraphics[width=\figwt]{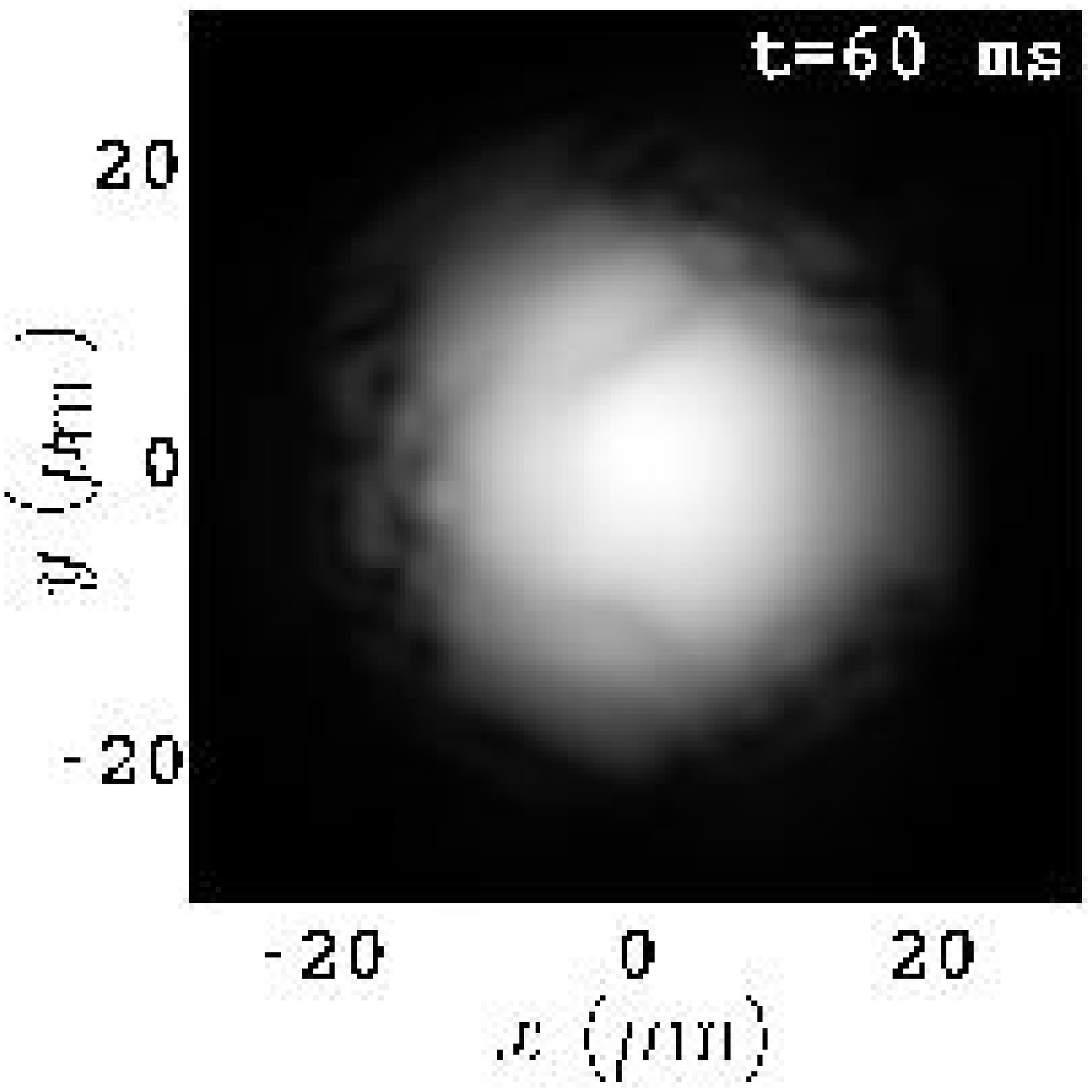}~~
\includegraphics[width=\figwt]{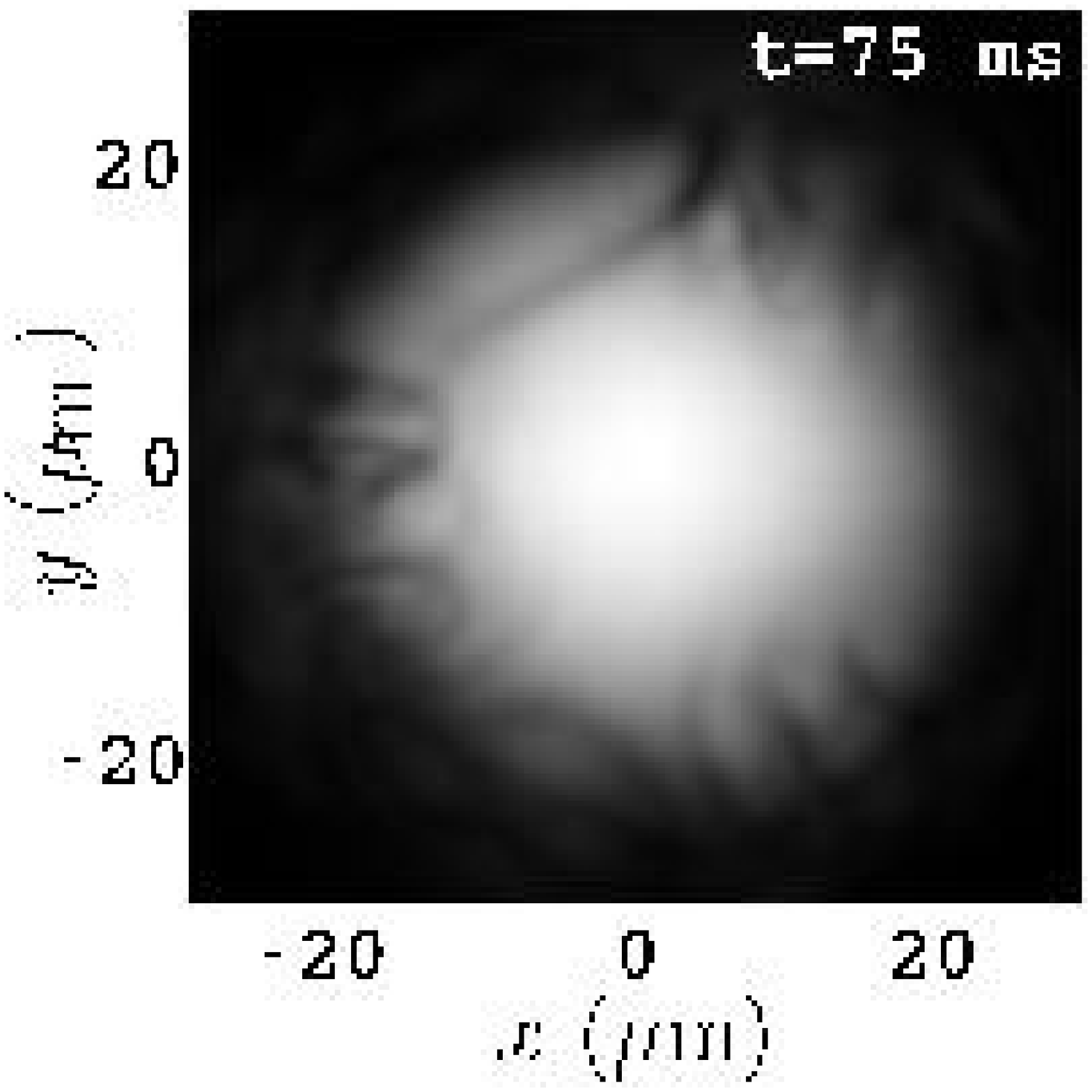}~~
\includegraphics[width=\figwt]{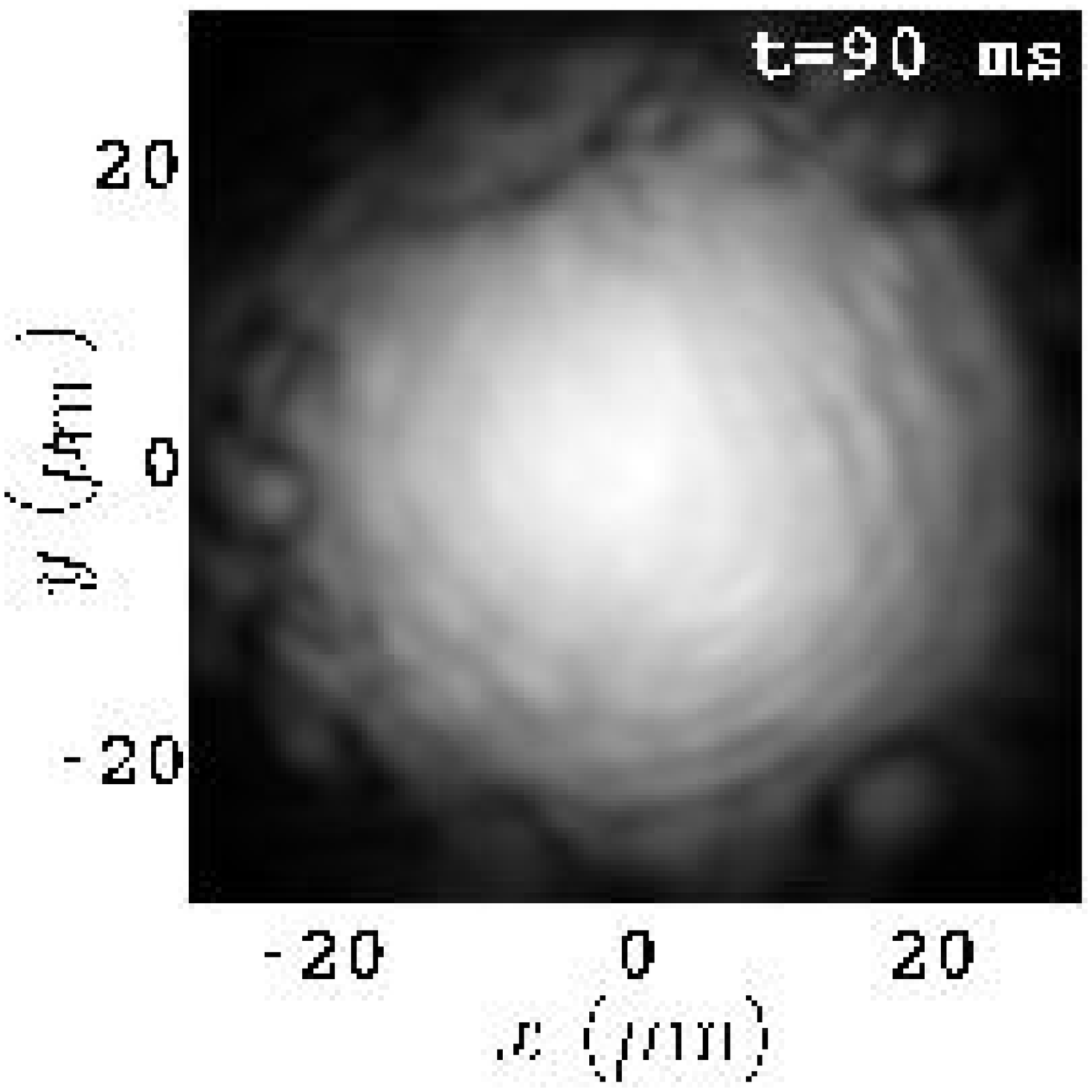}~~
\end{center}
\vskip-0.4cm
\caption{(Color online)
Same as in Fig.~\ref{3Da} for an initial phase distribution
corresponding to $\phi_k=k\pi/3$.
}
\label{3Dc}
\end{figure*}

%%%%%%%%%%%%%%%%%%%%%%%%%%%%%%%%%%%%%%%%%%%%%%%%%%%%%%%%%%%%%%%%%%%%%%%%%%%%%%
\subsection{Three-dimensional BECs.}
%%%%%%%%%%%%%%%%%%%%%%%%%%%%%%%%%%%%%%%%%%%%%%%%%%%%%%%%%%%%%%%%%%%%%%%%%%%%%%

As shown in the previous section, the 2D setting lends itself to a more detailed
examination of important features 
such as the parametric 
dependence on the ramping time and the relative phases of the fragments. 
Nevertheless, it is important to also consider some of the delicate points particular
to the 
3D nature of the experiments and the
observable quantities available within the experimental images.
For this reason, we now focus on the 3D setting, presenting results of the   
simulations 
relevant to the experiment of Ref.~\cite{bpa_prl}. The setup is the same as
in the previous section but we now use the full 3D space with
the {\em same} chemical potential as before.
Typical results are shown in Figs.~\ref{3Da}--\ref{3Dc}
for a ramping down time of the potential barrier of 25 ms.
Figure~\ref{3Da} corresponds to the case of equal initial phases
$\phi_k=0$, while Fig.~\ref{3Db} 
pertains to initial phases $\phi_k=2\pi k/3$, and Fig.~\ref{3Dc} to
phases $\phi_k=\pi k/3$.
The figures depict contour plots of the density (top rows) and
vorticity (middle rows), as well as a $z$-projection of the
density (bottom rows) as it would be observed in the laboratory.
The vortex structure is considerably more complex in the 3D scenario because
the vorticity does not show up as straight vortex lines but
rather as a complex web of vortex filaments in various directions.
As in the 2D case, there is the formation of a
vertical vortex line at the center of the cloud
(cf.~second row in Fig.~\ref{3Db}) for the appropriate relative phases
of the different fragments of the condensate with
the same conditions as before.
Nonetheless, in the 3D case, the central vortex line is prone
to bending as can be clearly seen in the later stages
of the dynamical evolution presented in the
second row of Fig.~\ref{3Db}. In fact, the
vortex bending is even clearly visible in the $z$-projection
(see third row of Fig.~\ref{3Db}).
These 3D numerical experiments are quite revealing in that the laboratory
experiments can only show projections of the density and thus
missing to a great extent are the intricate vortex line dynamics.
Importantly also, between the bending effect and the integrated
view used in the experimental images, it is possible for the
presence of vortex-like structures or filaments to be blurred
(as in the later stages of Fig.~\ref{3Db}) or entirely lost
(as in the later stages of Fig.~\ref{3Da} and Fig.~\ref{3Dc}).

As can be seen from Figs.~\ref{3Da}--\ref{3Dc}, the vorticity
emerges at the early stages of the merger ($t<20$ ms), through
vorticity sheets that nucleate some of the vortex line 
structures. Nonetheless, it is interesting to note that most 
of the vorticity is carried by vortex lines and vortex rings 
that are {\em horizontal} (except the notable case of 
the vertical vortex line depicted in the second row of 
Fig.~\ref{3Db}). This fact is also 
quite visible in the density contour plots for $t=60$ ms where 
the horizontal vortex lines ``pinch'' the cloud and create 
peripheral horizontal ridges around the cloud. 
It is also possible to observe some vorticity in the bulk
of the cloud that does not directly come from the phase
differences between the initial fragments, but from the
actual turbulence that is created by the fragment collision.
As an example, two small vortex rings are clearly visible in
Fig.~\ref{3Da} for $t=60$ ms (one close to the top of the
cloud and the other one 1/3 from the bottom).
We would like to stress the difficulty of capturing
the vorticity at the edge of the cloud (where most vorticity
is actually observed) in our numerical experiments. This is due to the
fact that the vorticity is defined as the curl of the fluid
velocity of Eq.~(\ref{fluid_vel}) that is normalized
by the density. The numerical effect is that close
to the periphery (where the density is small) the fluid velocity
corresponds to the ratio of small numbers which imposes
great numerical difficulties.
Nonetheless, by using a fine grid of $301\times 301\times 121$ we
are able to capture most of the delicate vorticity dynamics at
the periphery of the cloud consisting, mostly, of horizontal
vortex lines that are parallel to the periphery of the
cloud.

Another interesting phenomenon is the oscillation of the atomic cloud. 
The cloud starts with a larger horizontal extent compared
to the vertical one and after merger creates an almost spherical
cloud, which in turn elongates again in the horizontal direction
after the fragments go ``through'' each other. This behavior
repeats a few times until the cloud takes an approximate 
spherical shape (results not shown here).

We have also monitored the effects of damping due to the coupling 
of the condensed atoms to the thermal cloud. Equation~(\ref{gpe}) 
is obtained by supposing a dilute Bose gas
at a temperature close to {\em absolute zero}.  
However, at finite temperatures, but still smaller than
the critical temperature $T_c$ for condensation, a fraction of the
atoms are not condensed and form the so-called thermal cloud.
In turn, this thermal cloud induces a damping on the
dynamics of the condensed cloud.
We used the approach of phenomenological damping
\cite{PitaevskiiPhenomenology} described in Refs.~\cite{Ueda02,Ueda03}
that relies on replacing the $i$ in front of the time
derivative in Eq.~(\ref{gpe}) by $(i-\gamma)$, where $\gamma$
is the damping rate, and by renormalizing the solution at
each iteration to keep the initial mass (number of atoms) constant
during integration. We tested values of $\gamma$ in the
interval $[0.01,0.1]$ that contains the value of 0.03
estimated in Ref.~\cite{Choi} for a temperature $T = 0.1 T_c$.
The results of the phenomenological damping are, qualitatively,
very similar (results not shown here)
to the effects of ramping down the barrier
over longer time scales:
larger damping resulting in a stronger suppression of
the vorticity generated by the merger of the
different cloud fragments.

%%%%%%%%%%%%%%%%%%%%%%%%%%%%%%%%%%%%%%%%%%%%%%%%%%%%%%%%%%%%%%%%%%%%%%%%%%%%%%
\section{Comparison of Numerical and Experimental Results \label{SEC:expt}}
%%%%%%%%%%%%%%%%%%%%%%%%%%%%%%%%%%%%%%%%%%%%%%%%%%%%%%%%%%%%%%%%%%%%%%%%%%%%%%

The simulations described in the present work were directly aimed at developing a more thorough 
understanding of both the experimental results of Ref.~\cite{bpa_prl} and the dynamics of vortex 
formation during BEC merging and collisions. In this section, we thus briefly discuss the observed 
similarities and differences between the experimental and theoretical results.

\begin{figure}
\begin{center}
\includegraphics[width=8.5cm]{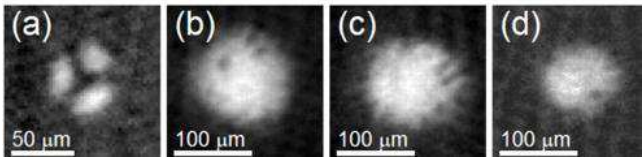}
\end{center}
\vskip-0.4cm
\caption{(a) In-situ phase-contrast image of three BECs trapped in a triple-well 
potential. (b)--(d) Absorption images of BECs after 56 ms of ballistic 
expansion. Each BEC was created by merging three BECs, as described in 
the text. 
}
\label{fig_expt}
\end{figure}

In the laboratory experiment, an optical potential was used to segment a harmonic trap before 
condensation was achieved; with additional evaporative cooling, three initially isolated and 
mutually independent (i.e., uncorrelated phases between the different fragments) 
condensates were created. A phase-contrast image of three such BECs is given 
in Fig.~\ref{fig_expt}(a), which can be directly compared with the simulated data of Fig.~\ref{3Da} 
(bottom row, left image). The initial condition of incoherent condensate fragments serves as the 
conceptual basis behind the motivation to impose and examine various relative phases between the 
fragments in this work's simulations. In the experiment, the optical barrier was ramped off approximately 
linearly over time scales between 50 ms and 3 s, significantly longer than the time scales 
considered in the simulations. At the end of the merging process, 
immediately after the barrier was completely removed, the fully merged BEC was released from the 
harmonic trap and allowed to ballistically expand for 56 ms to enable imaging of vortex cores. 
Example images of merged and expanded BECs are shown in Fig.~\ref{fig_expt}(b) through (d). 
These images can be compared with the simulated data of Figs.~\ref{3Da} through \ref{3Dc} (bottom rows); 
note that the simulations do not involve an expansion stage.

In both experimental results and in the simulations shown here, we note the following important similarities. 
First, it is clear that vortex cores may be formed during the merging process. Second, as noted in the experimental work, 
the vortex formation process should depend upon relative phases between the condensates. This conclusion is borne out by 
the present work. Specifically, the simple analysis of a slow merging process leading to a 25\% probability of vortex formation, 
as presented in the experimental work, matches the results of the simulations summarized in Fig.~\ref{diagram}. Finally, 
experimental and simulated results show that faster merging leads to more vortices initially created, but these vortices 
may self-annihilate with time by holding the fully merged BEC in the trap.

However, there is one notable quantitative difference between the experimental and numerical results. In the experimental work, 
for barrier ramp-down times longer than 1 s, single vortices were experimentally observed in approximately 25\% of the images 
obtained directly at the conclusion of the merging process. Multiple vortex cores were not observed under these conditions. 
For faster ramps, experimental images containing either single or multiple cores were more often obtained, with significantly 
more than 25\% of the images containing at least one vortex core. The results of the numerical data show that the slow merging 
limit (where multiple vortices cease to be created during merging) is reached for merging times that are much shorter than in 
the experiment. In other words, images with multiple vortices are seen in the experiment under conditions where the numerical 
results would suggest that a given BEC should have at most one vortex.

There are a few possible sources of this discrepancy. At first glance, it might appear that the spontaneous formation of 
vortices in BECs during evaporative cooling in an axi-symmetric \emph{harmonic} trap, as noted in Ref.~\cite{bpa_prl}, 
could play a role in higher percentage of vortices seen in the experiment. Such vortex formation processes 
can not be described by the GPE and are thus not observable in the simulations of this work. However, due to angular 
momentum damping and self-annihilation of vortices in the asymmetric local potentials in which the three BEC fragments grow, 
we believe that vortices that might be spontaneously created in one (or more) of the three BECs are unlikely to survive 
at rates that would affect the experimental observations of Ref.~\cite{bpa_prl}. This possible source for the quantitative 
discrepancy could be tested, for example, with simulation methods based on the stochastic GPE \cite{MattDavis}.

Perhaps a more likely source of the quantitative discrepancy may lie in the optical potential energy or shape; 
differences between the experiment and simulations regarding barrier heights, widths, and ramp-down trajectory might 
induce more vortices to form during merging. For example, if center-of-mass oscillations of the cloud were induced in 
the experiment, atomic fluid flow around the central portion of the optical barrier could induce formation of vortices 
and lead to increased vortex observation rates in the experiment. Such processes could be studied in future GPE simulations 
in order to further characterize dynamical processes that may be involved in vortex formation. It might also be possible 
that imperfections in the true optical barrier used in the experiment could pin vortices for a portion of the barrier 
ramp process, and significantly alter the vortex formation and annihilation process. Finally, much of the present work 
focused on the central portion of the trap, a region that encompasses the center of the merged BEC. In the experimental work, 
however, vortices were most often seen further from the BEC center.

%%%%%%%%%%%%%%%%%%%%%%%%%%%%%%%%%%%%%%%%%%%%%%%%%%%%%%%%%%%%%%%%%%%%%%%%%%%%%%
\section{Conclusions \label{SEC:conclusions}}
%%%%%%%%%%%%%%%%%%%%%%%%%%%%%%%%%%%%%%%%%%%%%%%%%%%%%%%%%%%%%%%%%%%%%%%%%%%%%%

We have 
studied the formation and subsequent evolution of vortex structures and filaments 
in a system directly simulating the experimental setup of Ref.~\cite{bpa_prl}. 
In particular, we have considered the case of three independent 
fragments (of variable initial relative phases) and how these 
merge upon the ramping down and eventual removal of the optical 
barrier that separates them. 
While there are many similarities between the numerical results and the experimental results 
of Ref.~\cite{bpa_prl}, the numerical simulations importantly show features and new dynamics 
not discussed or observed in the experimental work. 

The first part of our study concerns the simpler two-dimensional setting, where it is 
straightforward to observe the interference of the independent matter waves, the ensuing 
formation of vortices, as well as their motion within the cloud, as a function of different
parameters such as the ramping-down time or the initial relative phases between the fragments. 
Different diagnostics for the vorticity were developed in the process (such as the $z$-component
of the angular momentum, or the integrated velocity of the flow
throughout the cloud or near its center) and their dynamics was
explained based on the evolution simulations. Principal findings
of this part of the work included the formation of smaller numbers
of vortices as the ramping-down time was increased and the formation
of a single vortex in the core of the condensate for appropriate, 
discrete-vortex-like relations between the phases of the different fragments.

The second part of our work explored how the features found in the two-dimensional setup 
are generalized in a fully three-dimensional setting, and how these affect the measurement 
process through, e.g., the projection of the BEC density on the plane.
Key features of the latter dynamical evolutions involved the blurring 
of the vortex dynamics by the projection process coupled with 
the spontaneous vortex bending even when the different fragments 
have the appropriate phase relation to generate a vortex through 
their merging. In the 3D setting, the vorticity emerged in the 
form of vorticity sheets inducing vortex filaments (most often in 
a horizontal form) which led to pinching effects at the vortex cloud 
periphery and the formation of corresponding ridges in the atomic density profile. 

Our work is related 
to a physical mechanism that has previously been discussed in the context of 
topological defect formation and trapping during phase transitions, often referred to as the 
Kibble-Zurek (KZ) mechanism \cite{zurek,kibble}. For the case of cooling of an atomic gas through 
the BEC phase transition \cite{Anglin}, the KZ scenario involves the growth and subsequent merging 
of phase-incoherent regions of the atomic cloud, with vortices being trapped in the merging process 
as the BEC grows in size and atom number. The full KZ mechanism involves physics beyond the scope of 
the simulations of this work; however, our simulations show dynamical processes that may relate to the 
portion of the KZ mechanism involving the merging of phase-incoherent regions of condensed atoms. 
Our numerical results showing the phase relationships involved in vortex formation during merging 
are consistent with basic notions of this portion of the KZ mechanism.

There are 
many interesting questions to consider for future work in the present framework. Firstly, it is clear 
that further analysis of experimental data and further variations of simulation parameters will be needed 
in order to resolve the quantitative differences between experimental and numerical results, as discussed 
in Sec.~\ref{SEC:expt}. Perhaps additional light on this question (and an interesting
diagnostic in its own right) would be the examination of the integrated 
density along the $(x,z)$ plane which should perhaps detect some of the 
horizontal vorticity filaments illustrated herein. A particularly challenging (and more general) 
question along the same vein concerns the extent to which it may be possible to reconstruct the fully
three-dimensional cloud density from such projections.

Both from an experimental and from a theoretical point of view it would be 
interesting to extend the present considerations also to multi-component condensates in order to examine the
potential formation of vortex-like filaments and structures in settings similar to the ones presented herein (e.g., containing
fragments from different components). In the latter setting, there would exist an exciting interplay between the interference
mechanisms and the formation of the coherent structures, and the phase separation dynamics between the components; see e.g., the
recent experimental results of Ref.~\cite{our} and references therein. 

\section*{Acknowledgments}
We are thankful to Matthew Davis for providing a careful reading of 
the manuscript.
PGK and RCG acknowledge the
support of NSF-DMS-0505663. PGK also acknowledges support
from NSF-DMS-0619492 and NSF-CAREER. BPA acknowledges support 
from the Army Research Office and NSF-MPS-0354977.

\end{document}